\documentclass[useAMS,usenatbib]{mn2e}
\usepackage{epsfig}

\newcommand{\cgs}{erg~s$^{-1}$~cm$^{-2}~$}
\newcommand{\cgsd}{erg~s$^{-1}$~cm$^{-2}$~deg$^{-2}~$}
\newcommand{\thirt}{$13^H~$}
\newcommand{\xmm}{XMM-{\it Newton}}
\newcommand{\chandra}{{\it Chandra}}

\newcommand{\spitzer}{{\it Spitzer}}
\newcommand{\combo}{COMBO-17}
\newcommand{\cdfs}{CDFS}
\newcommand{\cdfn}{CDFN}
\newcommand{\xcdfs}{XMM-CDFS}
\newcommand{\ecdfs}{E-CDFS}
\newcommand{\eml}{\textsc{EMLDETECT}}
\newcommand{\ebox}{\textsc{EBOXDETECT}}
\newcommand{\detml}{\textsc{DET\_ML}}
\newcommand{\ueda}{{\em U03}}
\newcommand{\treister}{{\em T04}}
\newcommand{\gillia}{{\em GSH01A}}
\newcommand{\gillib}{{\em GSH01B}}
\newcommand{\betab}{$(\log N_H)^{\beta}$}
\newcommand{\betatwo}{$(\log N_H)^2$}
\newcommand{\betafive}{$(\log N_H)^5$}
\newcommand{\betaeight}{$(\log N_H)^8$}

\title[The distribution of absorption in the AGN population]{The distribution of absorption in AGN detected in the \xmm\ observations of the \cdfs}
\author[Dwelly]{T. Dwelly$^{1,2}$\thanks{E-mail:td@phys.soton.ac.uk},
M.J. Page$^{1}$\\ 
$^{1}$Mullard Space Science Laboratory, UCL, Holmbury St. Mary, Dorking, Surrey, RH5 6NT, UK.\\
$^{2}$School of Physics and Astronomy, University of Southampton, Southampton, SO17 1BJ, UK.}

\begin{document}

\date{\today}

\pagerange{\pageref{firstpage}--\pageref{lastpage}} \pubyear{2006}

\maketitle

\label{firstpage}

\begin{abstract}

We have used very deep \xmm\ observations of the \chandra\ Deep
Field-South to examine the spectral properties of the faint active
galactic nucleus (AGN) population.  Crucially, redshift measurements
are available for $84\%$ (259/309) of the \xmm\ sample.  We have
calculated the absorption and intrinsic luminosities of the sample
using an extensive Monte Carlo technique incorporating the specifics
of the \xmm\ observations.  Twenty-three sources are found to have
substantial absorption and intrinsic X-ray luminosities greater than
$10^{44}$~erg~s$^{-1}$, putting them in the ``type-2'' QSO regime.  We
compare the redshift, luminosity and absorption distributions of our
sample to the predictions of a range of AGN population models.  In
contrast to recent findings from ultra-deep \chandra\ surveys, we find
that there is little evidence that the absorption distribution is
dependent on either redshift or intrinsic X-ray luminosity.  The
pattern of absorption in our sample is best reproduced by models in
which $\sim 75\%$ of the AGN population is heavily absorbed at all
luminosities and redshifts.

\end{abstract}

\begin{keywords}
galaxies : active -- quasars : general -- X-rays : galaxies -- X-rays : diffuse background
\end{keywords}

\section{Introduction}

Recent X-ray spectral studies using \xmm\ and \chandra\
\citep{piconcelli03,civano05,mateos05a} have begun to unravel the
nature of the faint X-ray sources that make up the bulk of the
extragalactic X-ray background (XRB) below 10~keV. Most of these
sources are accretion powered AGN \citep{barger03,page03}.  However,
the population is far from homogeneous; the AGN exhibit a range of
spectral properties, both in X-rays, and at other wavelengths.
Unified models of AGN \citep[e.g.][]{antonucci93}, have been
reasonably successful in explaining the observational differences
between the various classes of AGN/Seyfert galaxies. The X-ray spectra
of the AGN can, in the majority of cases, be adequately described by a
power-law model, in many cases with some degree of absorption.  The
AGN with significant X-ray absorption are predominantly identified
with narrow emission line galaxies, and are seen as higher redshift,
higher luminosity analogues of nearby Seyfert 2 galaxies.  Population
synthesis models such as those of \citet{comastri95}, \citet{gilli01}
and \citet{ueda03} use a superposition of faint AGN to reproduce the
observed XRB. These models incorporate a large fraction of absorbed
AGN in order to reproduce the hard spectrum of the XRB.
\citet{gilli01} found that the XRB could be reproduced using a model
in which the absorbed and unabsorbed AGN shared a common intrinsic
X-ray luminosity function.  A similar AGN population model was found
to represent adequately the multi-wavelength properties of AGN in the
GOODS dataset \citep{treister04}.  In our previous work
\citep{dwelly05}, we compared the X-ray colour distribution of sources
detected in the \thirt\ \xmm\ deep field with the X-ray colour
distributions predicted by a number of model $N_H$ distributions. We
found that the best match to the data was made by a $N_H$ distribution
model in which absorbed and unabsorbed AGN share the same intrinsic
luminosity function, and where the number of AGN per unit $\log N_H$
is proportional to \betaeight.

However, the results from other studies based upon deep X-ray surveys
(e.g. \citealt{piconcelli03,ueda03,cowie03,barger03,barger05}), appear to
contradict some of the foundations of these simple synthesis models.
These studies all found that on average, the absorbed AGN which are 
optically identified lie at lower redshifts, and have lower
intrinsic luminosities, than their unabsorbed counterparts.

Clearly the relationship between absorption and luminosity in the AGN
population requires further examination.  Therefore, in order to
constrain the relationship between AGN luminosity, redshift and
absorption we have undertaken a study of the X-ray properties of AGN
in the \chandra\ Deep Field-South (\cdfs).  In this study we examine
the 500~ks \xmm\ European Photon Imaging Camera (EPIC) observations of
the \cdfs\ (hereafter \xcdfs) which provide a superb dataset for
measuring the spectral properties of faint X-ray sources.  The EPIC
imaging reaches to fluxes well below the break in the 2--5~keV source
counts, covers around 0.19~deg$^2$ and contains enough photons to
permit broad band X-ray spectral analysis for even the faintest
sources.  The studies of \citet{streblyanska04} and \citet{braito05}
have also used the \xcdfs\ dataset to investigate the X-ray spectra of
a number of the brighter AGN in the field.  However, until this work,
there has been no investigation of the X-ray properties of the entire
source population detected in the \xmm\ imaging.

The EPIC data are complimented by \chandra\ observations
\citep{giacconi02,lehmer05}, which provide sub-arcsecond positions for
the majority of the EPIC sources, allowing us to identify uniquely the
X-ray sources with optical counterparts.  What is more, the entire
EPIC field of view is covered by the \combo\ survey \citep{wolf04},
providing photometric redshift estimates for nearly all $R<24$ optical
counterparts.  We use an extensive Monte Carlo simulation process to
recover the $N_H$ and intrinsic luminosity of the AGN detected in our
sample. We also use the simulations to compare directly the
distribution of sources with the predictions of a number of AGN
population models.

This paper is laid out as follows.  In section \ref{sec:xmm_data} we
describe the reduction of the \xmm\ data, and the source detection
process.  In section \ref{sec:other_data} we detail how we have correlated
this with the other datasets available in the \cdfs.  In section
\ref{sec:nh_method} we introduce our Monte Carlo method for calculating
the absorption and luminosity of the sample, and
we demonstrate its fidelity.  In section \ref{sec:results} we present 
the distribution of absorption, luminosity and redshift
in the \xcdfs\ sample and compare it to the predictions 
of a number of AGN population models.
Finally, in section \ref{sec:discussion} we compare our
results to those found by other studies and discuss the implications
for AGN population models.

Throughout the paper we use a lambda-dominated flat cosmology with
$H_0 = 70$~km~s$^{-1}$Mpc$^{-1}$, $(\Omega_M,\Omega_{\Lambda }) =
(0.3,0.7)$.  $S_{E_a-E_b}$ denotes the flux of a source in the observed
$E_a$--$E_b$~keV band, corrected for Galactic absorption.  $L_{2-10}$ refers
to an object's {\em intrinsic} X-ray luminosity (that is, before
absorption), in the rest-frame 2--10~keV energy band.  $N_H$ is the
equivalent hydrogen column density in units of cm$^{-2}$.

\section{Observations and data reduction}

\subsection{The \xmm\ dataset}
\label{sec:xmm_data}
The \xmm\ data consist of two observations carried out in July 2001,
and six in January 2002.  The observations all have similar pointing
centres (approximately $RA = 03^h 32^m 27^s$, $Dec= -27\degr 48\arcmin
55\arcsec$), but the Jan 2002 observations have position angle rotated
$\sim 180\degr$ with respect to the July 2001 observations.  The \xmm\
data cover $\sim 0.19$~deg$^2$ (nearly twice the sky area of the 1Ms
\chandra\ observations), and total around 500~ks.  All three EPIC
detectors (MOS1, MOS2 and pn) were operated with the `Thin1' filters
and were in full frame mode.  The \xmm\ data were reduced using
standard Science Analysis Software (SAS, version 6.0) tasks, following
the method described by \citet{loaring05}.  After temporally filtering
periods of enhanced particle background from the event lists, we are
left with approximately 340~ks of pn, and 395~ks of MOS exposure time.
We notice an enhancement of the 0.2--0.5~keV background level for
CCD~\#5 of MOS1, therefore we have discarded all the data from this
chip in this energy range.  We note that this effect has been reported
recently by \citet{pradas05}.  We make an image, and an exposure map
for each of the EPIC cameras, in each of the
0.2--0.5, 0.5--2, 2--5 and 5--10~keV energy bands, and for each of the
eight observations.  We also produce out-of-time event images for the
pn camera in each of the four energy bands and for each observation
which are used later by the background fitting algorithm.
We have tied our coordinate system to the positions of (relatively)
bright point-like X-ray sources detected in the 1Ms \chandra\ imaging
of the field, taking into account the (-1.1\arcsec, 0.8\arcsec) offset
between the \chandra\ positions and optical counterparts
\citep{giacconi02}.  A summary of the \xmm\ observations is given in
table \ref{tab:obs_table}.

\begin{table}
\caption{Summary of the \xmm\ observations in the \cdfs, showing the
observation ID numbers, dates, exposure times, and telescope position
angles (PA).  Exposure times are given in kiloseconds for the pn and the
average of the two MOS detectors, and indicate the length of good time
remaining after removing periods of high background. }
\label{tab:obs_table}
\begin{tabular}{@{}lrrrr}
\hline
Observation & date       &  pn &      MOS  &       PA \\
\hline
0108060401  & 2001-07-27 &  13.4~ks   &  20.2~ks & 	 59\degr \\
0108060501  & 2001-07-28 &  31.7~ks   &  40.2~ks & 	 59\degr \\
0108060601  & 2002-01-13 &  42.0~ks   &  47.6~ks & 	239\degr \\
0108060701  & 2002-01-14 &  67.9~ks   &  73.3~ks & 	239\degr \\
0108061801  & 2002-01-16 &  36.5~ks   &  54.8~ks & 	239\degr \\
0108061901  & 2002-01-17 &  38.9~ks   &  42.6~ks & 	239\degr \\
0108062101  & 2002-01-20 &  40.6~ks   &  43.3~ks & 	239\degr \\
0108062301  & 2002-01-23 &  69.3~ks   &  73.5~ks & 	239\degr \\
\hline
Total       &            &  340.3~ks  & 395.5~ks &        \\
\end{tabular}
\end{table}

\begin{figure}
\includegraphics[angle=0,width=80mm]{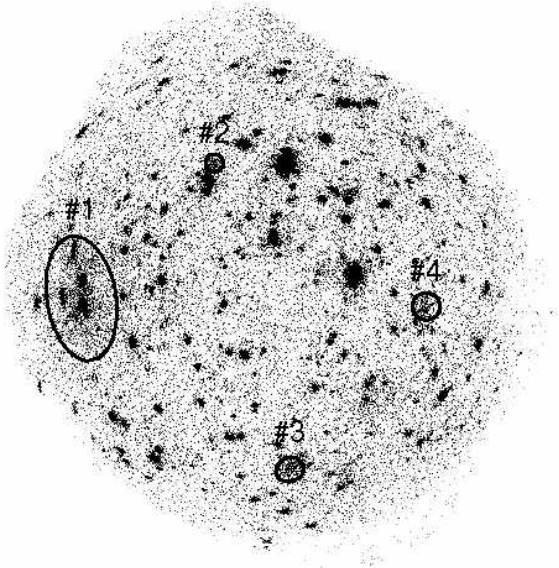}
\caption{The combined \xmm\ EPIC MOS+pn 0.5--2~keV image, background
subtracted, and displayed on a linear scale. The field of view of EPIC
is approximately 30\arcmin\ in diameter.  The four regions that we
excluded due to the presence of extended emission are shown with
numbered ellipses. These are likely to be galaxy clusters and
are discussed in Appendix \ref{app:clusters}.  }
\label{fig:soft_image}
\end{figure}

\begin{figure}
\begin{center}
\includegraphics[angle=0,width=80mm]{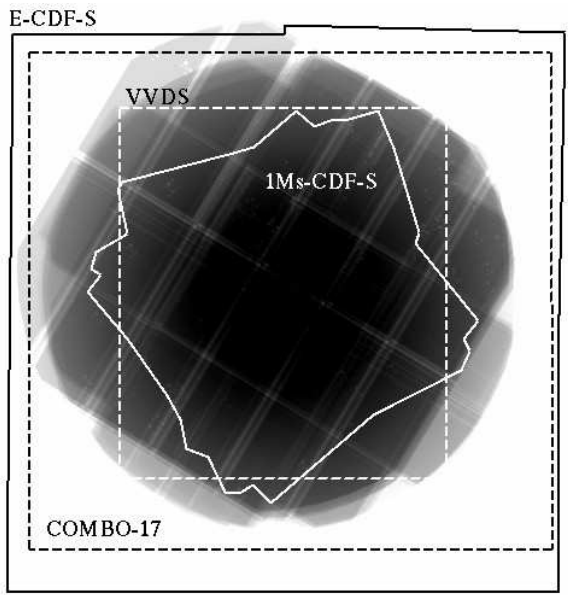}
\end{center}
\caption{The combined \xmm\ EPIC MOS+pn 0.5--2~keV exposure map. The peak
pn-equivalent exposure time is $\sim 540ks$.  We show the regions
covered by the \combo\ survey (large black dashed rectangle), the
1Ms \chandra\ imaging (white polygon), and extended \chandra\ field
(large black polygon), and the approximate area covered by the VVDS (smaller white
dashed rectangle). }
\label{fig:expmap_coverage}
\end{figure}

Visual inspection of the 0.2--0.5 and 0.5--2~keV images reveals four
regions of large scale ($>1\arcmin$ diameter) diffuse emission, the
locations of which are shown in figure \ref{fig:soft_image}.  The most
likely origin of this emission is from highly ionised gas in galaxy
groups or clusters, the study of which is outside the scope of this
AGN paper.  The background fitting algorithm we have employed is not
designed to remove diffuse emission on these scales. This unsubtracted
diffuse emission will affect our measurement of the X-ray spectral
properties of any AGN lying in these four regions, therefore we have
excluded them from further analysis.  The total sky area removed
amounts to $\sim 28$~arcmin$^2$ (4\% of the total \xmm\ sky coverage).

We have used the iterative background fitting and multi-band source
searching method to detect sources in the EPIC images. This process is
described in detail by \citet{loaring05}, but for
completeness we give a summary here.  Our method uses the SAS source
searching routines \ebox\ and \eml, together with an iterative
background fitting algorithm to detect sources simultaneously in the
multi-band EPIC images. In order to maximise the sensitivity, the
final source searching and source characterisation is carried out on
composite images (one per energy band), summed over all observations
and summed over all three EPIC detectors (MOS1, MOS2 and pn). In order
to take account of the different effective areas, exposure times, and
chip layouts/gaps of the EPIC detectors, we generate a composite
exposure map.  The contribution to this map from the MOS detectors was
scaled according to the MOS/pn response ratio.  The relative
sensitivity of the MOS and pn detectors in each energy band was
calculated with the XSPEC spectral fitting package \citep{arnaud96}
using standard on-axis EPIC pn and MOS response matrices.  A similar
process was used to determine the absolute pn count-rate to flux
conversion factors in each energy band.  For these calculations, we
assume a power-law spectrum with photon-index 1.7, corrected for
Galactic absorption of $8 \times 10^{19}$~cm$^{-2}$, \citep{rosati02}.
For the purposes of the final source detection process (in which an
off-axis dependent PSF model is used), the position of the optical
axis in the combined images is set to be the pn exposure-weighted mean
of the pointings of the eight separate observations.

For each candidate detection, the source searching routine \eml\
reports a multi-band detection likelihood parameter, \detml, where
$\detml = -lnP$.  $P$ is the probability (taking account of all four
energy bands), that a random background fluctuation would occur within
the detection element with an equal or greater number of source counts
than the candidate detection.  At the default minimum \detml\ level of
5.0, the ``raw'' \eml\ sourcelist contains 435 detections. It is not
necessary for these sources to be individually detected in all four
energy bands. At this low detection threshold, we expect a number of
spurious detections to contaminate the faint end of the \xmm\ sample;
our method for dealing with these is discussed in the next section.
A small number of the detections have very poorly
determined positions ($\sigma_{pos} > 5\arcsec$), or have poorly
determined extent (the 90\% error on the measurement of the extent is
greater than the extent itself), and so we remove these from our
sourcelist.  
We define three hardness ratios,

\begin{eqnarray}
HR1 & = & \frac{R_{0.5-2}-R_{0.2-0.5}}{R_{0.5-2}+R_{0.2-0.5}} \nonumber \\ 
HR2 & = & \frac{R_{2-5}-R_{0.5-2}}{R_{2-5}+R_{0.5-2}}\\
HR3 & = & \frac{R_{5-10}-R_{2-5}}{R_{5-10}+R_{2-5}} \nonumber
\end{eqnarray}
where $R_{E_{a}-E_{b}}$ is the combined MOS+pn source count rate, corrected for
vignetting, in the $E_a$ to $E_b$ (keV) energy band.

\section{Matching to \chandra\ and optical catalogues}
\label{sec:other_data}
\subsection{Cross correlation with \chandra\ observations of the field}
The original 1Ms \chandra\ imaging of the \cdfs\ covers the central
part of the \xmm\ field of view (FOV) to great depth
\citep[][see fig. \ref{fig:expmap_coverage}]{giacconi02,alexander03}.  
Recently, the \chandra\ sky coverage
of the \cdfs\ has been increased by a mosaic of four 250~ks \chandra\
pointings: the Extended \chandra\ Deep Field-South (\ecdfs)
\citep{lehmer05}.  We used the higher positional accuracy of the
\chandra\ observations to aid unambiguous optical identification of
the \xmm\ sources.  We matched the \xcdfs\ detections to sources in a
combined \chandra\ catalogue constructed from the catalogues of
\citet{giacconi02} and \citet{lehmer05}.  We have taken 
into account the (-1.1\arcsec, 0.8\arcsec) offset
between the \chandra\ positions and optical counterparts
in the \citet{giacconi02} catalogue.  For those \chandra\ sources
which appear in both the \citet{giacconi02} and \citet{lehmer05}
catalogues, we mainly used the positions from the former. The point spread
function of the \xmm\ EPIC detectors is strongly off-axis angle
dependent \citep{gondoin00}.  At large off-axis angles the azimuthal
component of the PSF becomes rather extended, whereas the radial
component remains relatively constant.  Therefore, we match the
\xcdfs\ detections to \chandra\ counterparts using an ellipsoidal
region.  The semi-major axis of this ellipse is increased from
5\arcsec\ for sources at the centre of the field, up to a maximum of
10\arcsec\ for sources at off axis angles greater than 15\arcmin.  The
semi-minor axis is kept constant at 5\arcsec, and is oriented parallel
to the line joining the source position and the nominal optical axis of the
EPIC-pn detector.  In addition, for the few \xcdfs\ detections which
\eml\ determines to be slightly extended, we increase both semi-axes
of the search ellipse by the measured extent.  This choice of X-ray
position matching criteria is discussed further in Appendix
\ref{app:pos_diffs}.

\begin{figure}
\begin{center}
\includegraphics[angle=0,width=80mm]{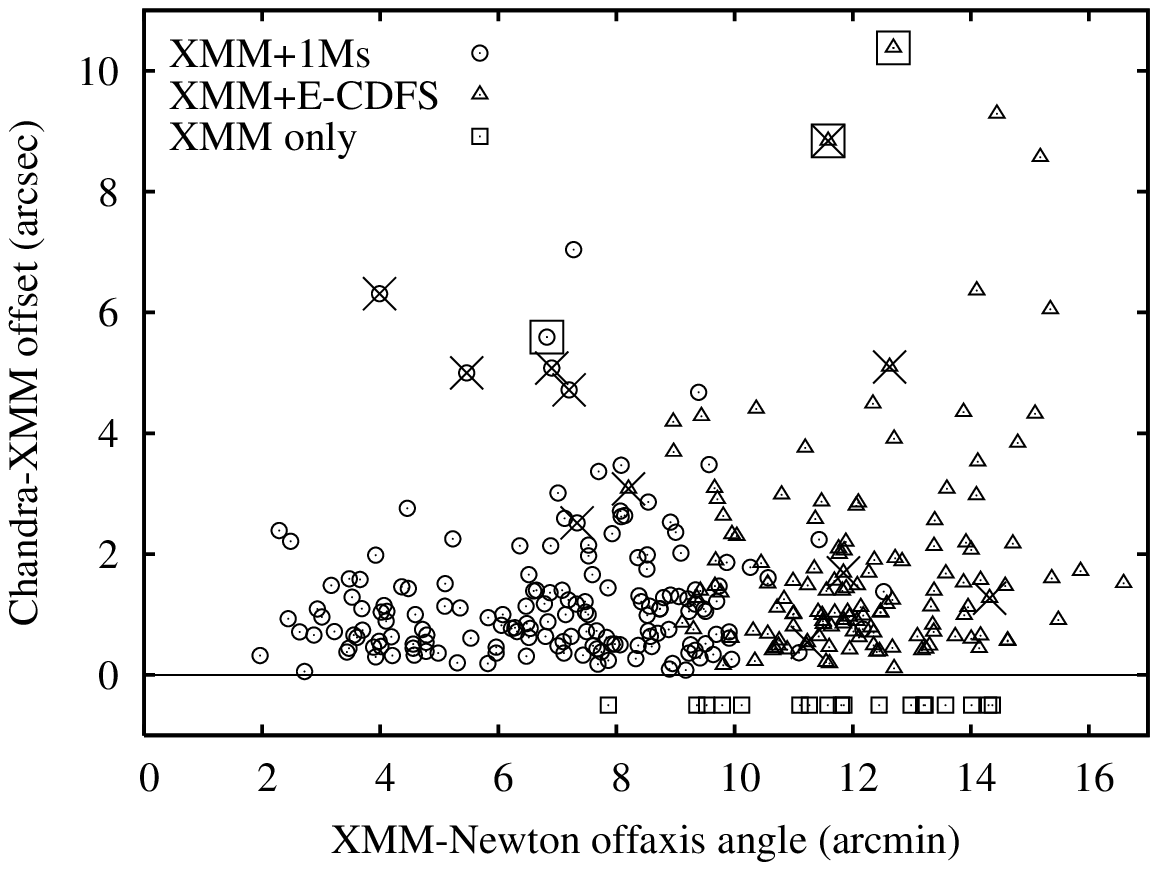}
\includegraphics[angle=0,width=80mm]{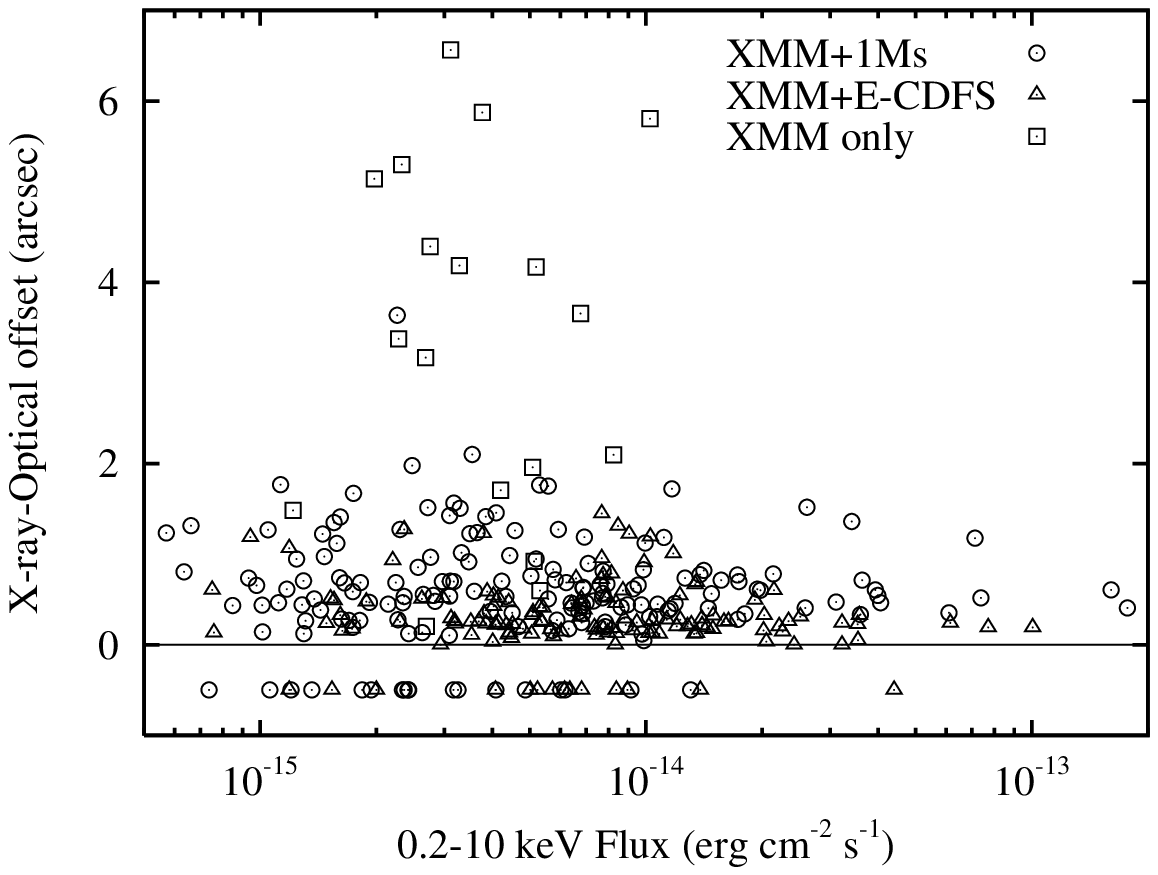}
\end{center}
\caption{{\em Upper panel:} The differences between the positions
determined using \xmm\ and \chandra\ as a function of \xmm\ off-axis
angle.  The \xcdfs\ sources with \chandra\ counterparts are shown with
circles for matches to \chandra\ 1Ms sources, and triangles for
matches to \ecdfs\ sources. Those \xmm\ sources manually matched to a \chandra\ 
source are highlighted with boxes.  \xcdfs\ sources which do not have
\chandra\ counterparts are shown with small square symbols at offset =
-0.5..  The ``confused'' \xmm\ detections are marked with crosses.
{\em Bottom panel:}
The differences between the X-ray position and the position of the
optical counterpart, as a function of \xmm\ 0.2--10~keV flux.  The
\xcdfs\ sources which have \chandra\ counterparts are shown with
circles (1Ms matches) and triangles (\ecdfs\ matches), and those which
do not have \chandra\ counterparts are shown with small square
symbols.  \xcdfs\ sources having no optical counterpart within the
matching region are placed at offset =-0.5. }
\label{fig:xmm_cha_offset}
\end{figure}

Using these positional criteria, we find that 330 of the 431 \xcdfs\
detections are matched to \chandra\ sources; 185 of these matches are
to sources in the 1Ms \chandra\ catalogue, and 145 are to sources in
the \ecdfs\ catalogue. For the \xcdfs\ sources with no \chandra\
counterpart, we have manually examined the \xmm\ and \chandra\
images. We find that in 15 cases, there is a nearby \chandra\
counterpart just outside the matching ellipse. We adopt the \chandra\
positions for these sources.  Figure \ref{fig:xmm_cha_offset} shows the
positional offsets between these matched \xcdfs\ and \chandra\
sources.

The determination of the \xmm\ detection likelihood limit is a balance
between the desire to include as many sources in our sample as
possible, against the need to minimise the number of spurious
detections.  We could simply reject all \xmm\ detections that are not
matched to \chandra\ sources.  However, whilst the headline flux
limits achieved by the \chandra\ observations are fainter than those
for the \xcdfs, the coverage is not uniform over the \xmm\ FOV.  What
is more, the relative sensitivity of the \xmm\ and \chandra\ detectors
varies with energy.  In particular, \xmm\ EPIC is much more sensitive
than \chandra\ at very high photon energies ($> 5$~keV), and at very
low photon energies ($< 0.5$~keV), so sources having either very hard
or very soft spectra will be preferentially detected with \xmm.  The
\xmm\ and \chandra\ observations of the \cdfs\ span approximately four
years, therefore intrinsic variability on timescales of several months
to years could also account for sources appearing in some catalogues
and not others. For these reasons, and because this study is based
primarily upon \xmm\ data, we curtail our \xcdfs\ sourcelist purely on
the basis of the \xmm\ detection likelihood.  However, we set the
level of this such that approximately $90\%$ of the \xmm\ detections
have \chandra\ counterparts.  At a detection likelihood threshold of
8.5, we find that there are 335 \xmm\ detections and 302 ($90.1\%$) of
these have at least one \chandra\ counterpart.

There are 16 cases where a \xcdfs\ detection has more than one
\chandra\ source inside (or very close to) the matching ellipse.  In
order to determine whether these are genuinely confused sources, we
have manually inspected the \xmm\ images, the 1Ms \chandra\
images\footnote{http://www.astro.psu.edu/users/niel/hdf/hdf-chandra.html}
\citep{alexander03}, and the \ecdfs\
images\footnote{http://www.astro.psu.edu/users/niel/ecdfs/ecdfs-chandra.html}
\citep{lehmer05}.  In one case, the ``confusion'' appears to be the
result of a single real astrophysical source appearing in both the 1Ms
and \ecdfs\ \chandra\ catalogues. We chose the \ecdfs\ source in this
case.  We find that in the other cases, the \xmm\ detection is matched
to a clearly separated pair of \chandra\ sources.  However, for four
of these, there is a large brightness contrast between the two
\chandra\ sources, and so we do not consider the \xcdfs\ source to be
``confused''.  We have removed from our sample the remaining eleven
truly ``confused'' \xmm\ detections because their \xmm\ determined
properties are superpositions of more than one real astrophysical
source.  The small number of detections that we have had to remove
demonstrates that source confusion plays only a small role ($\sim 3\%$
of sources) in (high Galactic latitude) \xmm\ surveys of several
hundred kiloseconds.

\subsection{Optical counterparts and redshifts}
\label{sec:optical_counterparts}

The \cdfs\ has been the target for a number of space and ground based
deep optical/infrared imaging campaigns, which cover different parts
of the field to different depths
\citep[e.g.][]{arnouts01,wolf04,giavalisco04}.  A large amount of
VLT-FORS time has been expended on optical spectroscopy of
counterparts to X-ray sources detected in the 1Ms \chandra\ survey
\citep{szokoly04}.  \citet{zheng04} used these spectroscopic
identifications together with optical and NIR measurements to estimate
redshifts for virtually all of the X-ray sources in the 1Ms \chandra\
catalogue of \citet{giacconi02}.  Initially, we adopted the
\citet{zheng04} estimates for all of the 167 \xcdfs\ detections
matched to sources in the 1Ms \chandra\ catalogue.  However, as noted
by \citet{barger05}, a number of the \citet{zheng04} optical
counterparts have relatively large offsets from the X-ray source
positions.  We have manually examined the \chandra\ {\em vs} optical
positions for the sources where the optical position stated by
\citet{zheng04} is more than 2\arcsec\ from the 1Ms \chandra\
position, or where there is more than one \combo\ source within
2\arcsec\ of the 1Ms \chandra\ position.  We have drawn upon the
\chandra\ 1Ms images \citep{alexander03}, the \ecdfs\ images
\citep{lehmer05}, the \combo\ optical
images\footnote{http://www.mpia-hd.mpg.de/COMBO/combo\_CDFSpublic.html}
\citep{wolf04}, and the GEMS/GOODS ACS
images\footnote{ftp://archive.stsci.edu/pub/hlsp/gems/}
\citep{caldwell06}, to choose the most likely optical counterparts.
For sixteen cases (i.e. 10\% of those sources that were examined), we
decided that an alternative optical source was a more likely
counterpart to the X-ray source.  All of these alternative optical
counterparts are closer to the X-ray position than the counterpart
chosen by \citet{zheng04}, and most are optically fainter. The ID
numbers of these sources in \citet{zheng04} are 3, 17, 23, 25, 36,
61, 64, 70, 97, 99, 213, 517, 528, 548, 591 and 641.  Where possible,
(four cases) we have used the \combo\ redshift estimates for our
preferred counterpart. Otherwise we consider the X-ray source to be
optically unidentified.  We note that for another six of the
\citet{zheng04} sources, the correct spectroscopic redshift is quoted,
but an incorrect optical position is stated.  These were all cases
where the X-ray sources had multiple optical counterparts listed in
\citet{szokoly04}.

We note that several of the \xcdfs\ detections matched to 1Ms \chandra\
sources have faint ($R>25$) optical counterparts for which photometric
redshifts have been calculated by \citet{mainieri05b}.  However, in
each case, the \citet{mainieri05b} redshift estimate is in agreement
with the \citet{zheng04} value within the errors. So for simplicity
and consistency, we prefer to adopt the \citet{zheng04}
redshift estimate.

For the \xcdfs\ detections having a counterpart in the \ecdfs\
\chandra\ catalogue, and where an optical counterpart is found, we
adopt the optical counterpart position from
\citet{lehmer05}.  We use the \combo\ photometric redshift if there is
an object in the \combo\ catalogue within 1\arcsec\ of the optical
position given by \citet{lehmer05}. We find that in several cases, the
optical counterpart has also been spectroscopically identified
(i.e. it appears in the VVDS catalogue of \citealt{lefevre04}, or the
\citealt{szokoly04} field galaxy list), and so for these \xcdfs\
sources we adopt the spectroscopic identifications.

For the 33 \xcdfs\ detections having no \chandra\ counterpart, we have
attempted to assign an optical counterpart from the \combo\ catalogue.
Our starting point was to choose the optically brightest source inside
the variable matching ellipse discussed earlier. We then manually
examined the X-ray and optical images to determine if this was the
correct choice.  For most of these sources, we adopted the initial
choice of counterpart. However, for two sources, we chose an
alternative optical counterpart because of a co-location with an
enhancement in the \ecdfs\ image. For 13 of the \xcdfs\ sources
without \chandra\ counterparts, the \xmm\ detection is most likely due
to diffuse emission from a group or cluster. That is, there are
several galaxies having similar photometric redshifts located close to
the \xmm\ position.  We have excluded these detections from our final
\xcdfs\ sample, as they are unlikely to be AGN.  There is one \xcdfs\
detection located away from the centre of a bright ($R \sim
17$) face-on spiral galaxy.  
The soft X-ray colours of this source,
together with its non-detection by \chandra\ suggest that it is likely
to be due to diffuse emission, and so we remove this detection from
the sample.  Finally, there were two \xcdfs\ sources located at the
edge of the \xmm\ FOV, outside the \ecdfs\ and \combo\ coverage. For
simplicity, we have removed these two sources from our sample.
The lower panel of figure 
\ref{fig:xmm_cha_offset} shows the X-ray-optical position differences 
for all the \xcdfs\ sources as a function of X-ray flux. 

We note that the \xcdfs\ sample contains several low redshift sources
with fluxes that imply very low luminosities ($L_{2-10} \sim
10^{40}$~erg~s~${-1}$).  It is therefore feasible that our sample
contains a small number of ultra luminous X-ray (ULX) sources.
\citet{lehmer06} have recently used \chandra\ deep field data to look
for low X-ray luminosity sources lying off-axis in low redshift,
optically bright galaxies.  The \citet{lehmer06} sample contains 8
objects within the area covered by the \xmm\ observations of the
\cdfs, but only one of these (J033234.73-275533.8 at $z=0.038$) is associated with
an \xcdfs\ source.  We exclude this object from our sample because it
is unlikely to be an AGN.

\begin{figure*}
\begin{center}
\includegraphics[angle=270,width=160mm]{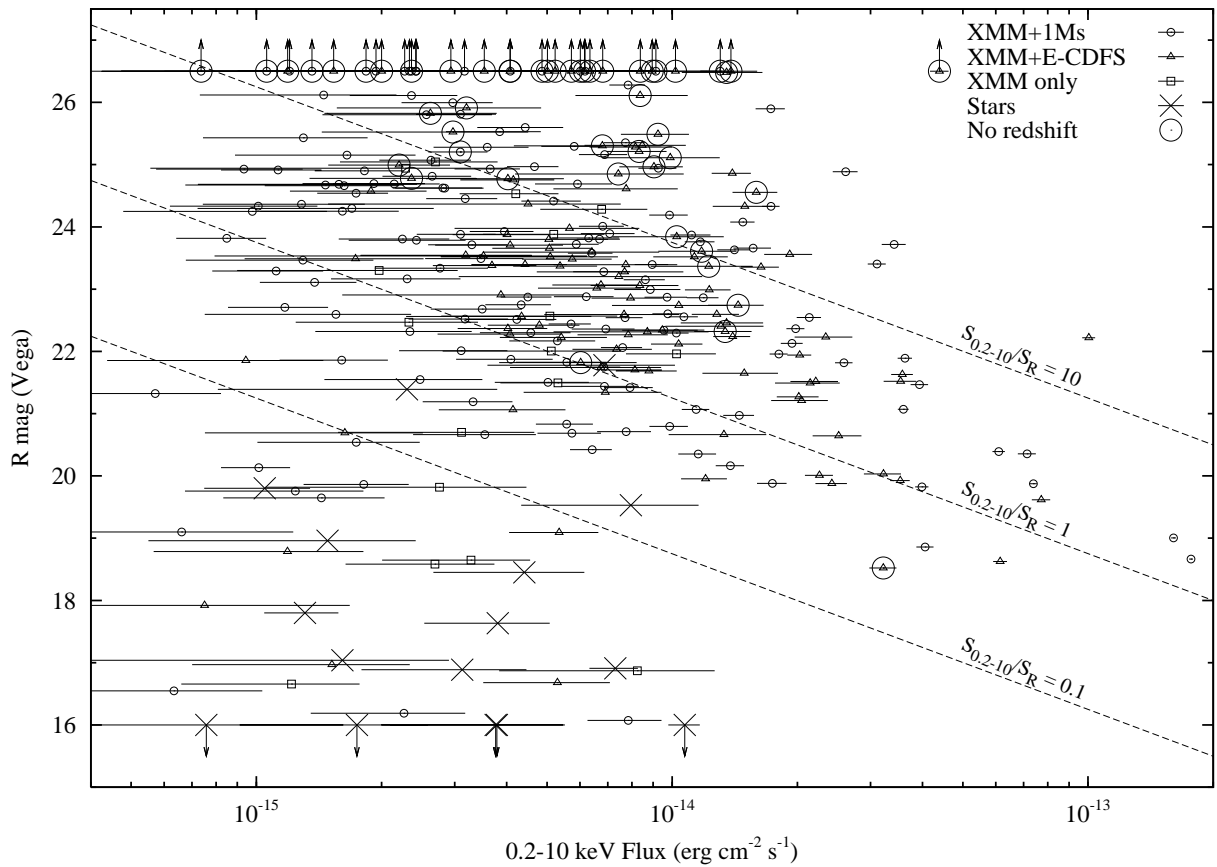}
\end{center}
\caption{Distribution of 0.2--10~keV flux {\em vs} $R$ band optical
counterpart magnitudes for the point like sources in the \xcdfs\
sample. Objects identified as Galactic stars are indicated with
crosses. \xcdfs\ detections with \chandra\ counterparts are shown with
small circles (1Ms matches), and small triangles (\ecdfs\ matches).
\xcdfs\ detections without \chandra\ counterparts are shown with small
boxes.  \xcdfs\ detections with no optical counterpart, or a
counterpart fainter than $R=26.5$ are marked with upward pointing
arrows. Very optically bright counterparts are placed at $R = 16$, and
marked with downward facing arrows.  Objects with no redshift estimate
are highlighted with large circles.  Optical magnitudes are taken from
\citet{wolf04}, \citet{giacconi02}, and \citet{lehmer05}.  Dashed
lines show several values of X-ray-to-optical flux ratio
$S_{0.2-10}/S_R$, where the optical flux $S_R$ is calculated from the
R-band magnitude: $S_R = 10^{-5.5 - 0.4R}$~\cgs\ \citep{barger03}.
}
\label{fig:flux_vs_R}
\end{figure*}

In summary, after applying our \xmm\ detection likelihood and
positional criteria, and after removing the detections which are
confused or unlikely to be point sources, there are 309 sources in the
\xcdfs\ sample.  Of these, 291 have \chandra\ counterparts, 278 are
matched to optical counterparts, and 259 (84\%) have optical
spectroscopic identifications and/or photo-z estimates.  Fifteen of
these are associated with Galactic stars, and one with a candidate ULX
in a low redshift galaxy.  Figure \ref{fig:flux_vs_R} compares the
optical magnitudes of the \xcdfs\ sources to their 0.2--10~keV X-ray
flux. Note that over a third (109/309) of the \xcdfs\ sources are
optically faint ($R>24$).

\section{Estimating the intrinsic properties of the sample using X-ray colours and Monte Carlo simulations}
\label{sec:nh_method}
It has been shown by several authors \citep[e.g.][]{mainieri02,
dellaceca04,perola04,dwelly05} that X-ray hardness ratios can be
utilised to determine the spectral properties of faint \xmm\ sources,
and in particular, the amount of absorption.  This approach relies on
the observed source belonging to some assumed family of spectral
types, for which the spectral parameters can be deduced.  Spectral
analyses of relatively bright AGN have shown that nearly all can be
broadly described by a spectral model consisting of a primary
power-law with slope $\Gamma \sim 1.9$, attenuated with some absorbing
column of neutral material \citep[e.g.][]{piconcelli03,page06}.  A number
of additional spectral model components are sometimes required to
provide the best fits to the highest signal to noise AGN spectra,
although these extra components are generally much less important than
the primary power-law component when considering broad band X-ray
colours.  However, in our previous work \citep{dwelly05}, we found
that we were able to provide a better match to the X-ray colours of
AGN detected in the \thirt\ deep \xmm\ field by including an
unabsorbed cold reflection component to the AGN spectral model; we use
this as our baseline spectral model. The most important effect of the
reflection component is to harden the spectrum at high energies ($E >
5$~keV), and thus it reduces the amount of absorption needed to
explain the X-ray spectrum of AGN with hard-sloped spectra.

The traditional approach to estimate the absorption in faint X-ray
detected AGN, is to fit the observed multi-band X-ray hardness ratios
to a model spectrum using a spectral fitting package such as XSPEC.
However, there is a large degree of degeneracy in such an approach
because of the number of fitted spectral parameters ($N_H$ and/or
$\Gamma$), compared to the limited number of data points (it is
typical for authors to use just a single hardness ratio measure
between the 0.5--2 and 2--10~keV energy bands).  We have devised a
novel Monte Carlo approach, in which we deduce the intrinsic
properties of sources in the \xcdfs\ sample by comparing them to the
``output'' properties of a library of simulated AGN.  This method
allows us to take account of the scatter of sources in multiple $HR$
space (which may be strongly asymmetric, and is dependent on the
observations), and so allows a rigorous estimation of confidence
intervals.  What is more, we can use our simulation method to compare
directly the source distributions predicted by various synthesis
models, against those seen in our sample. We first summarise the
method used to generate a simulated library of sources, and then
describe the absorption and intrinsic luminosity estimation processes.

This approach allows us to treat all the identified sources in the
\xcdfs\ sample in a consistent, uniform fashion, as opposed to
examining the brighter sources using a different method to the fainter
sources.

\subsection{Monte Carlo simulations of the AGN populations}
\label{sec:simulations}
A detailed description of our Monte Carlo simulation process is given
in \citet{dwelly05}. Here we give a short summary.  A population of
AGN is randomly generated according to a model of the intrinsic X-ray
luminosity function (XLF). To calculate the expected number of input
AGN per field, the model XLF is integrated over the ranges
$0.015<z<5.0$, and $10^{40} < L_X < 10^{48}$~erg~s$^{-1}$ 
(extrapolating from published models where necessary).  For each
of these AGN we assign a random redshift and luminosity, where the
probability of a source having a particular value of $z$ and $L_X$ is
taken from the XLF model.  Each AGN is then randomly assigned values
of $N_H$ and $\Gamma$ according to the respective model distributions.
The $z$, $L_X$, $N_H$ and $\Gamma$ for each source are converted to
multi-band \xmm\ EPIC count-rates using the spectral model together
with the EPIC response matrices.  We adopt a spectral model consisting
of a primary transmitted component and a component reflected from
neutral material.  The primary component is modelled as a power-law
with an exponential cutoff ($E_{\mathrm{cutoff}}$ = 400~keV), absorbed
by a neutral column of material. The reflection component is
calculated from the {\em pexrav} model of \citet{magdziarz95}, and we
set the reflecting material to cover $\pi$ steradians, to be inclined
at $30$ degrees to the viewing direction, and to have solar
abundances.

We then simulate how this population would appear if it had been
observed in the same manner as for the \xmm\ observations of the
\cdfs.  We generate simulated images separately for every combination
of the four energy bands, three EPIC cameras, and eight observations,
totalling 96 images per simulated field.  These images incorporate the
effects of the EPIC point spread function and detector response, use
realistic exposure maps, and include a background at the same level as
is observed in the real data. We sum these images over all
observations and for the MOS1, MOS2 and pn cameras to produce one
image per energy band.  We mask out the four regions in the simulated
images which are affected by diffuse emission in the real data.
Background maps are calculated independently for each of the 96
images, (as for the real data) and then summed to produce one
background map per energy band.  We use the tasks \ebox\ and \eml\ to
search the images for sources in the four energy bands simultaneously.
The resultant output sourcelists are curtailed using the same
detection likelihood and positional accuracy criteria as for the real
\xmm\ dataset.

We use the off-axis angle dependent matching ellipse described in
section \ref{sec:other_data} to match output detections to input
sources.  To exclude output detections which are affected by
confusion, we flag all output detections which are matched to two or
more input sources having comparable (within a factor of five)
full-band count-rates. This step mimics the method we used to find
confused sources in the real \xcdfs\ sample.

\subsection{Constructing the library of simulated sources}
\label{sec:library}
We have used the Monte Carlo method to generate a large reference
library of simulated sources, and have used the following constituents
to the AGN population model.  We use the Luminosity Dependent Density
Evolution (LDDE) XLF model of \citet{ueda03}. The latter was fitted
only over the $10^{41.5} \le L_{2-10} \le 10^{46.5}$~erg~s$^{-1}$
range and to $z = 3$.  In order to cover the range of luminosities
expected in the \xcdfs\ sample, we have extrapolated this XLF model
down to $L_{2-10} = 10^{40}$~erg~s$^{-1}$, and out to $z=5$.  We use a
model $N_H$ distribution in which the number of AGN per unit $\log
N_H$ is proportional to \betaeight, and is independent of luminosity
and redshift.  To recreate the intrinsic scatter in the spectral
slopes of AGN \citep{piconcelli03,page06}, we assign a randomly
selected $\Gamma$ to each simulated source.  These spectral slopes are
randomly chosen from a Gaussian distribution centred on $\Gamma=1.9$,
with $\sigma_{\Gamma} = 0.2$, with the additional constraint that $1.2
< \Gamma < 2.6$.  We adjusted the absolute normalisation of the model
XLF such that the sky density of sources with 0.5--2~keV flux $\ge 2
\times 10 ^{-15}$~\cgs\ is similar to that measured in the \xcdfs\
sample (after removing confirmed non-AGN sources).  A total of 2000 fields 
worth of simulations were carried out
to generate the simulated source library; enough to ensure that
redshift, luminosity and $HR$ space is well populated with simulated
sources, but not so large as to consume a prohibitive amount of
processing time.  We have verified that this population model broadly
reproduces the source counts of the extragalactic sources in the
\xcdfs\ sample.

\subsection{Models of the AGN $N_H$ distribution}
\label{sec:nh_models}
During this study, we test the predictions of a number of model $N_H$
distributions, which are described below.

The ``\betab'' $N_H$ distribution models: In these models the number
of AGN with absorption $N_H$, per unit $\log N_H$, is proportional to
\betab, and is not dependent on redshift or luminosity.  We have
tested three variations by setting the parameter $\beta$ to 2, 5, and
8. A similar parameterisation of the $N_H$ distribution was introduced
in the XRB synthesis models of \citet{gandhi03}.

The ``\treister'' $N_H$ distribution model: \citet{treister04},
introduce an $N_H$ model in which the number of AGN having a
particular value of absorption is based on a model in which the
density of the obscuring torus decreases with distance away from its
plane.  The torus geometry in this model is independent of redshift
and luminosity.

The ``\gillia'' and ``\gillib'' $N_H$ distribution models:
\citet{gilli01} investigated the ability of two absorption
distribution models to reproduce both the shape of the XRB and the
AGN source counts below 10~keV.  In both models, the distribution of
$N_H$ within the absorbed AGN was taken to be the same as that
observed in nearby Seyfert 2 galaxies \citep{risaliti99}. In model
``A'', the number ratio between AGN with $N_H \ge 10^{22}$~cm$^{-2}$
and those with $N_H<10^{22}$~cm$^{-2}$ is fixed to be 4. In model
``B'' the ratio increases with redshift; at $z=0$ the ratio is 4, and
at $z\ge1.32$ the ratio is 10.

The ``\ueda'' $N_H$ distribution model: \citet{ueda03} fitted a
luminosity dependent model to the distribution of $N_H$ in their AGN
sample. In this model, the fraction of AGN having
$N_H>10^{22}$~cm$^{-2}$ decreases linearly with luminosity, from $\sim
0.6$ of AGN with $L_{2-10} \le 10^{43.5}$~erg~s$^{-1}$, to $\sim 0.4$
of AGN with $L_{2-10}=10^{45}$~erg~s$^{-1}$.

\subsection{Absorption estimation technique}
\label{sec:nh_est_method}
The measured properties of the \xcdfs\ sources which we use to
estimate the absorption are the redshift ($z$), the vignetting
corrected count rate in the 0.2--10~keV band ($R_{0.2-10}$), and three
hardness ratios ($HR1$, $HR2$, and $HR3$). The absorption and
luminosity of each real source is estimated from those simulated
library sources which have similar values of $z'$, $R'_{0.2-10}$,
$HR1'$, $HR2'$, and $HR3'$.  The following process is carried out for
each optically identified source in the \xcdfs\ sample.

We first select the objects from the simulated library which have
similar redshift ($|z - z'| \leq (1 + z) \times 0.1$), a similar full
band count-rate, $0.5 < R'_{0.2-10}/R_{0.2-10} < 2$, and have similar
hardness ratios to the real \xcdfs\ source ($|HR1 - HR1'| \leq 0.1$,
$|HR2 - HR2'| \leq 0.1$, and $|HR3 - HR3'| \leq 0.2$). The weaker
constraint on $HR3$ reflects the poorer counting statistics at harder
energies.  We compensate for the possible influence that the shape of
the baseline $N_H$ distribution used to generate the simulated library
may have on the estimation process.  Statistical weights are
calculated by counting, for a large number of bins in $N_H$, the
numbers of library sources which satisfy the redshift and countrate
criteria.  The weight of each selected object is the inverse of the
number counted in the $N_H$ bin in which it lies.  A sliding box
technique is then used to estimate the absorption in the real source
from the $N_H$ values and weights of the selected objects.  We choose
the $N_H$ value where the sum of the statistical weight inside a
sliding box of width 0.25~dex is maximised.  The confidence interval
is taken to be the range of $N_H$ about this peak which contains 68\%
of the statistical weight of the selected objects.

\subsection{Intrinsic luminosity estimation technique}
We estimate the intrinsic, rest-frame 2--10~keV luminosity
($L_{2-10}$) of the real sources using a similar technique as that
used to estimate absorption.  For each \xcdfs\ source, we select the
subset of sources from the simulated library which have similar
redshifts, count rates, and hardness ratios (using the same criteria as
before).  We can account for the differences between the
$z,R_{0.2-10}$ of the real \xcdfs\ source, and the $z',R'_{0.2-10}$ of
each of the selected library sources: the $L'_{2-10}$ of the library
sources are corrected by factors of $R_{0.2-10}/R'_{0.2-10}$ and
$d^2_{L}(z')/d^2_{L}(z)$ (where $d_{L}$ is the luminosity distance).
We then take the median of the corrected $L'_{2-10}$ of the selected
subset of simulated sources as our estimate of the intrinsic
luminosity of the real source.  The confidence interval is given by
the range of corrected $L'_{2-10}$ about the median value which
contains 68\% of the simulated subset.

\subsection{Fidelity of the $N_H/L_X$ estimation technique}
We have measured the efficacy of our absorption/luminosity estimation technique by
quantifying both how well it is able to estimate the
$N_H/L_X$ values of individual sources, as well as how well it can recover
an $N_H/L_X$ distribution of a population of sources.

\subsubsection{Ability to recover $N_H/L_X$ of individual sources}
\label{sec:individual}
We constructed a test population of simulated AGN using the method
described in \ref{sec:library}. The equivalent of one hundred \xcdfs\
fields were generated.  For each test source, we made an estimate of
absorption and intrinsic luminosity using our $N_H/L_X$ estimation
technique, in the same way as we would for the real sources.  The
estimated $N_H/L_X$ values for each test source are then compared to
the input parameter values.

\begin{figure}
\begin{center}
\includegraphics[angle=0,width=41.5mm]{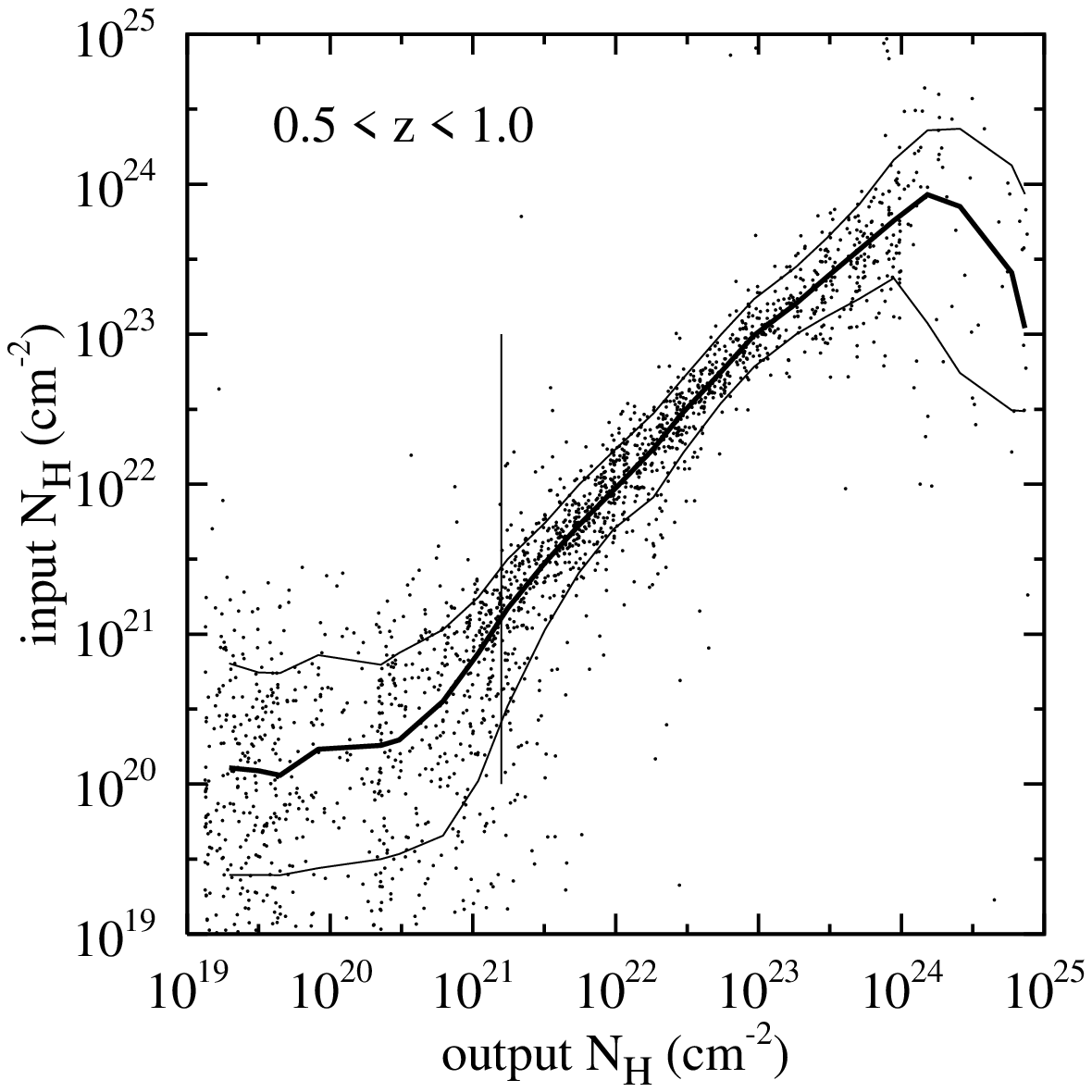}
\includegraphics[angle=0,width=41.5mm]{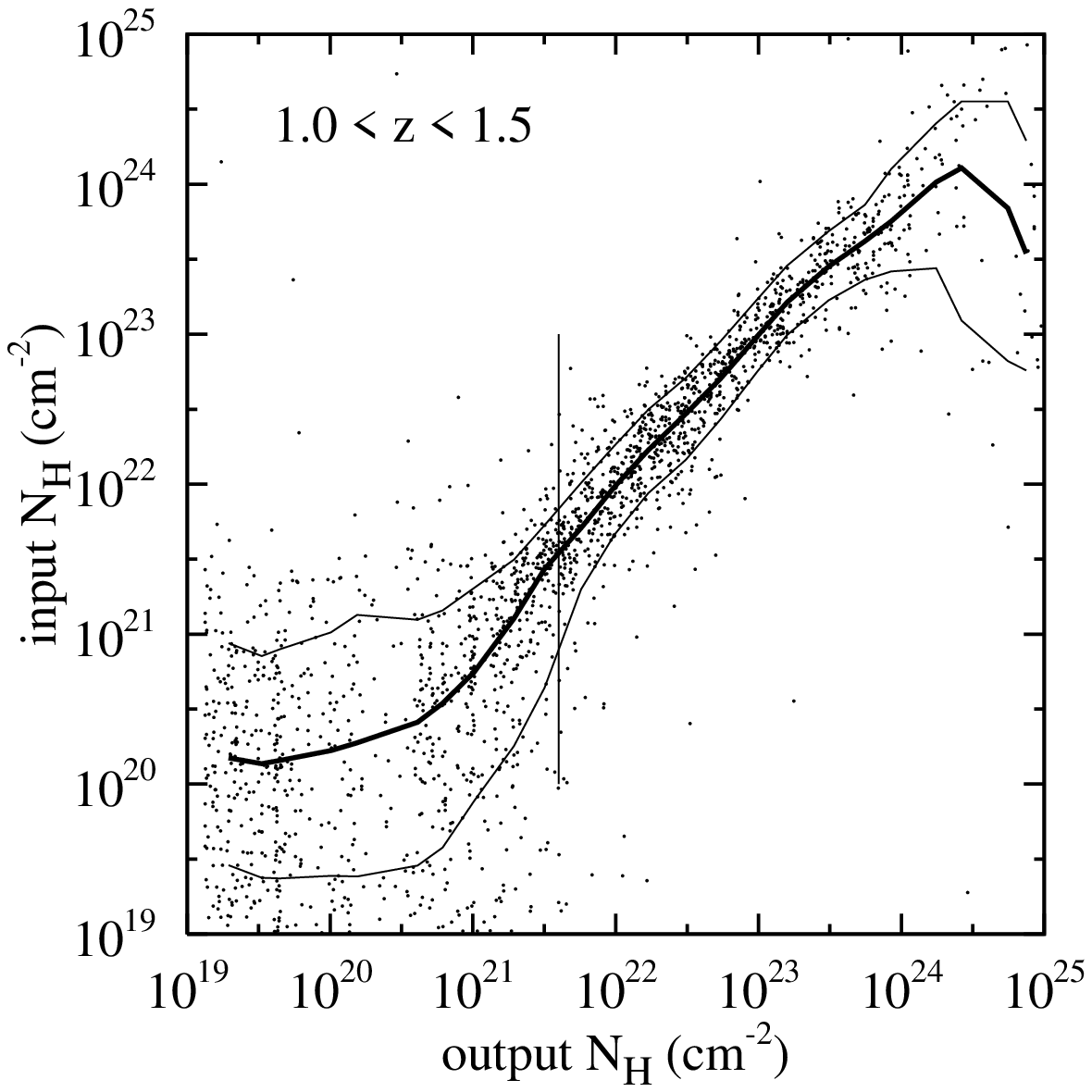}
\includegraphics[angle=0,width=41.5mm]{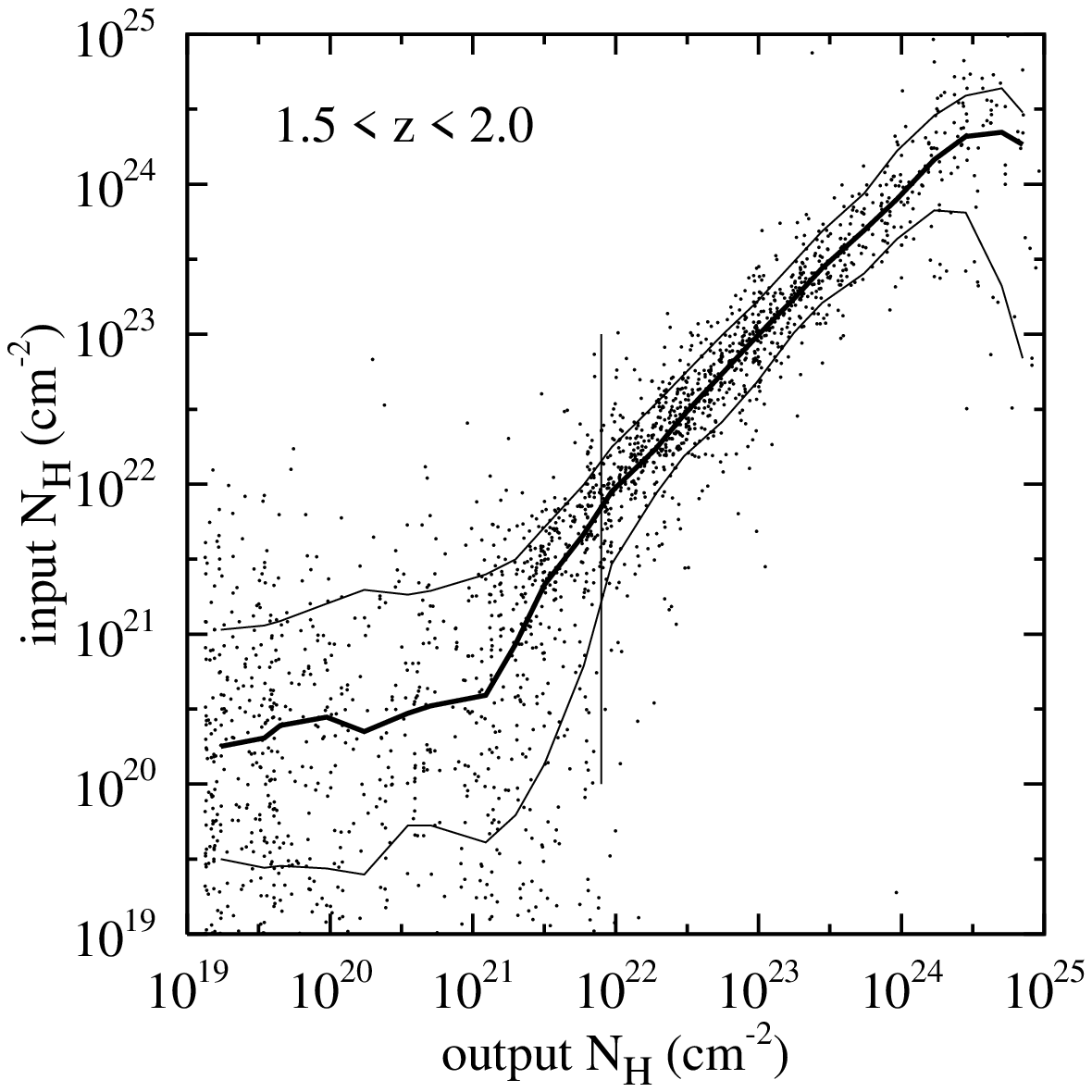}
\includegraphics[angle=0,width=41.5mm]{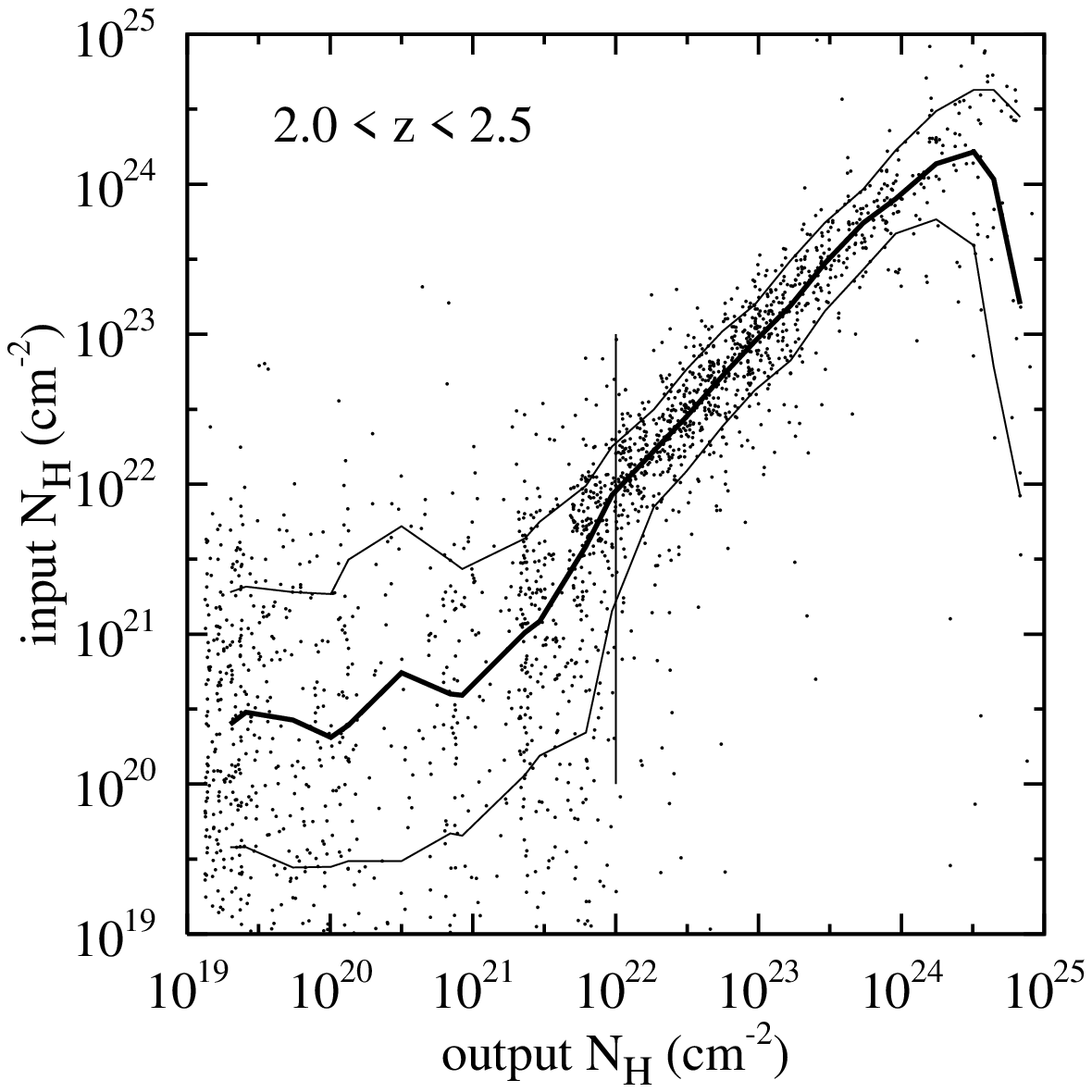}
\end{center}
\caption{An illustration of the fidelity of our absorption estimation
technique evaluated from a population of simulated test sources. The panels 
show the degree of scatter of output $N_H$ about input $N_H$ values for test sources in 
four redshift bins. The bold curves show the median input
$N_H$ as a function of output $N_H$. The thin curves
show the degree of scatter (they contain 68\% of the test
sources).  The vertical lines show the lowest output $N_H$ value
for which this scatter is less than $\pm 0.5$~dex in each redshift bin.  }
\label{fig:nh_det_vs_nh_inp}
\end{figure}

Figure \ref{fig:nh_det_vs_nh_inp} shows the relationship between the
estimated and input $N_H$ for test sources in a number of redshift
ranges.  The technique recovers the input $N_H$ values very well for
test sources having moderate to heavy absorption. However, for low
absorbing columns, the scatter increases rapidly.  This becomes
increasingly apparent at higher redshifts, as more and more of the
absorption is shifted out of the EPIC bandpass.  We have calculated
the level below which less than 68\% of test sources have $N_H$ estimates within
0.5 dex of the input value.  This ranges from $10^{21.1}$~cm$^{-2}$
for sources in the $0<z<0.5$ redshift range, up to $10^{22.6}$~cm$^{-2}$ for
the $3<z<4$ range.

The estimation technique becomes less accurate at very high levels of
absorption. This is not unexpected; for all but the highest redshift
AGN having this level of absorption, virtually all the flux has been
removed below 5~keV.  Thus $HR1$ and $HR2$ contain little information,
meaning that we have less diagnostic power to determine the amount of
absorption. What is more, at these high column densities, the effects
of Compton scattering, which are not included in our spectral model,
will become significant in the spectra of the real sources.  We are
therefore cautious about the exact $N_H$ for any sources estimated to
have an absorbing column of greater than $10^{24}$~cm$^{-2}$, but we
can be confident that such objects are very heavily absorbed.
However, because of X-ray selection effects, we expect our sample to
contain rather few of these very heavily absorbed AGN.

\begin{figure}
\begin{center}
\includegraphics[angle=270,width=80mm]{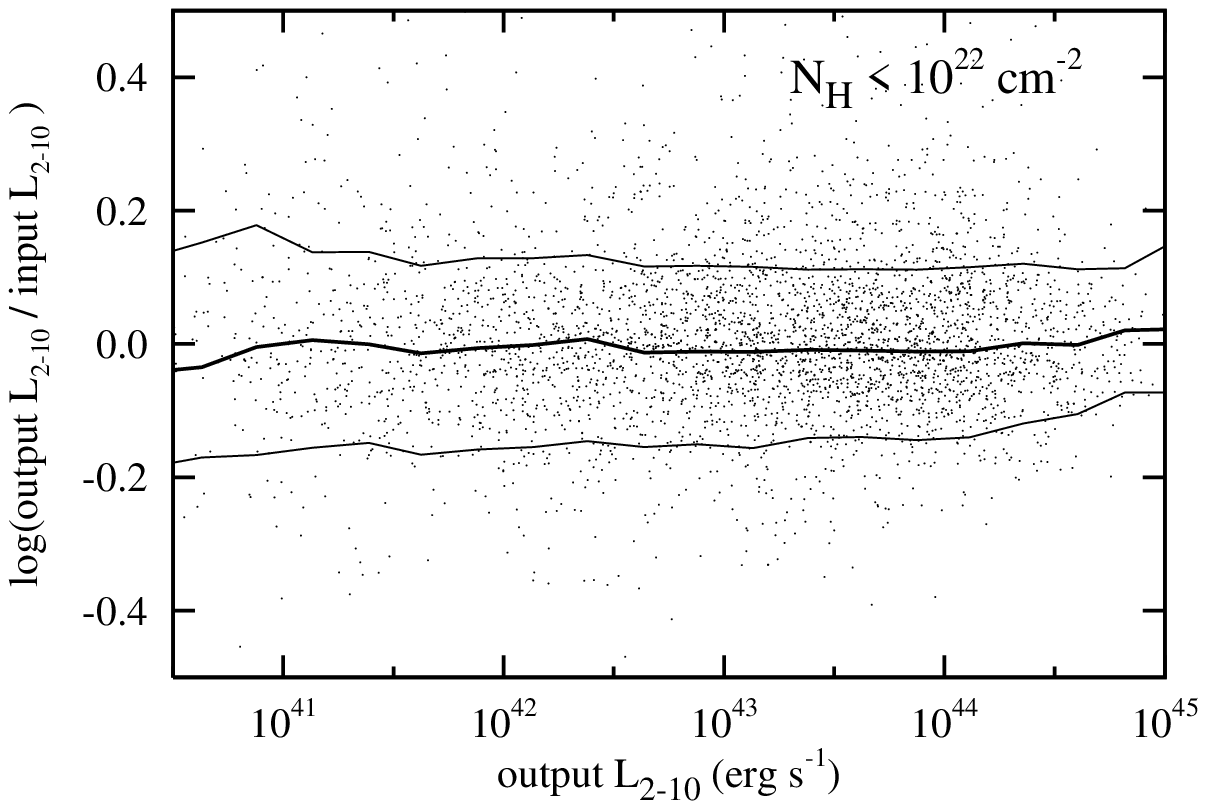}
\includegraphics[angle=270,width=80mm]{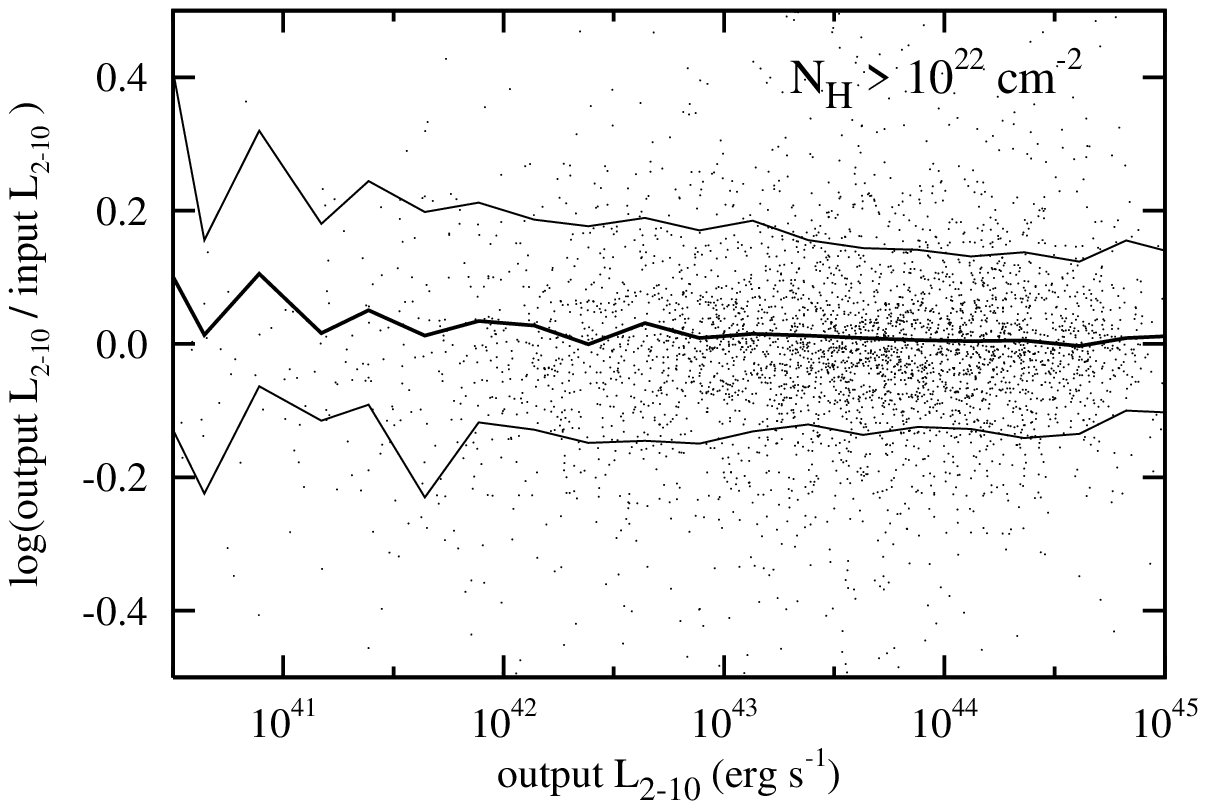}
\end{center}
\caption{An illustration of the fidelity of the intrinsic luminosity
estimation technique, showing the difference between the estimated and
input intrinsic luminosities of the test sources as a function of
their estimated intrinsic luminosity.  The lines show the median input
luminosity and the levels which contain 68\% of the test sources as a
function of estimated intrinsic luminosity. The upper panel is for
test sources with $N_H < 10^{22}$~cm$^{-2}$, and the lower panel for
test sources with $N_H > 10^{22}$~cm$^{-2}$.}
\label{fig:lum_det_vs_lum_inp}
\end{figure}

In figure \ref{fig:lum_det_vs_lum_inp} we show the relationship
between estimated and input intrinsic luminosity for the test
sources. The high fidelity of the technique is evidenced by the
low scatter of points about the one-to-one relation (less than $\pm
0.2$ dex for most of the luminosity range).

\subsubsection{Ability to recover $N_H/L_X$ distributions of a population of sources}
We have investigated how well the estimation technique can recover an
input model absorption distribution. In addition, we have checked that
the initial choice of AGN population model used to generate the
simulated source library does not have a major effect on the estimated
$N_H/L_X$ distributions.

\begin{figure}
\begin{center}
\includegraphics[angle=270,width=80mm]{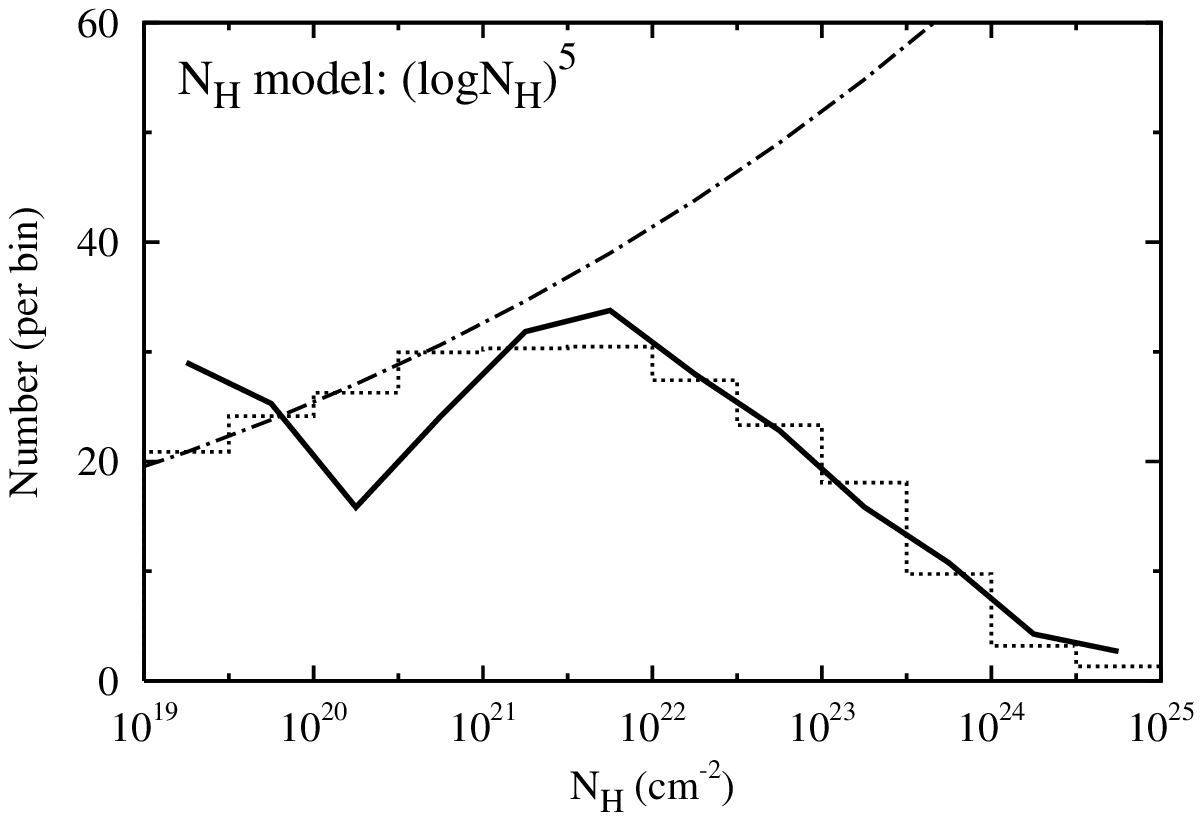}
\includegraphics[angle=270,width=80mm]{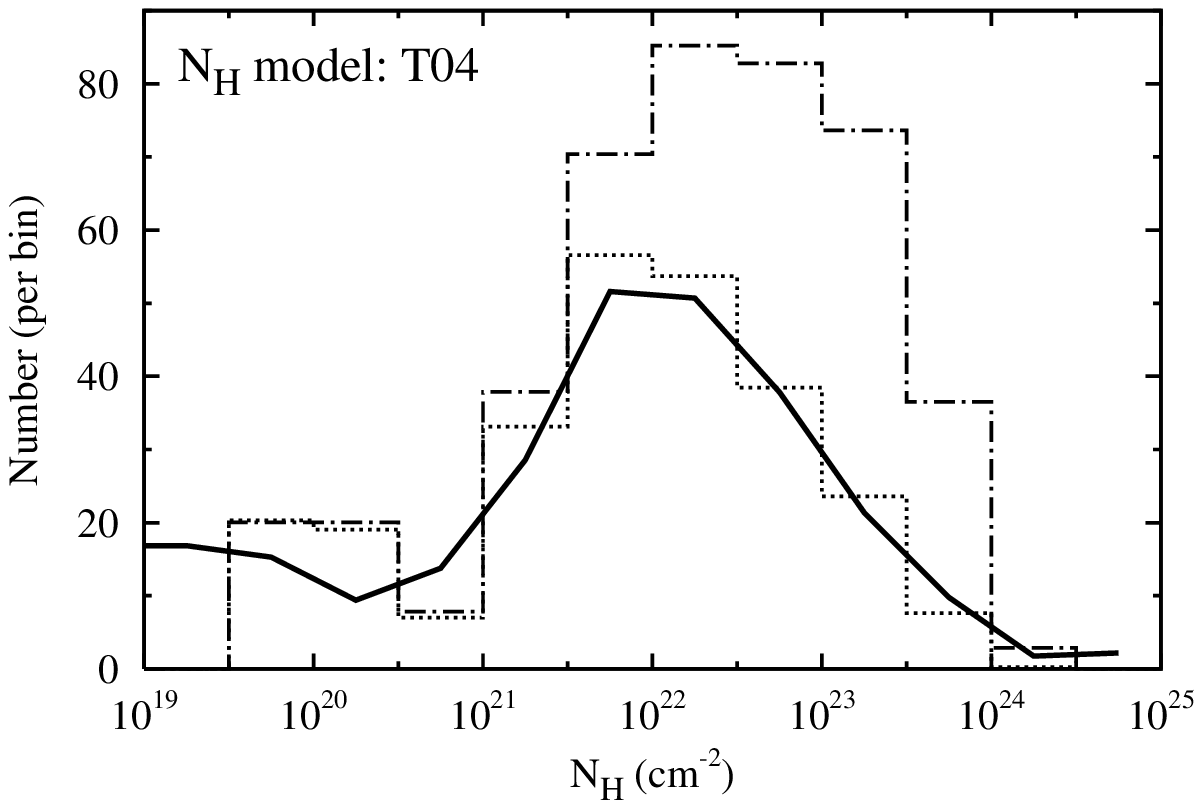}
\includegraphics[angle=270,width=80mm]{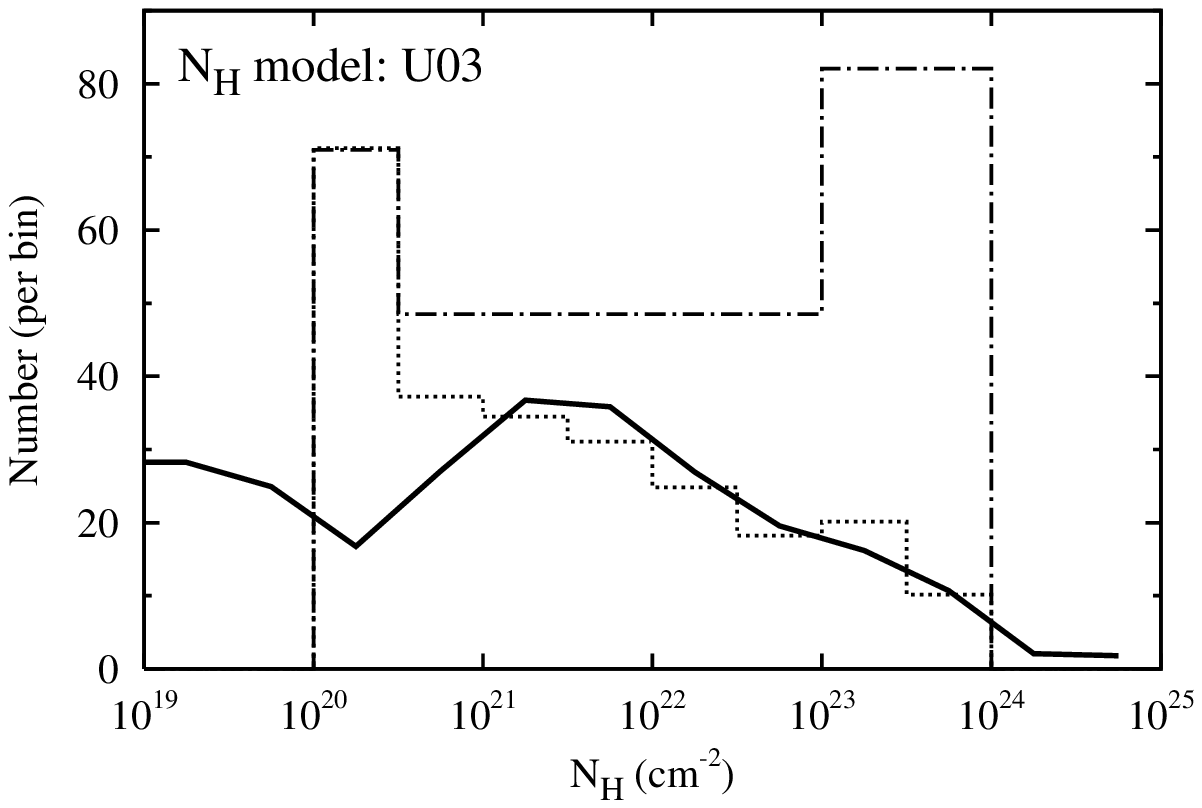}
\end{center}
\caption{The ability of the $N_H$ estimation process to recover a the
distribution of $N_H$ in a population of test sources. We show three
test populations generated according to the XLF model of
\citet{ueda03}. The upper panel shows the \betafive\ model $N_H$
distribution, the centre panel shows the \treister\ $N_H$
distribution, and the bottom panel shows the luminosity dependent
\ueda\ $N_H$ distribution.  In each panel, the dot-dash solid line
shows the shape of the input $N_H$ model The dotted line is the
distribution of $N_H$ in the test sources output from the Monte Carlo
simulations.  The solid line is the distribution of $N_H$ which is
recovered by applying our $N_H$ estimation technique to these output
test sources.  }
\label{fig:nh_dist_est}
\end{figure}

We extend the method of section \ref{sec:individual} to generate
several simulated test populations, each of which is based upon a
different input $N_H$ model.  The $N_H$ distribution of each
population is then recovered using our $N_H$ estimation technique.
Figure \ref{fig:nh_dist_est} shows a comparison of the input and
recovered absorption distributions for three different input $N_H$
models. We show the input distribution of absorption in the test
model, as well as the distribution in the sources that are output by
the Monte Carlo simulation process. Heavily absorbed sources are
less likely to be detected by the \xmm\ observations than sources with
lower absorbing columns, this selection effect results in the
differences between the input and output distributions (see section
\ref{sec:selection}).  Above column densities of $\sim
10^{21}$~cm$^{-2}$, the distribution of absorption recovered by our
$N_H$ estimation method is a good approximation to the ``true''
$N_H$ distribution in the output test population.  There are
disparities at low absorbing columns, where the estimation
method can provide only weak constraints on the $N_H$ values of the
test sources. However, the {\em total} number of output test sources 
having estimated $N_H \le 10^{21}$~cm$^{-2}$ is
consistent with the total number having ``true'' $N_H \le 10^{21}$~cm$^{-2}$.

\begin{figure}
\begin{center}
\includegraphics[angle=270,width=80mm]{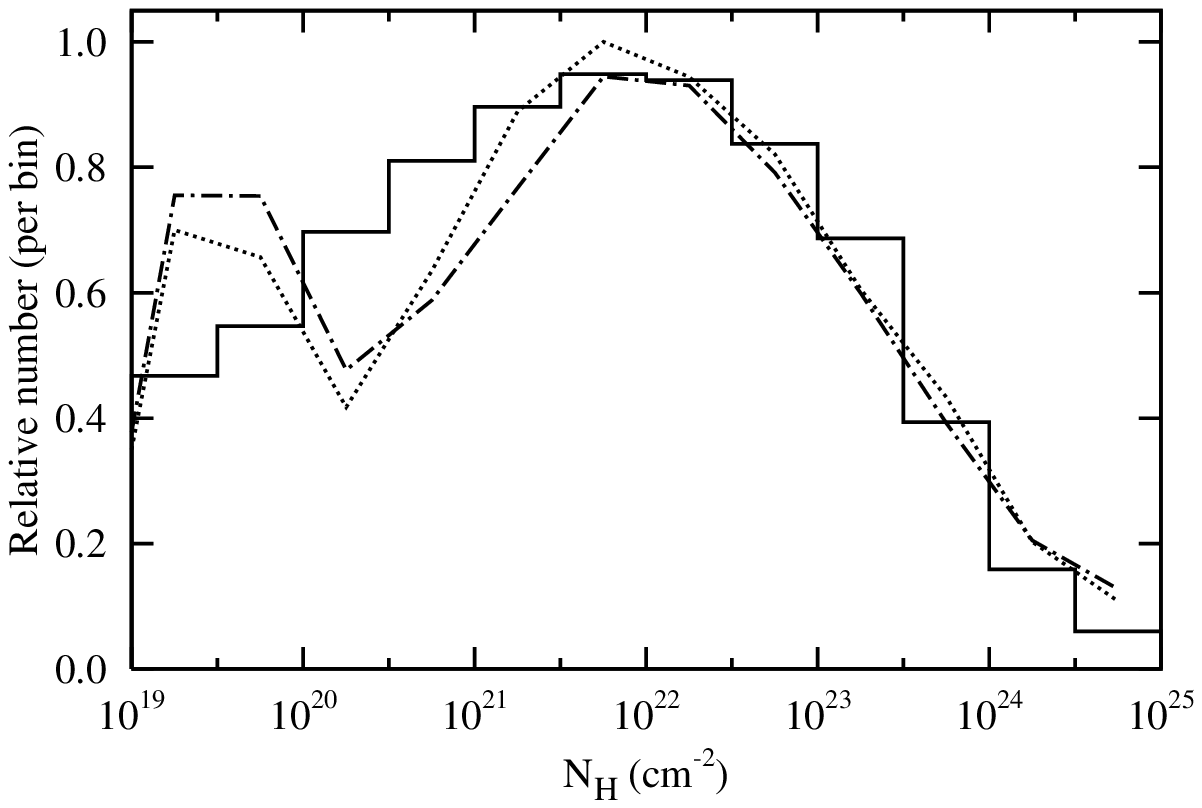}
\includegraphics[angle=270,width=80mm]{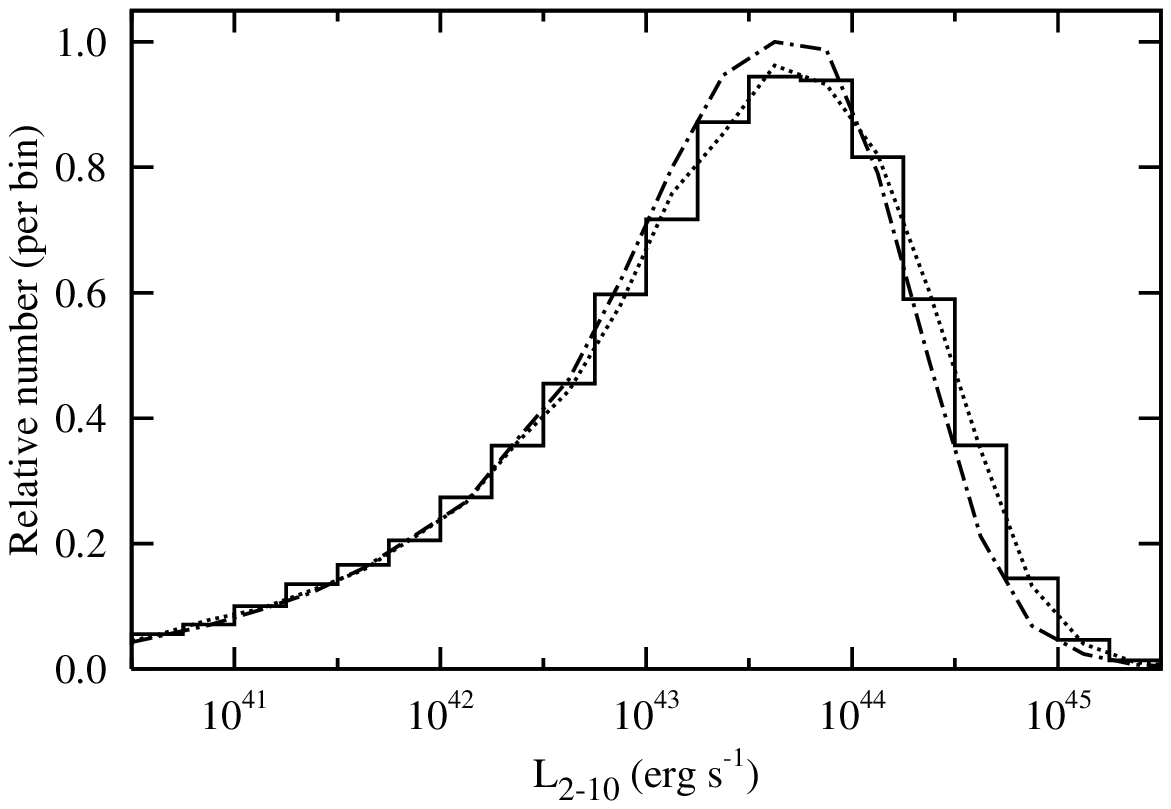}
\end{center}
\caption{Plots demonstrating the independence of our $N_H/L_X$
estimation technique from the X-ray luminosity function model used to
generate the simulated source library. In the top panel, the solid
line shows the distribution of ``true'' $N_H$ for a population of test
sources, in this case generated using the \betaeight\ model
together with the model XLF of \citet{ueda03}. The dotted line is the
$N_H$ distribution estimated when we use the simulated source library
generated from the model XLF of \citet{ueda03}. The dot-dash line
shows the estimated distribution when we use a second source
library which was generated from the ``LDDE1'' model XLF of
\citet{miyaji00}. The lower panel shows the equivalent plot for 
the input and estimated distributions of intrinsic luminosity.}
\label{fig:nh_lum_compare_libraries}
\end{figure}	

We have investigated whether the luminosity function used to construct
the simulated source library has an effect on the outputs of our
$N_H/L_X$ estimation process.  We first generated a test population of
AGN distributed in redshift/luminosity space according to the XLF
model of \citet{ueda03}, and with an absorption distribution following
the \betaeight\ model.  We then used our $N_H/L_X$ estimation technique
to recover the absorption and luminosity of these test sources. This
was carried out twice, firstly using the original simulated source
library (generated according to the model XLF of \citealt{ueda03}), and
secondly using a new simulated source library in which the sources are
distributed according to the ``LDDE1'' XLF of Miyaji, Hasinger \&
Schmidt (2000). Note that the ``LDDE1'' XLF model is defined in the
observed 0.5--2~keV band, with no correction of luminosities for
absorption.  For the purposes of this study we convert the 0.5--2~keV
observed frame luminosities of the \citet{miyaji00} model to rest
frame 2--10~keV luminosities assuming a mean power law spectrum with
slope $\Gamma = 1.9$.  This conversion does assume that the
\citet{miyaji00} sample is predominantly AGN unabsorbed in the X-rays
(see section 3.1 of \citealt{miyaji00}).  Figure
\ref{fig:nh_lum_compare_libraries} shows the resultant $N_H$ and $L_X$
distributions recovered using the two different source libraries.  The
differences between the recovered $N_H$ distributions are relatively
small: for accurately measurable absorbing columns
($10^{21.5}<N_H<10^{24.5}$~cm$^{-2}$) they agree to better than
10\%. We compare this to the Poisson noise of $\ge 15\%$ in the 0.5
dex wide bins of $N_H$ measured in the \xcdfs\ sample.  We conclude
therefore that our $N_H$ estimation technique is not strongly
dependent on the AGN population model chosen to generate the simulated
source library.  However, we note that for $L_{2-10} >
10^{44}$~erg~s$^{-1}$, there is a significant difference between the
luminosity distributions recovered by the two simulated source
libraries (which have markedly different redshift/luminosity
distributions).  In order to mitigate this effect, when applying our
estimation technique to the real \xcdfs\ sample, we should use a
simulated source library in which the sources have a broadly similar
luminosity and redshift distribution to the sources in the sample.
Therefore, for the remainder of this study, we have used a simulated
source library that is generated according to the model XLF of
\citet{ueda03}, as described in section \ref{sec:library}.

\begin{figure*}
\begin{center}
\includegraphics[width=55mm,angle=270]{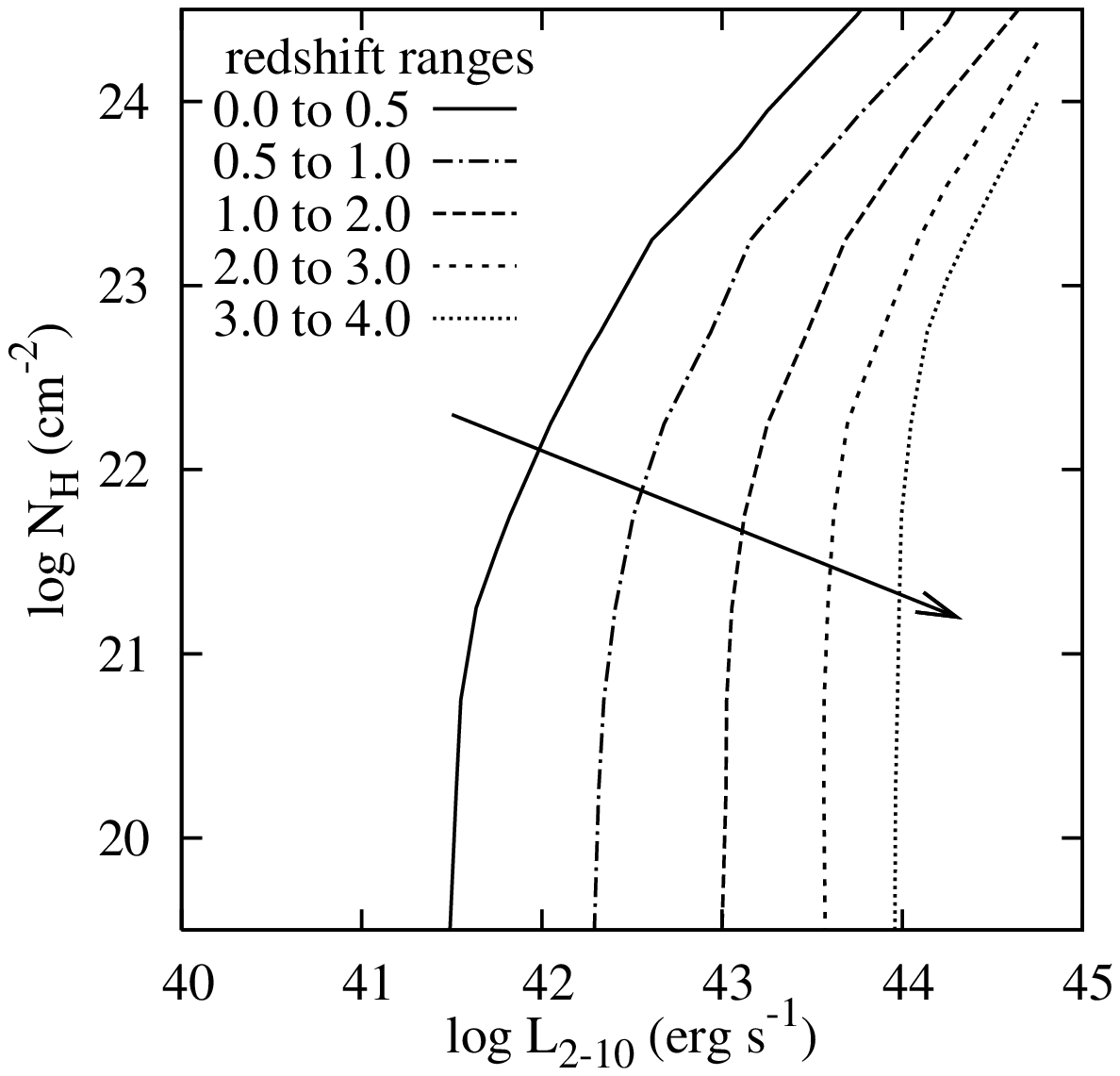}
\includegraphics[width=55mm,angle=270]{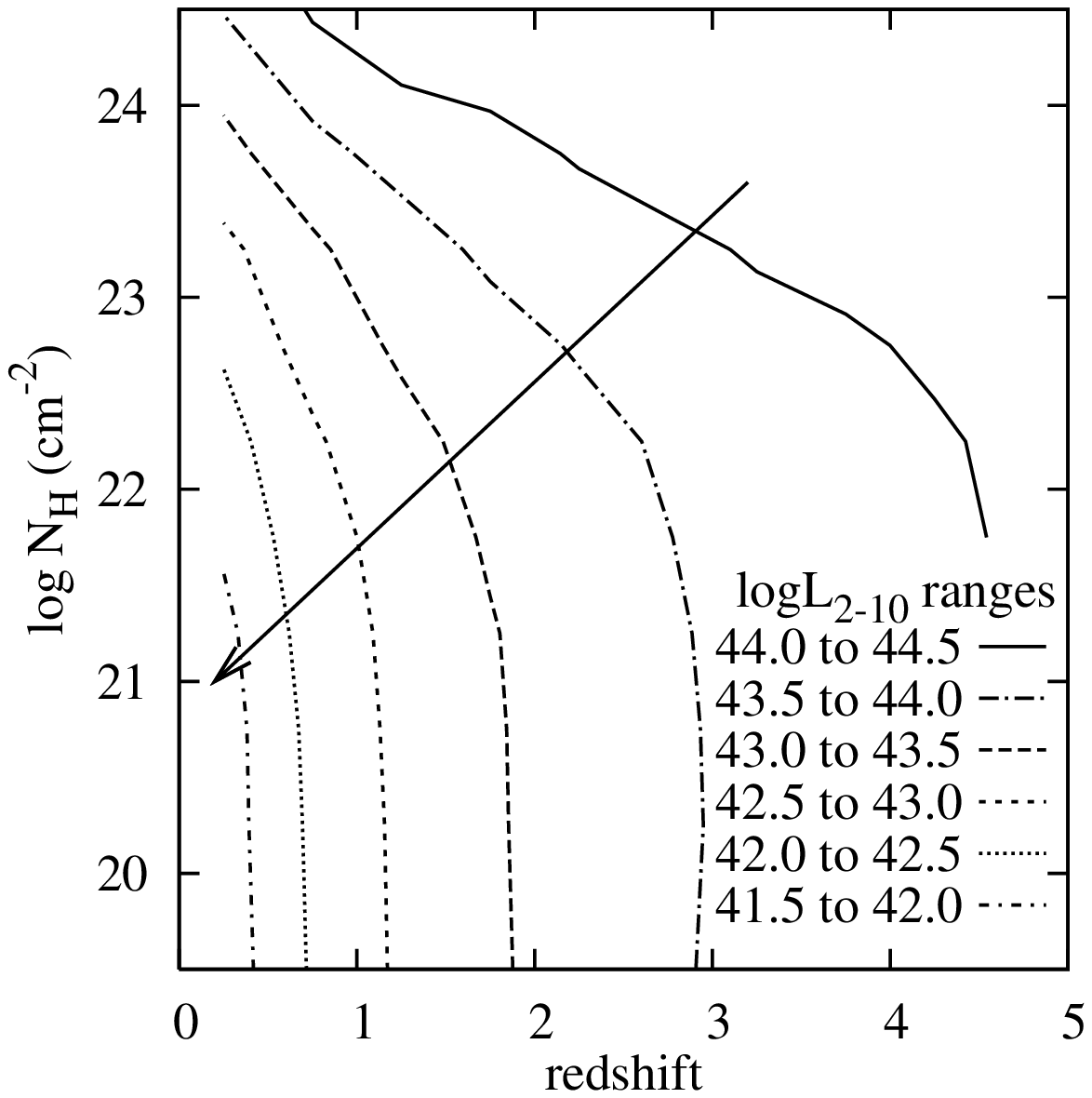}
\includegraphics[width=55mm,angle=270]{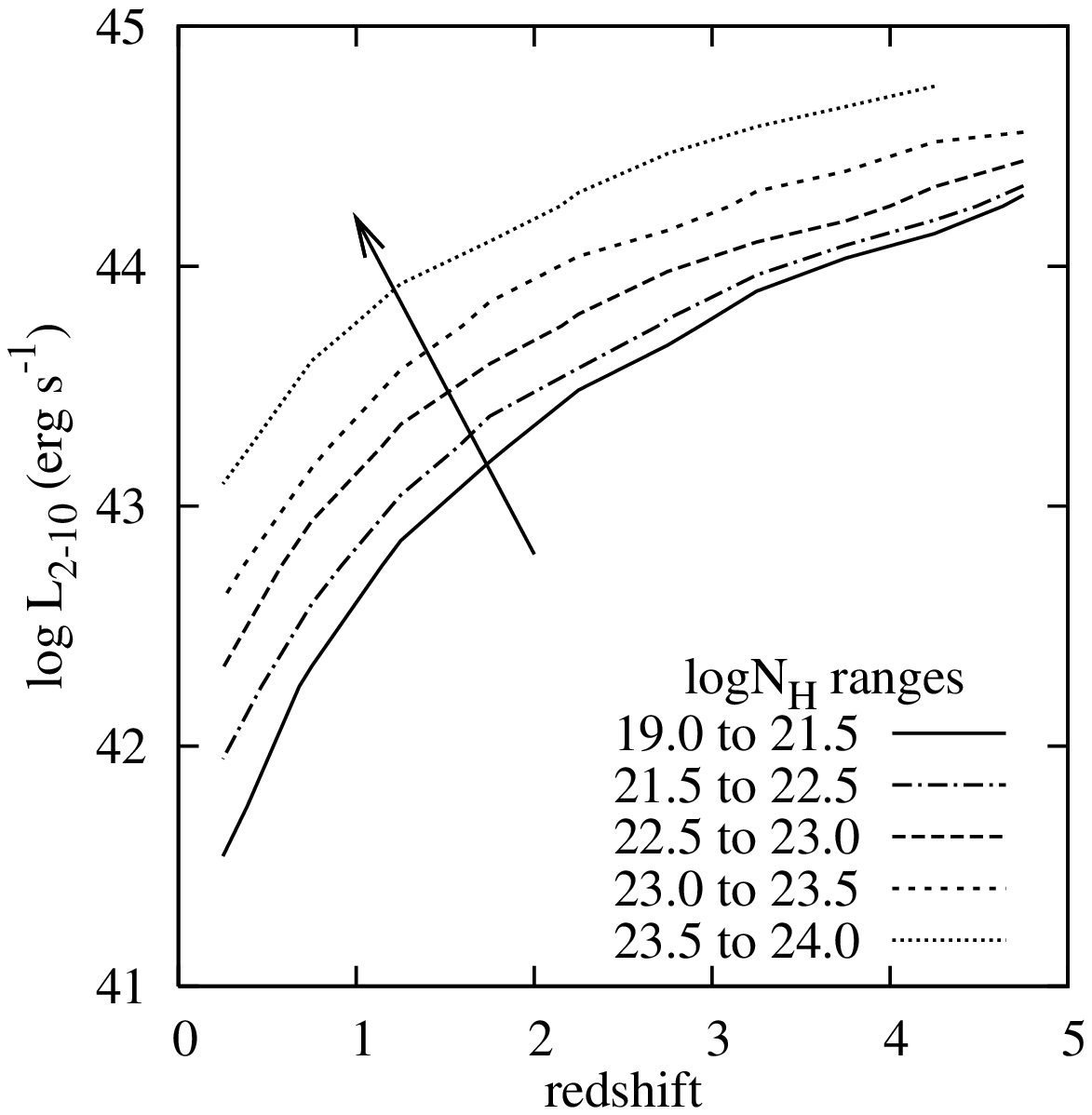}
\end{center}
\caption{ The X-ray completeness of the \xmm\ observations as a
function of absorption, intrinsic luminosity, and redshift.  The
contours are the boundaries of the regions where at least 50\% of the
input test population is detected in the simulated images.  Left
panel: The 50\% completeness limits as a function of luminosity and
absorption, for a number of redshift bins.  Centre panel: The 50\%
completeness limits as a function of redshift and absorption, for a
number of luminosity bins.  Right panel: The 50\% completeness limits
as a function of redshift and luminosity, for a number of absorption
bins.  In each panel, the arrow serves as a guide to the eye, and
indicates the direction of increasing X-ray completeness. }
\label{fig:completeness}
\end{figure*}

\subsection{X-ray completeness of the \xcdfs\ sample}
\label{sec:selection}

The probability that an AGN in the \cdfs\ will be detected in the
\xmm\ observations depends on the object's redshift, luminosity,
absorption, and position in the FOV.  In order to quantify the
selection function in the \xcdfs\ sample, we have compared the input
and output sources in a large simulated population generated using our
Monte Carlo process.  The X-ray completeness is simply the ratio of
the number of output detections to the number of input sources, and is
calculated for a number of bins in redshift, luminosity and
absorption.  Figure \ref{fig:completeness} shows the regions in
luminosity, absorption, and redshift space where at least half of the
input sources have output detections. The \xcdfs\ observations are
capable of detecting at least half the members of any population of
luminous ($L_{2-10} \ge 10^{44}$~erg~s$^{-1}$) obscured QSOs at
$z\sim2$ even if they are absorbed with large column densities ($N_H
\sim 10^{23}$~cm$^{-2}$).

\subsection{Direct comparison of the sample with model AGN populations}
\label{sec:model_populations}
We wish to compare directly the $N_H,L_X,z$ distribution observed in
the \xcdfs\ sample with the distributions predicted by various
AGN population models.  To accomplish this we have used our Monte
Carlo process to simulate a number of model AGN populations, then have
applied our $N_H/L_X$ estimation technique to recover the $N_H/L_X$
distributions of the model populations. In this way we incorporate
both the complex X-ray selection effects of the \xmm\ observations,
and account for the limitations of the $N_H/L_X$ estimation
technique. Therefore, the output simulated source distributions from
this process can be compared like-with-like to the real \xcdfs\
sample.  We have compared the \xcdfs\ sample with the predictions made
by the seven different $N_H$ model distributions described in section 
\ref{sec:nh_models}.  

Simulated populations are generated for each of these $N_H$ models,
according to both the LDDE XLF model of \citet{ueda03} as well as the
``LDDE1'' XLF model of \citet{miyaji00}. As before, the XLF models are
extrapolated to low luminosities ($L_{2-10} = 10^{40}$~erg~s$^{-1}$),
and high redshifts ($z=5$). For each of the fourteen combinations of
$N_H$ model and XLF model, the absolute normalisation of the XLF is
adjusted in order that the simulated integral 0.5--2~keV source counts
above $2 \times 10^{-15}$~\cgs\ match the integral source counts in
the extragalactic \xcdfs\ sample. We simulate 100 fields worth of
sources for each combination of $N_H$ model and XLF model. Finally, we
apply the $N_H/L_X$ estimation process to recover the absorption and
luminosity distributions of the simulated output populations.

\section{Results}
\label{sec:results}

\subsection{Applying the $N_H/L_X$ estimation technique to the \xcdfs\ sample}
We find that our technique is able to evaluate the absorption and
luminosity in the vast majority of the optically identified
extragalactic sources in the \xcdfs\ sample.  However, we find that
because of their very soft spectra, two AGN are not matched to any
objects in the simulated source library. We discuss the properties of
these sources in Appendix \ref{app:unusual}.  There is one \xcdfs\
source which we find to have a 2--10~keV luminosity less than
$10^{40}$~erg~s$^{-1}$. For the purposes of all the comparisons made
in this section, we have excluded this source because it lies outside
the luminosity range simulated in the model AGN populations.

\subsection{Source counts in the \xcdfs}
\begin{figure}
\begin{center}
\includegraphics[angle=0,width=80mm]{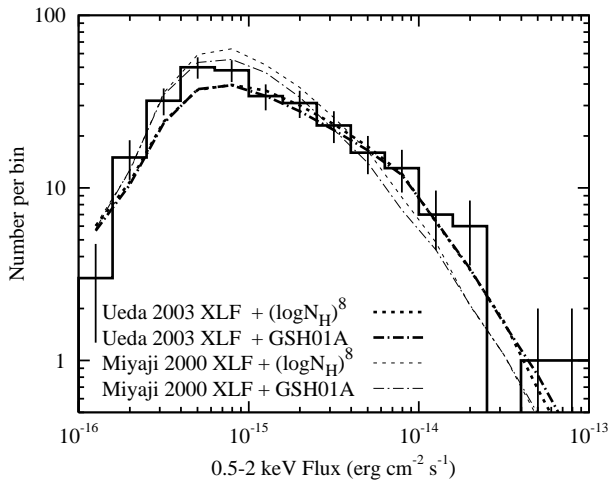}
\end{center}
\caption{The differential source counts
for the extragalactic \xcdfs\ sample as a function of flux in the 0.5--2~keV 
energy band.  For comparison, we show the source counts 
predicted by several different simulated model AGN populations.
The two dashed curves show the predictions when we combine the \betaeight\
model firstly with the XLF model of \citet{ueda03}, and secondly
with the ``LDDE1'' XLF model of \citet{miyaji00}.  The predicted
source counts from model populations generated according to the
\gillia\ $N_H$ model are also shown (dot-dashed curves). At faint fluxes,
the distributions are dominated by sources
which have their strongest detections in the other energy bands.
}
\label{fig:nofs_flux}
\end{figure}

In figure \ref{fig:nofs_flux} we show the differential 0.5--2.0~keV
source counts for the extragalactic (including unidentified) sources
in the \xcdfs\ sample, and compare them to the predictions of several
simulated model AGN populations. The shape of the predicted source
count curves is dependent predominantly on the form of the XLF model
rather than on the $N_H$ distribution.  We see that the XLF model of
\citet{miyaji00} predicts a 0.5--2~keV source count distribution which
is rather steeper than that found in the \xcdfs.  At 0.5--2~keV fluxes
above $10^{-15}$~\cgs, the source counts predicted by the
\citet{ueda03} XLF model are a good match to the shape of the source
counts in the \xcdfs\ sample. However, the \citet{ueda03} XLF model
under-predicts the observed numbers of sources at fainter fluxes.  A
Kolmogorov--Smirnov comparison of the predicted 0.5--2~keV source
count distributions with the observed \xcdfs\ distribution finds that
though the \citet{ueda03} XLF is favoured ($P_{KS} = 0.3$), the
\citet{miyaji00} XLF model cannot be completely rejected ($P_{KS} =
0.06$).  We remind the reader that for each model of the AGN
population, the XLF normalisation has been adjusted such that the
integral source counts in the simulated population match the 0.5-2~keV
extragalactic source counts measured in the \xcdfs\ sample.  (see section
\ref{sec:model_populations}).

\subsection{The redshift and luminosity distributions in the \xcdfs\ sample}

\begin{figure}
\begin{center}
\includegraphics[angle=0,width=80mm]{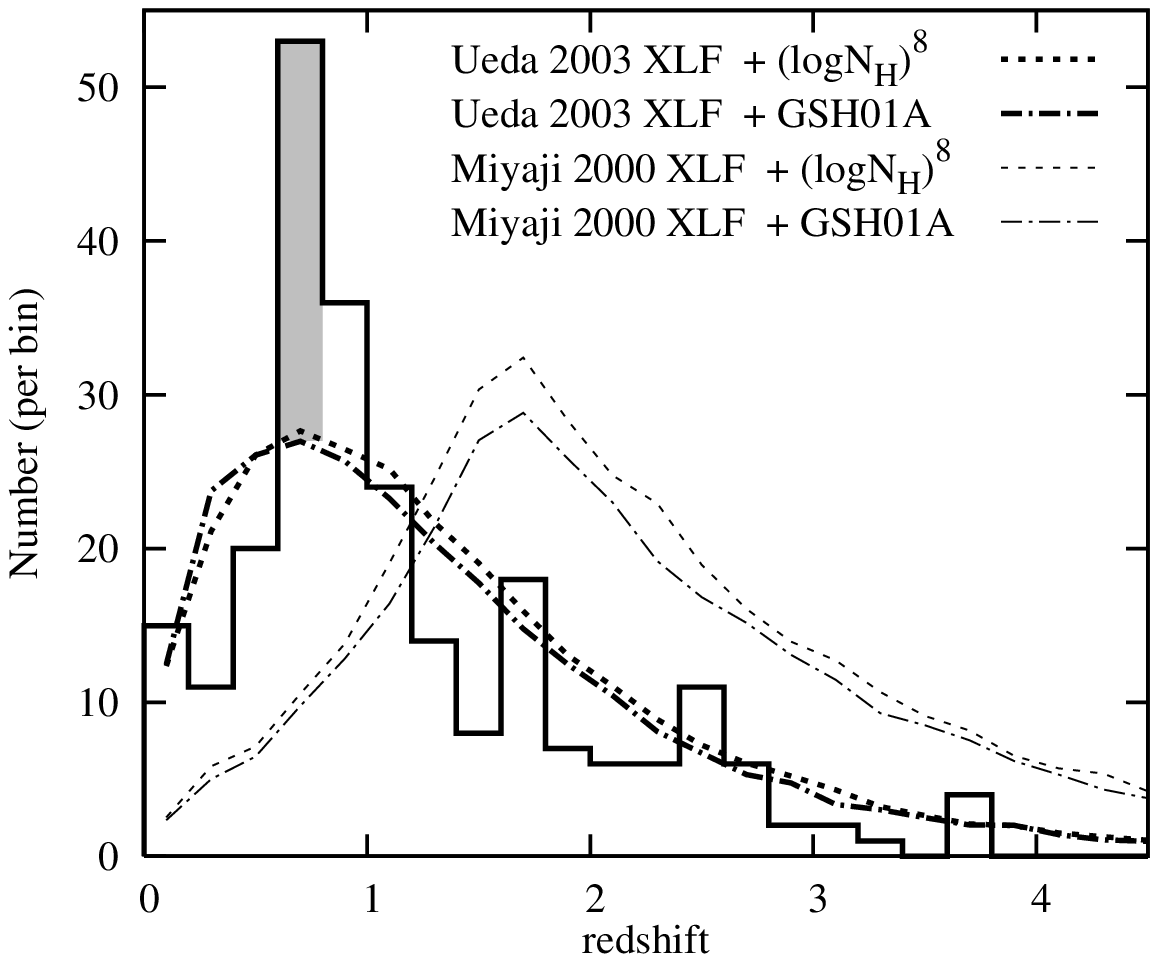}
\end{center}
\vspace{1mm}
\begin{center}
\includegraphics[angle=0,width=80mm]{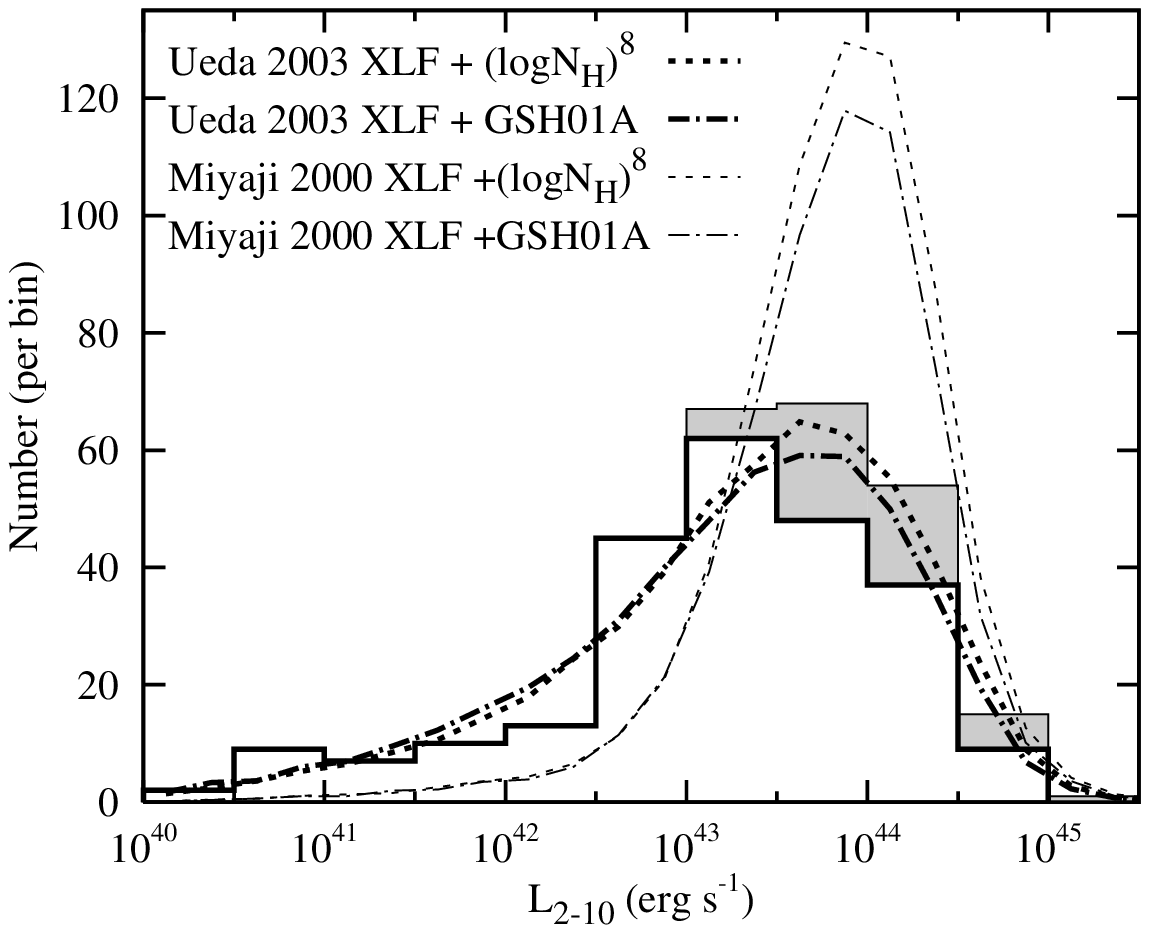}
\end{center}
\caption{Distribution of redshifts (upper panel) and intrinsic
2--10~keV luminosities (lower panel) of identified extragalactic
sources in the \xcdfs\ sample (solid histogram).  For comparison, we
show the distributions predicted by coupling the XLF model of
\citet{ueda03} with the \betaeight\ and \gillia\ $N_H$ distribution
models (dashed and dot-dashed thick curves respectively).  The thin
curves show the rather different predictions made by combining the
same $N_H$ distributions with the ``LDDE1'' XLF model of
\citet{miyaji00}.  There are two large-scale structures identified in
the \cdfs\ at $z=0.67$ and $0.73$ \citep{gilli03}. At least 26 sources
in the \xcdfs\ sample lie at these redshifts, as indicated by the
shaded area in the upper panel. The shaded area in the lower plot
shows the luminosity distribution of the 50 optically unidentified
sources in the \xcdfs\ sample if they are assumed to lie at $z=2$.}
\label{fig:z_lum_histos}
\end{figure}

\begin{figure}
\begin{center}
\includegraphics[angle=0,width=80mm]{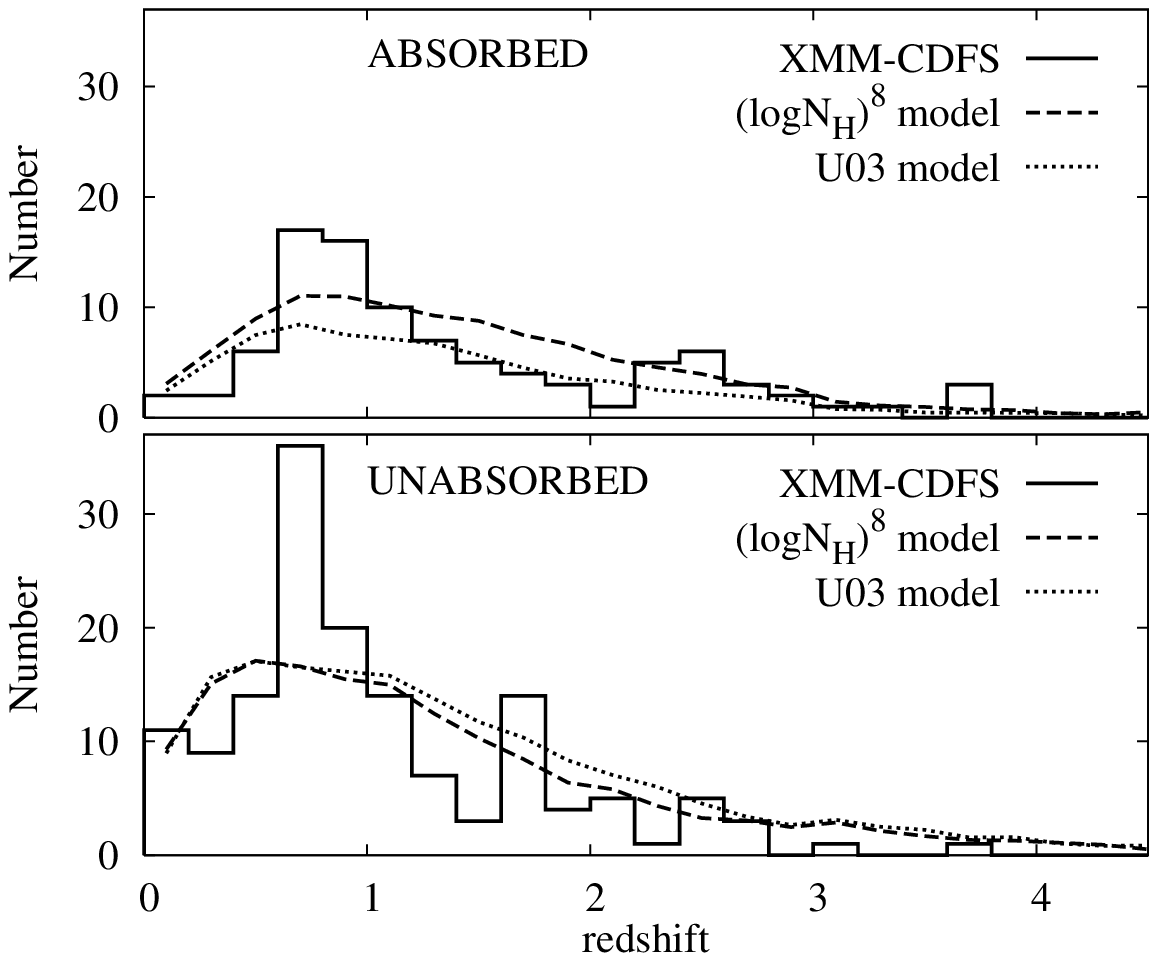}
\end{center}
\vspace{0mm}
\begin{center}
\includegraphics[angle=0,width=80mm]{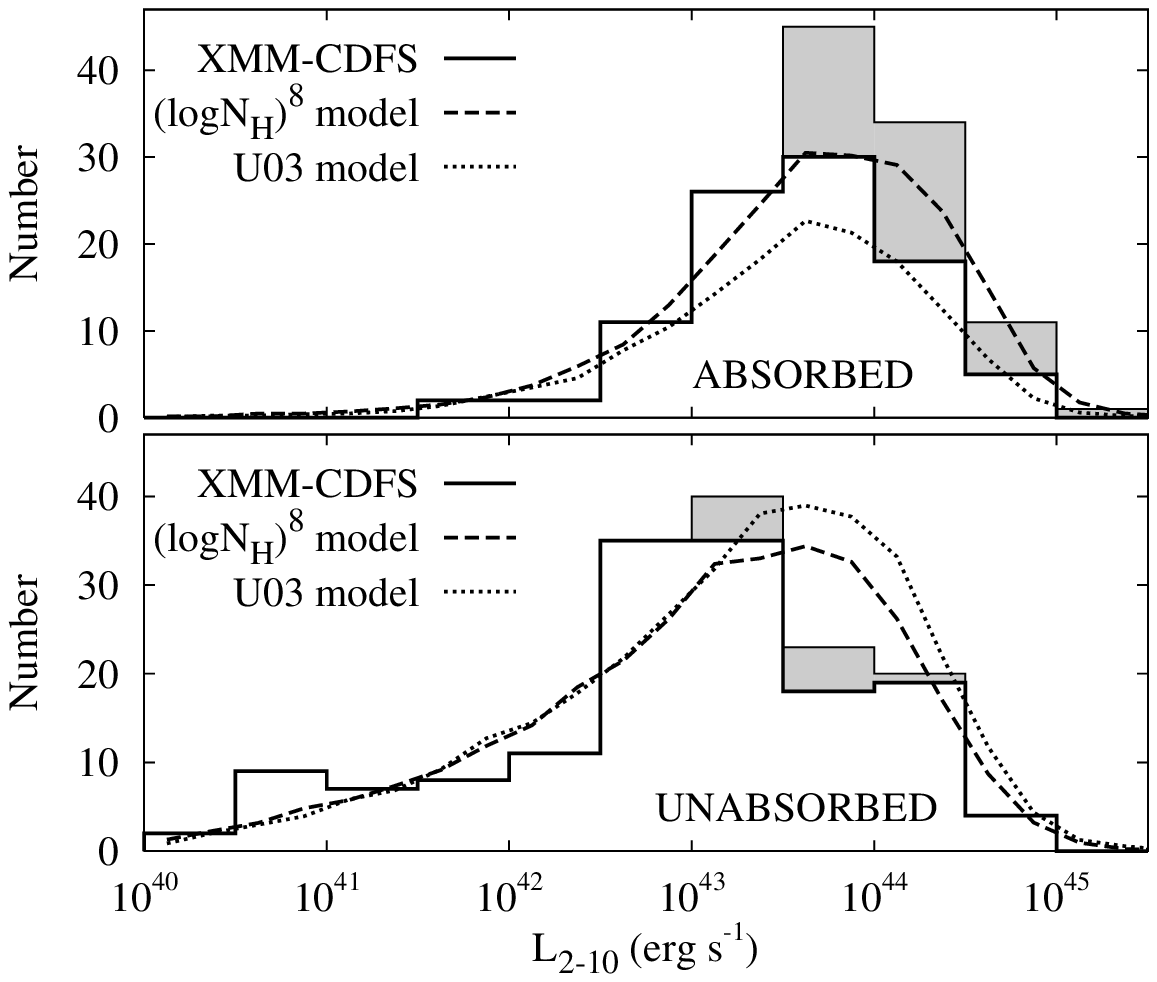}
\end{center}
\caption{Similar to fig.\ref{fig:z_lum_histos}, but with the AGN
divided into ``absorbed'' and ``unabsorbed'' objects.  Here we set the
threshold for an AGN to be considered ``absorbed'' to be
$10^{22}$~cm$^{-2}$ for sources at $z<3$, and $10^{22.6}$~cm$^{-2}$
for sources at $z>3$. In each panel the solid histogram shows the
\xcdfs\ sources, and the curves show the distributions predicted by
coupling the XLF model of \citet{ueda03} with the \betaeight\ and
\ueda\ $N_H$ distribution models. The shaded areas in the lower two
panels show the luminosity distribution of the optically unidentified
sources in the \xcdfs\ sample if they are assumed to lie at $z=2$.}
\label{fig:z_lum_histos_ltgt22}
\end{figure}

Figure \ref{fig:z_lum_histos} shows the redshift distribution of the
optically identified \xcdfs\ sample compared with the redshift
distributions for several of the simulated model populations.  For
clarity, we have shown only the results for the \betaeight\ and
\gillia\ models, because rather similar distributions are found in the
other $N_H$ models.  It is clear that the redshift distribution
predicted by the XLF of \citet{ueda03} is a far closer match to the
redshift distribution of the \xcdfs\ sample than the prediction from
the XLF model of \citet{miyaji00}.  Figure \ref{fig:z_lum_histos}
shows that the same holds true for the luminosity distribution.
We remind the reader that the \citet{miyaji00} luminosity function is
defined in the observed 0.5--2~keV band, and that we have converted to
intrinsic rest frame 2--10~keV luminosities assuming an unabsorbed AGN
X-ray spectrum (a $\Gamma=1.9$ powerlaw). This is of course a
simplification; some of the AGN in the \citet{miyaji00} sample may be
X-ray absorbed (and hence have hard spectra), and some of the AGN
may have very soft X-ray spectra. But this scatter of slopes is
unlikely to be a significant issue, and certainly not sufficient to
explain the large differences between the redshift and luminosity
distributions predicted by the \citet{miyaji00} XLF model and those
seen in the \xcdfs\ sample.

In figure \ref{fig:z_lum_histos_ltgt22} we show redshift and
luminosity distributions separately for the absorbed and unabsorbed
AGN in the \xcdfs\ sample, and compare these to the predictions 
of the \betaeight\ and \ueda\ $N_H$ models.

\subsection{The $N_H$ distribution in the \xcdfs\ sample}
\label{sec:nh_dist_kstest}
\begin{figure}
\begin{center}
\includegraphics[angle=0,width=80mm]{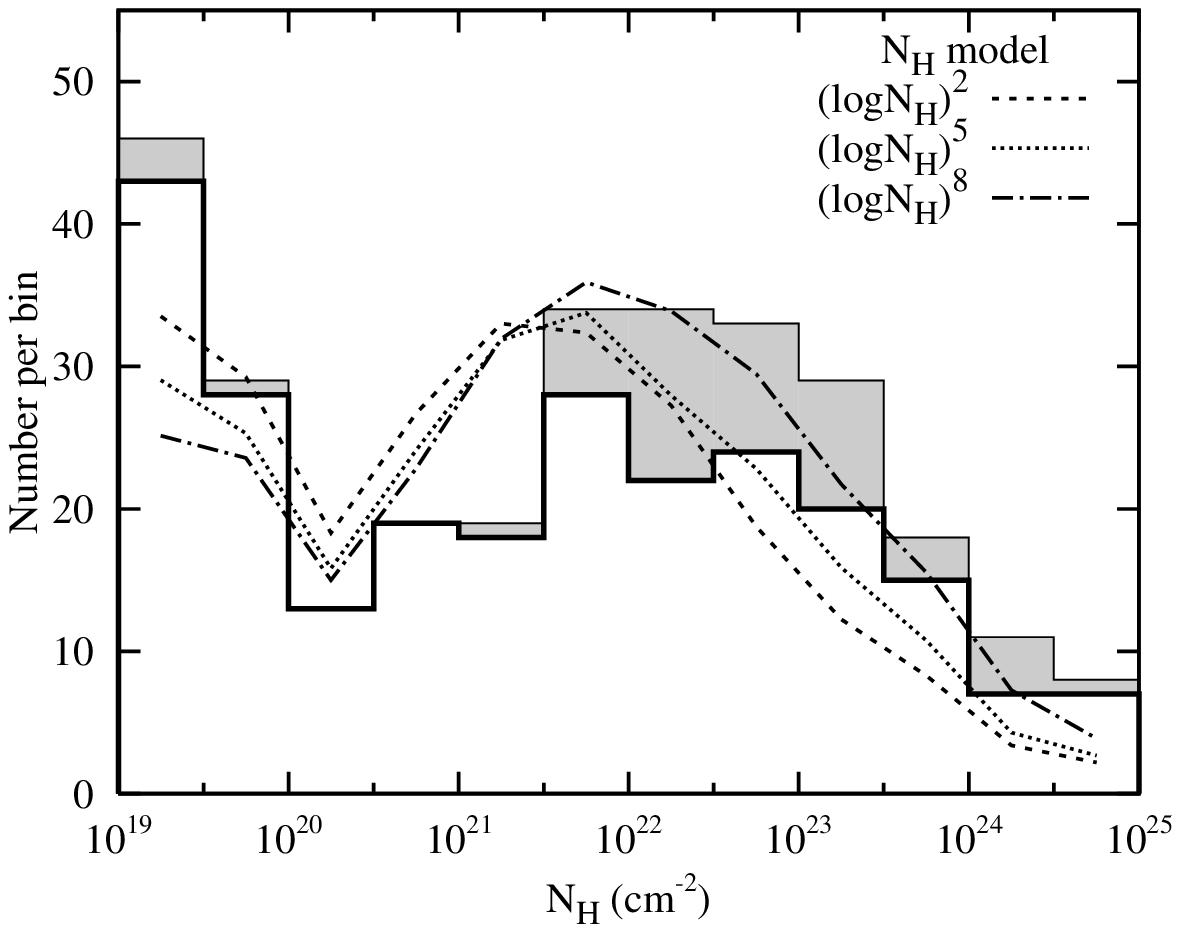}
\includegraphics[angle=0,width=80mm]{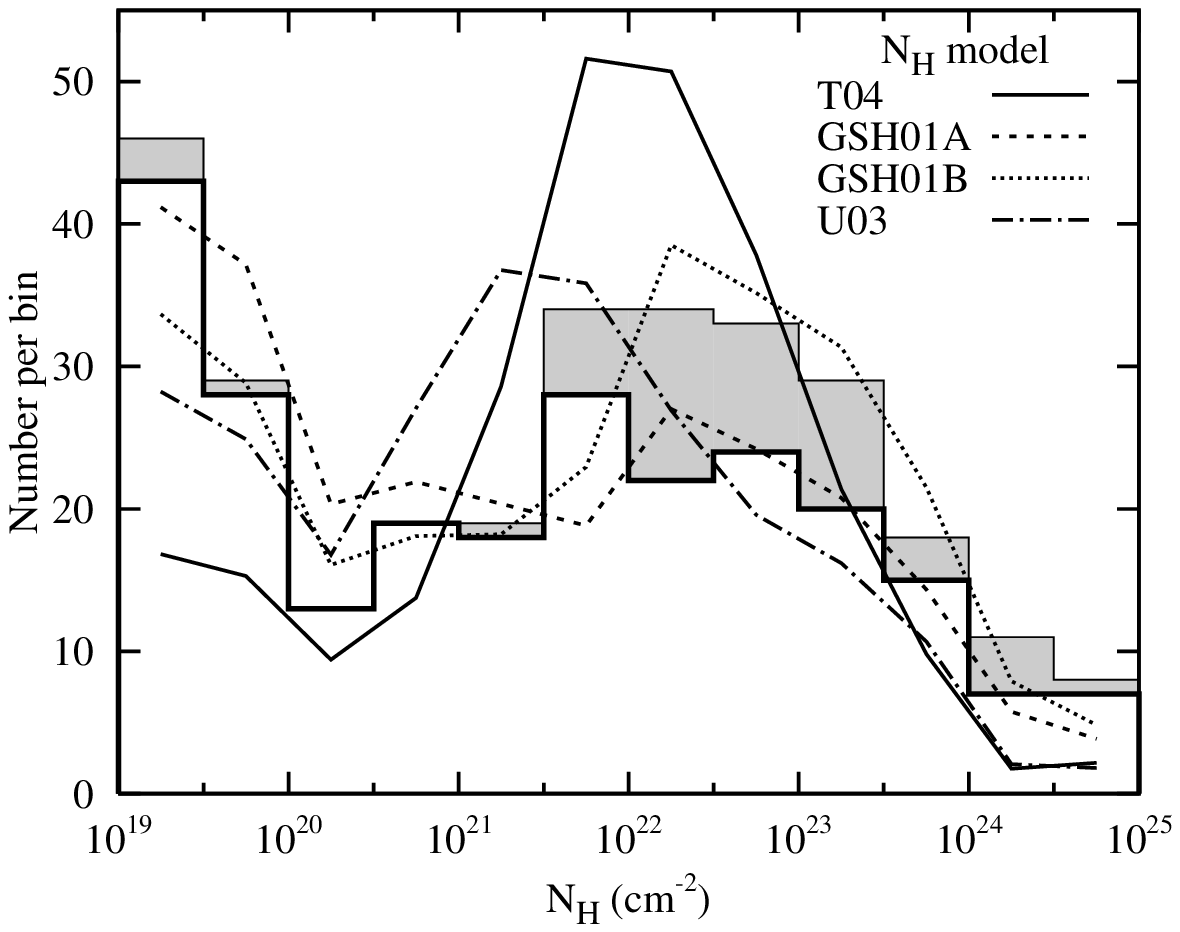}
\end{center}
\caption{The distribution of absorption in the optically identified
\xcdfs\ sample (histogram) in comparison to the predicted
distributions from the seven $N_H$ models (curves). The model $N_H$
distributions are for simulated populations generated according to the
XLF model of \citet{ueda03}. The shaded area shows the $N_H$
distribution of the optically unidentified sources if they are assumed
to lie at $z = 2$. For clarity, the model distributions are displayed
in two groups.}
\label{fig:nh_histo}
\end{figure}

In figure \ref{fig:nh_histo} we show the distribution of absorption in
the optically identified sources in the \xcdfs\ sample determined using
our $N_H$ estimation technique. The measured distribution is compared
to the predicted distributions from the seven simulated $N_H$
models. 

The high fidelity of our $N_H$ estimation process means that the
recovered absorption distribution in the \xcdfs\ sample contains more
information than just the relative numbers of absorbed and unabsorbed
AGN; we can compare the shape of the absorption distribution measured
in the \xcdfs\ sample with the shapes of the distributions predicted
by the simulated model AGN populations.  We have used the
Kolmogorov--Smirnov (KS) test to make this comparison, the results of
which are shown in table \ref{tab:stat_test}.  This test clearly
discriminates between the models, with the \treister\ $N_H$
distribution being the most strongly rejected.

\subsection{The absorbed fraction in the \xcdfs\ and its dependence on luminosity and redshift}
\label{sec:absorbed_fraction}

\begin{figure}
\begin{center}
\includegraphics[angle=270,width=81mm]{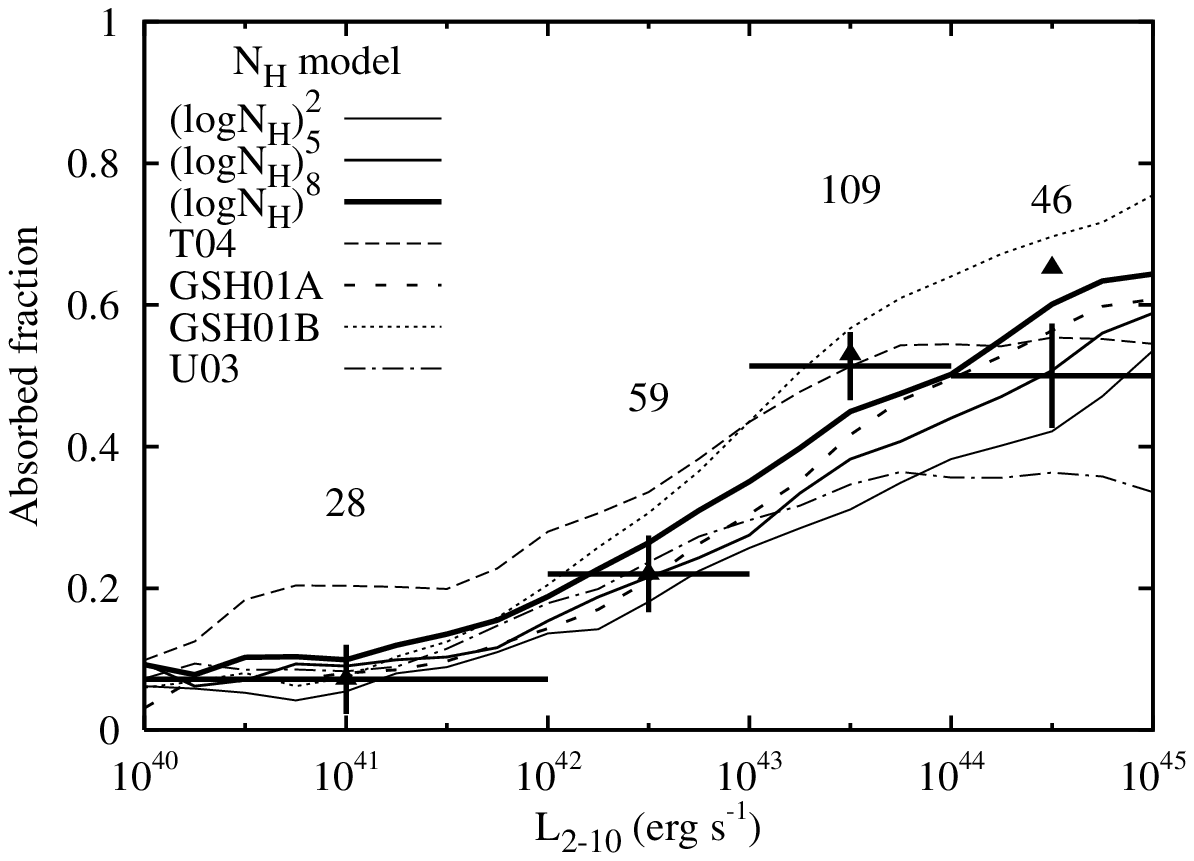}
\includegraphics[angle=270,width=80mm]{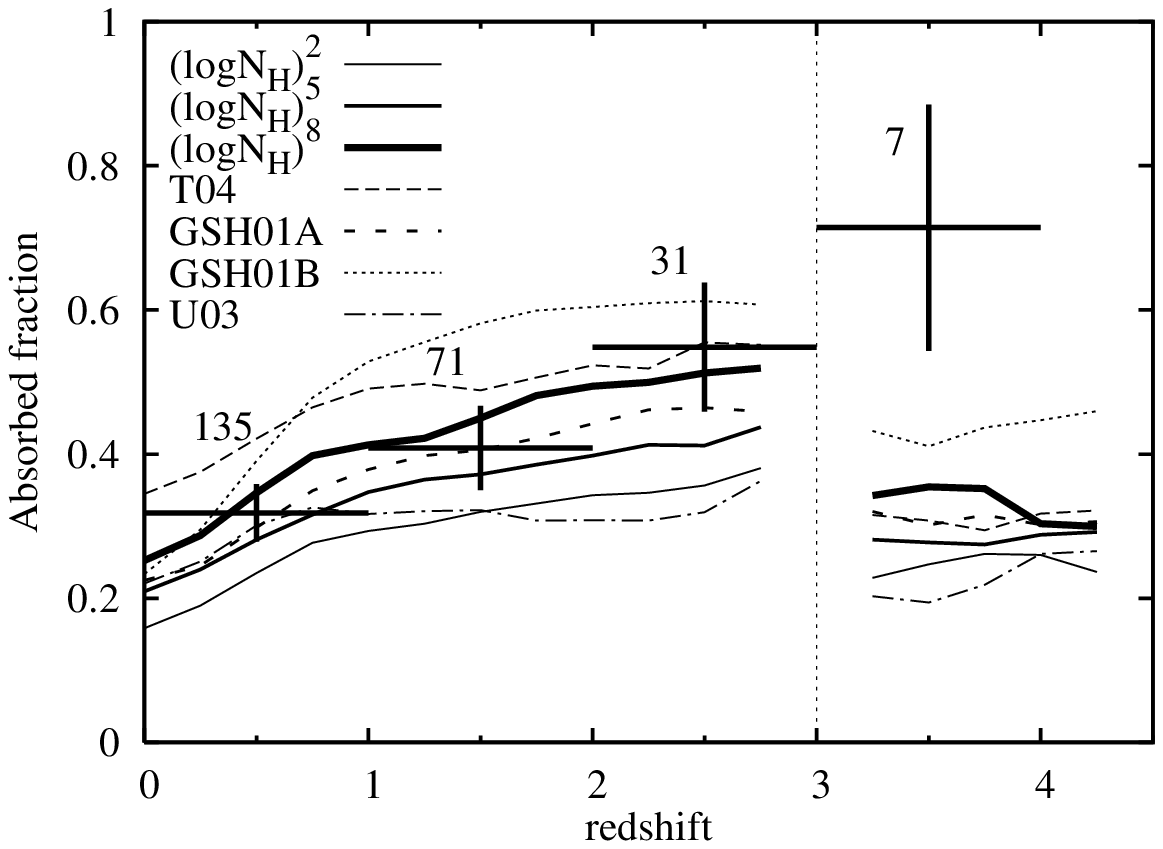}
\end{center}
\caption{The fraction of \xcdfs\ sources with significant absorption,
as a function of intrinsic 2--10~keV luminosity (upper panel), and
redshift (lower panel).  We set the threshold for a source to be
considered ``absorbed'' to be $10^{22}$~cm$^{-2}$ for sources at
$z<3$, and $10^{22.6}$~cm$^{-2}$ for sources at $z>3$.  Error bars
show the result for the \xcdfs\ sample: horizontal bars show the range
over which the absorbed fraction has been calculated, and the vertical
error-bars show the binomial error estimates,
$\sqrt{f_{abs}(1-f_{abs})/N}$, where $f_{abs}$ is the absorbed
fraction, and $N$ is the number of objects in the bin. The total numbers of \xcdfs\ 
sources in each bin are indicated on the plot. The curves
show the absorbed fractions predicted by the $N_H$ distribution
models, calculated in bins of width 0.5~dex in luminosity (upper
panel), and width 0.5 in redshift (lower panel).  In the upper panel,
we also show the absorbed fraction in the \xcdfs\ sample if all
unidentified sources are placed at $z=2$ (triangles).}
\label{fig:absorbed_fraction}
\end{figure}

In order to characterise the degree of luminosity and/or redshift
dependence of the absorption distribution, we have examined the
relative numbers of absorbed and unabsorbed AGN in the \xcdfs\ sample,
and compared it to the predictions of the simulated model populations.
We set the threshold for a source to be considered ``absorbed'' to be
$10^{22}$~cm$^{-2}$ for sources at $z<3$, and to be
$10^{22.6}$~cm$^{-2}$ for sources at $z>3$.  When we tested the
reliability of the $N_H$ estimation technique (see section
\ref{sec:individual}), we found that for sources in the $3<z<4$ range,
$10^{22.6}$~cm$^{-2}$ was the lowest level of absorption for which the
scatter of output $N_H$ about input $N_H$ was less than 0.5~dex.  The
total absorbed fraction in the identified \xcdfs\ sample is $0.39 \pm
0.03$, if the unidentified \xcdfs\ sources are assumed to lie at z=2,
then the absorbed fraction becomes $0.45 \pm 0.03$. These respective
values can be considered as lower and upper bounds on the ``true''
total absorbed fraction.  For comparison, in table
\ref{tab:absorbed_fractions} we show the {\em total} absorbed
fractions predicted by each of the population models.  When combined
with the XLF model of \citet{ueda03}, the \betaeight,\treister, and
\gillia\ $N_H$ models predict absorbed fractions consistent with these
bounds. For each $N_H$ model, the populations generated using the
\citet{miyaji00} XLF have a slightly higher absorbed fraction than for
the \citet{ueda03} XLF. This is because the former XLF model predicts
more objects at high redshifts (see fig. \ref{fig:z_lum_histos}), 
where heavily absorbed AGN are selected against less strongly.

\begin{table}
\caption{The absorbed fraction predicted using the AGN population
models for the \xcdfs\ sample.  Predictions are shown for AGN
populations generated according to each of the seven $N_H$ models
tested in this study, using the \citet{ueda03} XLF model and the
\citet{miyaji00} LDDE1 XLF model. The total absorbed fraction in the
optically identified \xcdfs\ sample is $0.39 \pm 0.03$. If the
optically unidentified \xcdfs\ sources are assumed to lie at $z=2$,
this fraction becomes $0.45\pm0.03$.}

\label{tab:absorbed_fractions}
\begin{tabular}{@{}lcc}
\hline
$N_H$ model & \multicolumn{2}{c}{XLF model} \\
            & \citet{ueda03} & \citet{miyaji00} \\
\hline
\betatwo   & 0.283     &   0.297   \\
\betafive  & 0.335     &   0.362   \\
\betaeight & 0.409     &   0.422   \\
\treister  & 0.460     &   0.468   \\
\gillia    & 0.364     &   0.395   \\
\gillib    & 0.492     &   0.535   \\
\ueda      & 0.301     &   0.311   \\
\hline
\end{tabular}
\end{table}

Figure \ref{fig:absorbed_fraction} shows the absorbed fraction as a
function of both redshift and intrinsic luminosity compared to the
distributions predicted from the Monte Carlo simulations of the $N_H$
models.  We see later that the optically unidentified sources, which
are not included in fig. \ref{fig:absorbed_fraction}, have on average,
harder X-ray colours than the identified sources, and so are likely
to increase the absorbed fraction.

For $L_{2-10} < 10^{43.5}$erg~s$^{-1}$ there is an apparent {\em
positive} correlation between intrinsic luminosity and the absorbed
fraction in the \xcdfs\ sample, as well as for each of the seven
$N_H$ models.  For the $N_H$ models which are not dependent on
luminosity, this can only be due to the selection function of the
\xmm\ observations.  For the luminosity dependent \ueda\ $N_H$ model,
we see a levelling out of the predicted absorbed fraction above
$L_{2-10}>10^{43.5}$erg~s$^{-1}$. This marks the transition from a
regime in which the observed absorbed fraction is determined primarily
by the \xcdfs\ selection function, to a regime in which the shape of
the underlying $N_H$ distribution becomes more important.  In table
\ref{tab:stat_test} we show the results of $\chi^2$ tests which
quantify how well the $N_H$ models reproduce the redshift and
luminosity dependence of the absorbed fraction in the \xcdfs\ sample.

\subsection{Distribution of the \xcdfs\ sample in $z$,$N_H$,$L_{2-10}$ space}
\label{sec:z_nh_lx_dist}
\begin{figure*}
\begin{center}
\includegraphics[angle=270,width=43mm]{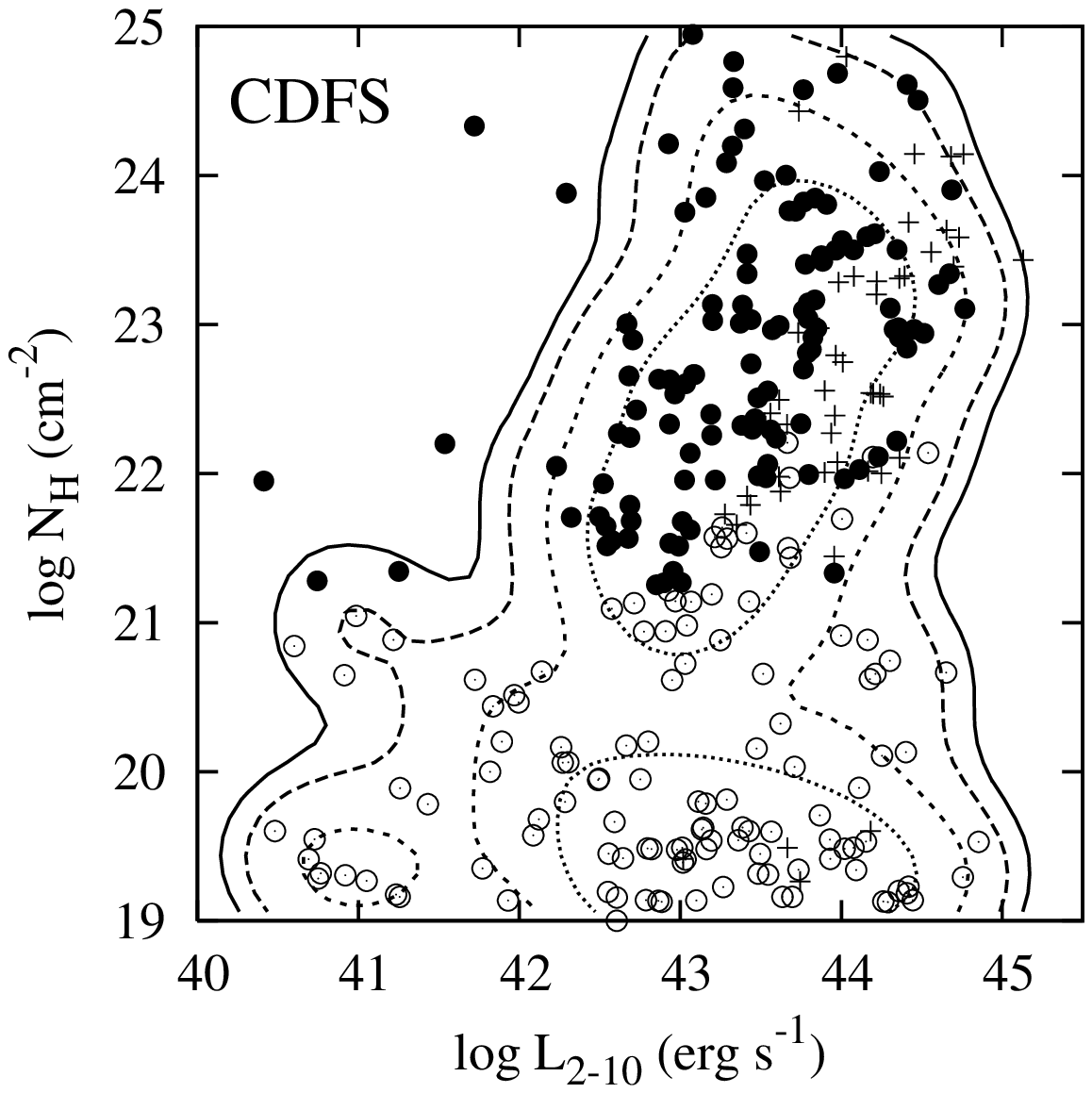}
\includegraphics[angle=270,width=43mm]{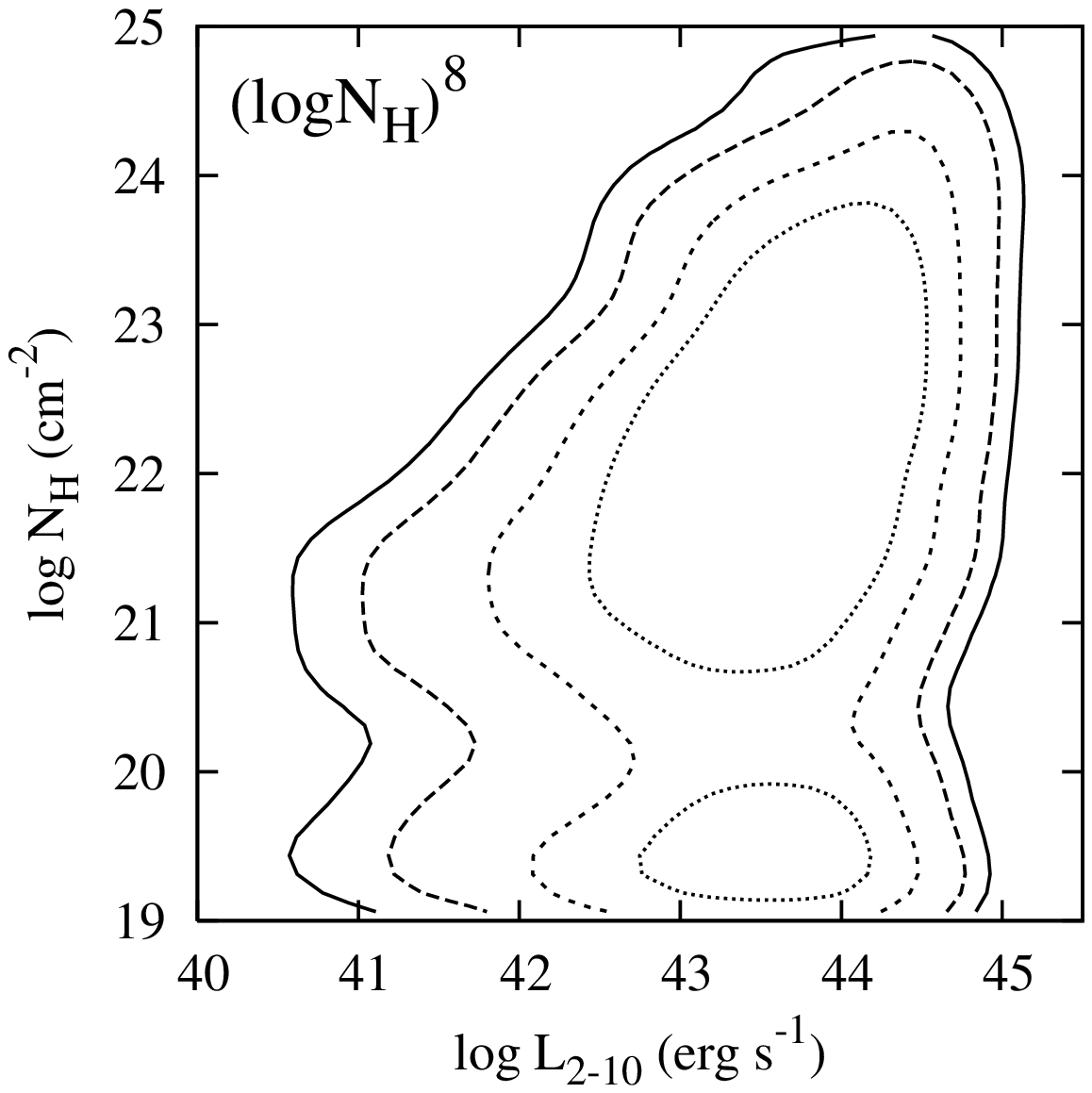}
\includegraphics[angle=270,width=43mm]{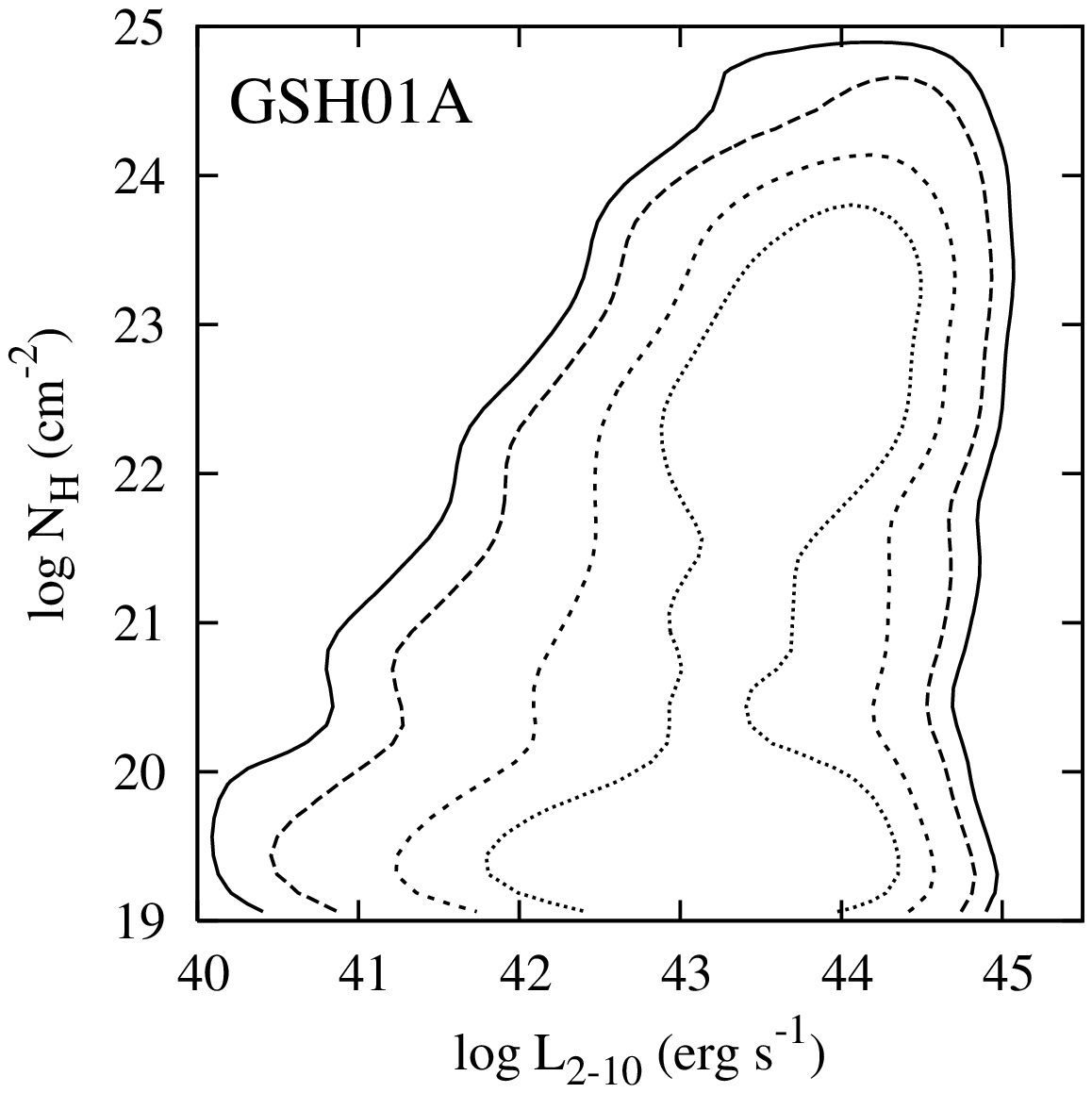}
\includegraphics[angle=270,width=43mm]{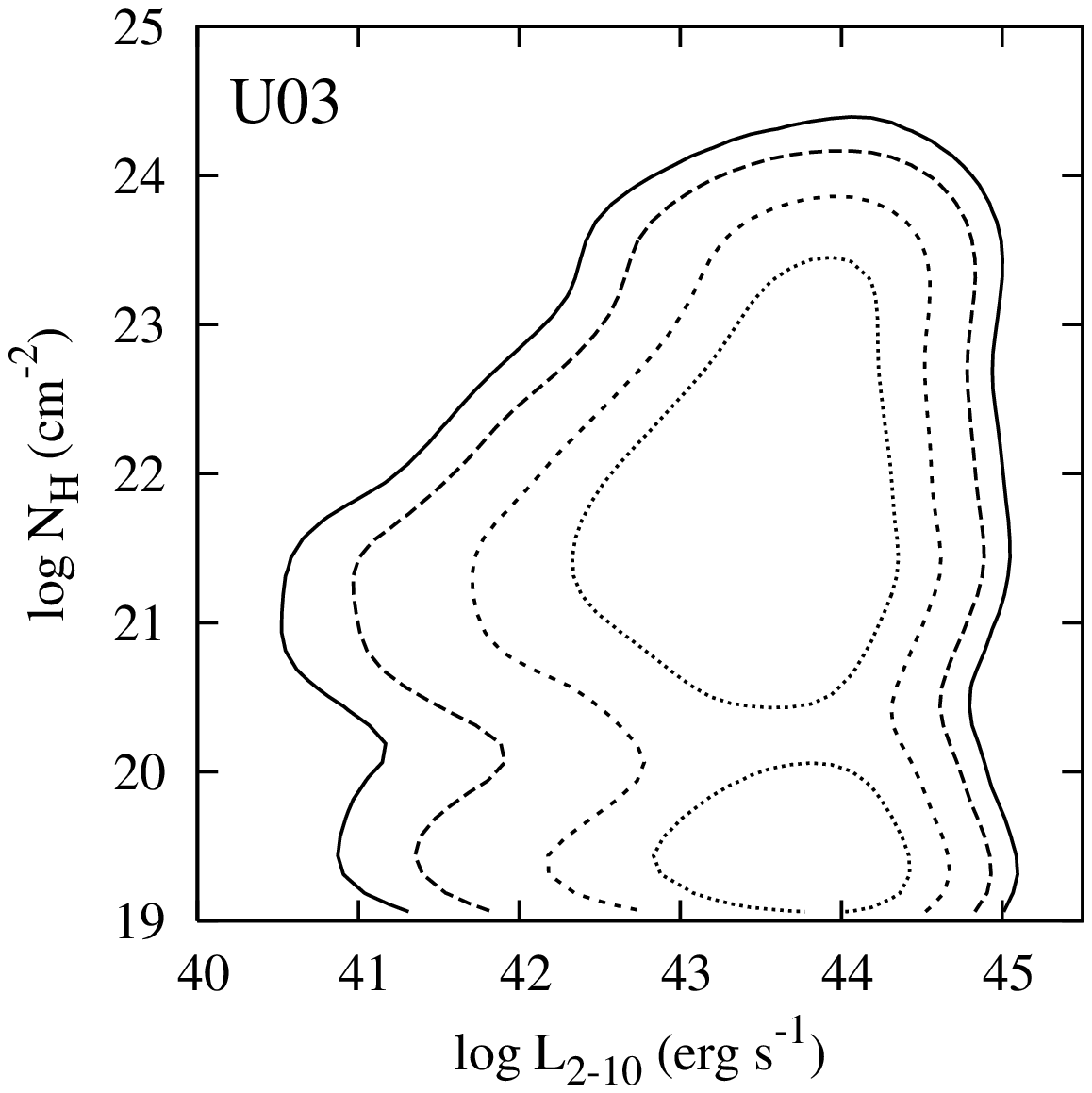}
\includegraphics[angle=270,width=43mm]{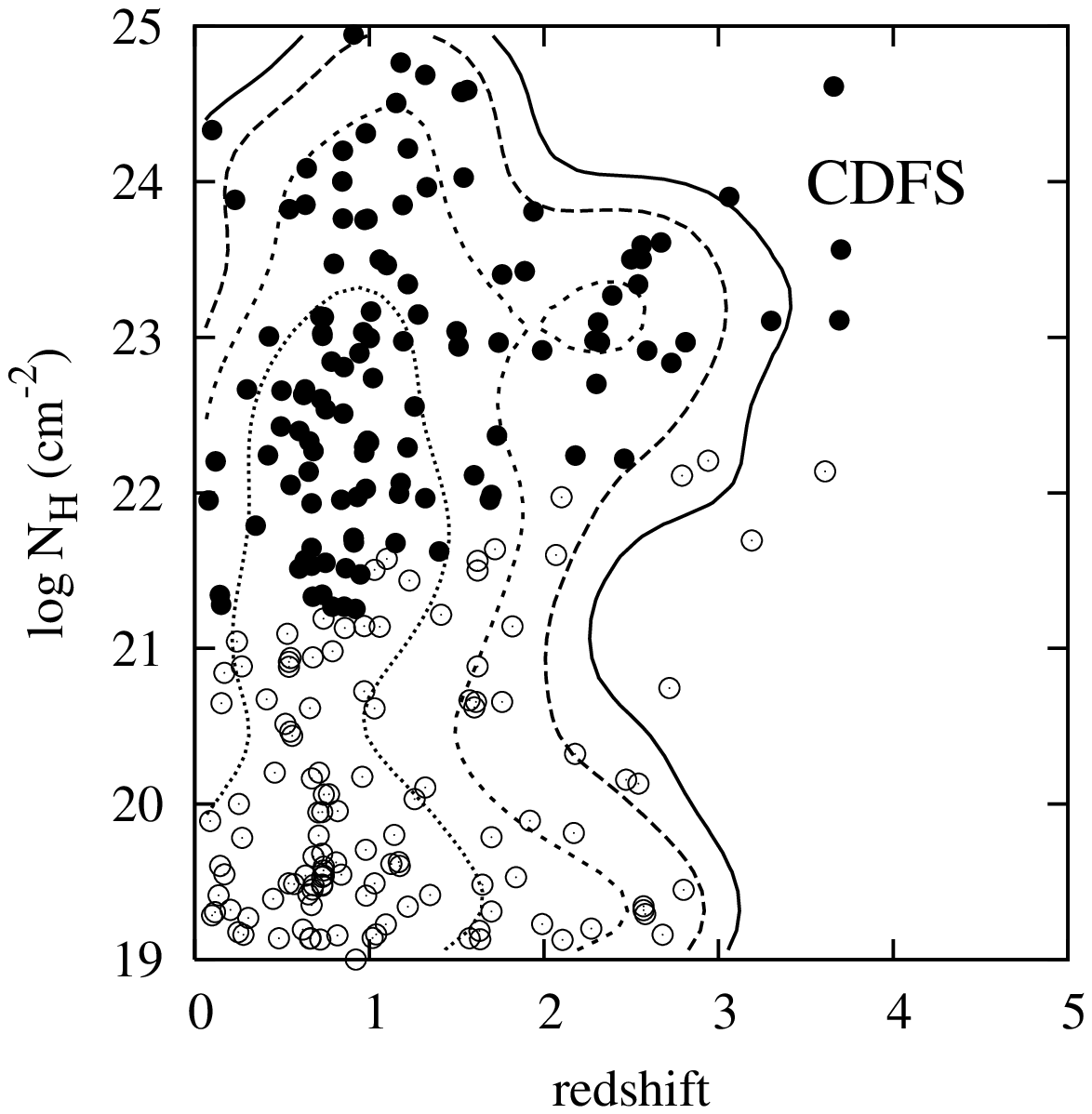}
\includegraphics[angle=270,width=43mm]{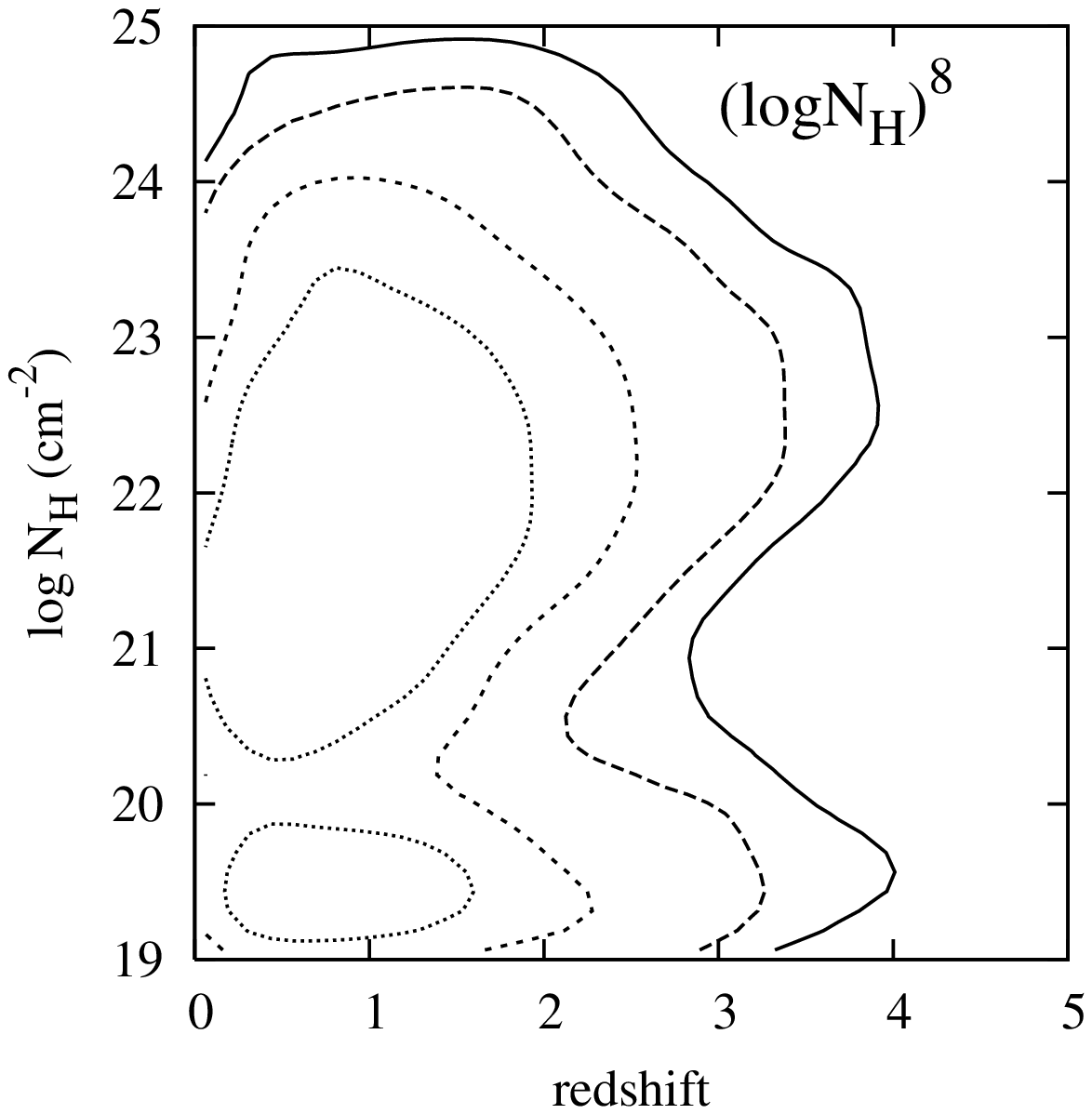}
\includegraphics[angle=270,width=43mm]{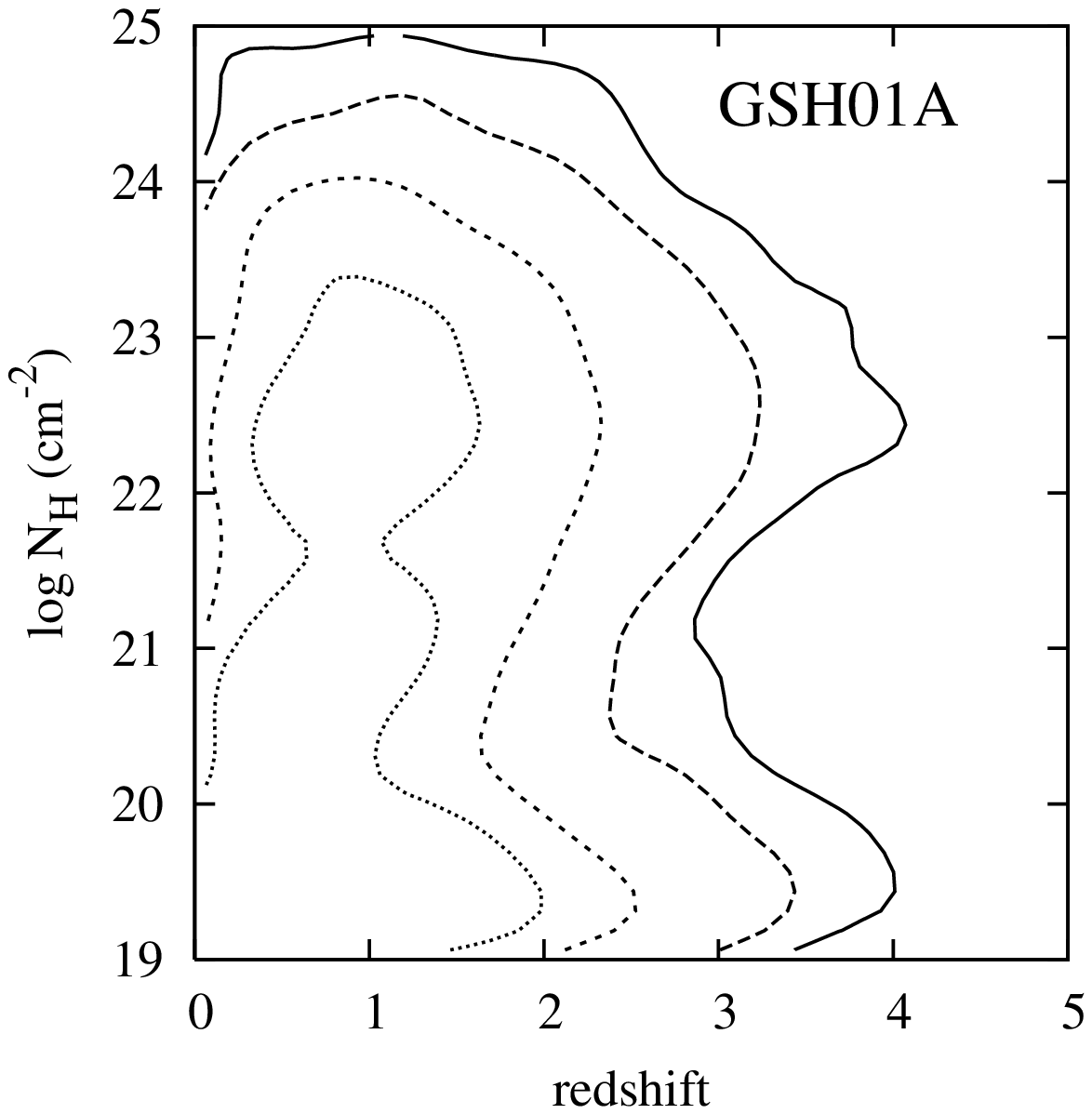}
\includegraphics[angle=270,width=43mm]{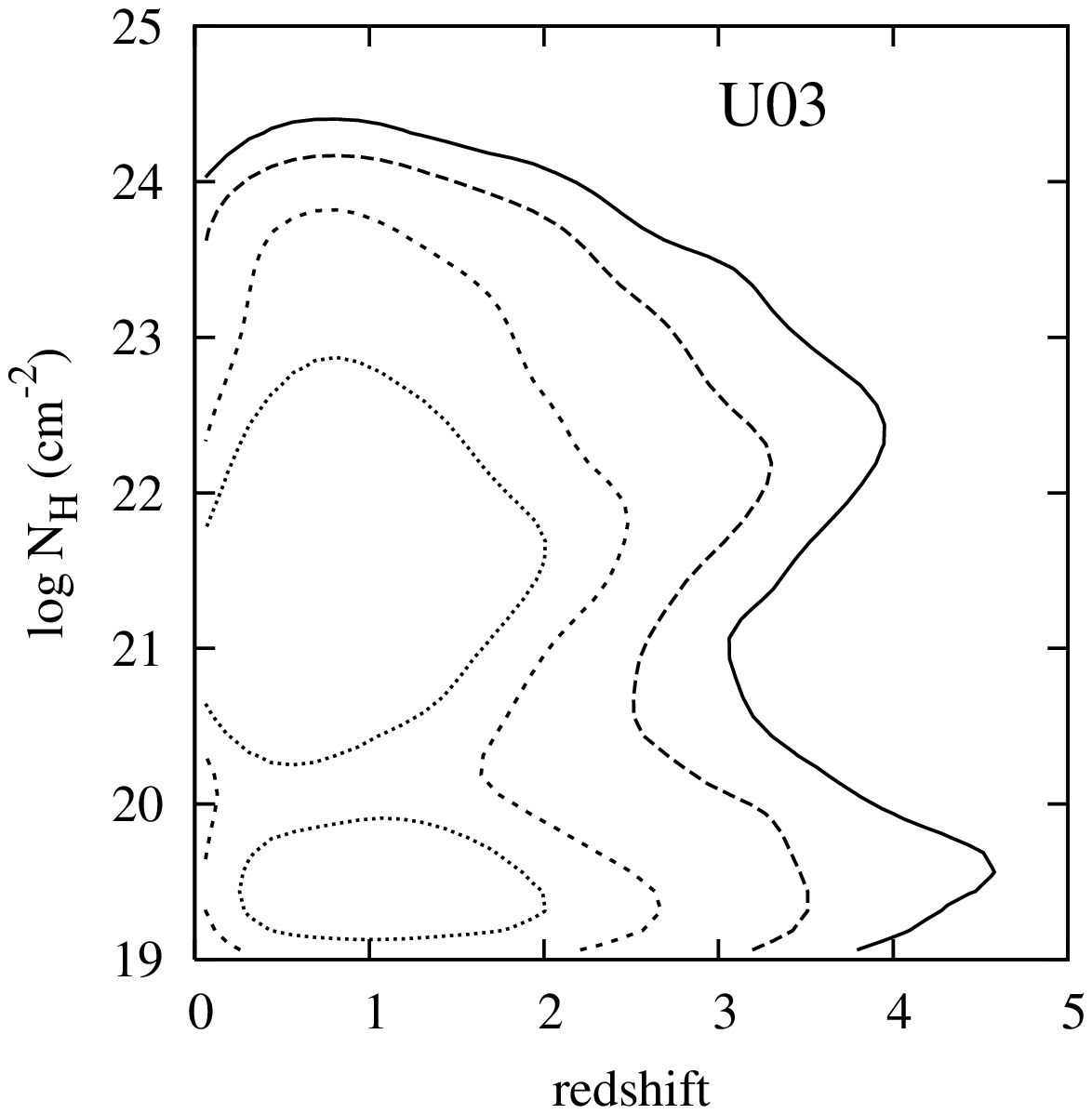}
\end{center}
\caption{Distributions of absorption {\em vs} intrinsic luminosity
(upper row), and {\em vs} redshift (lower row).  The distributions
found in the \xcdfs\ sample are shown in the left-hand column. 
In the other columns, we show the predictions from three of 
the model $N_H$ distributions.  The contours enclose the
regions occupied by $95\%$ (solid line), $90\%$ (long dashed line),
$75\%$ (short dashed line), and $50\%$ (dotted line) of the
sources. In order to construct the contours, the distributions have
been smoothed with a 2-D Gaussian. The widths used are 0.25 in
$\log L_{2-10}$, $z$ and $\log N_H$ for the model populations, and 0.4 for
the \xcdfs\ sample.  The identified sources of the \xcdfs\ sample are
shown with circles. Open circles mark the sources which have
absorbing columns smaller than the level where the estimation
technique is accurate within 0.5 dex (see section \ref{sec:individual}). 
In the top left panel, crosses show the luminosity/absorption of the 
unidentified \xcdfs\ sources if they are assumed to lie at $z=2$. }
\label{fig:nh_vs_Lz_sample_sims}
\end{figure*}

The most stringent test of the AGN population models is to see how
well they reproduce the distribution of \xcdfs\ sources simultaneously
in absorption,luminosity and redshift space. Figure
\ref{fig:nh_vs_Lz_sample_sims} shows the distribution in $L_X$, $z$
and $N_H$ of the \xcdfs\ sources and the predictions from three of the
$N_H$ models.  By using the three-dimensional Kolmogorov--Smirnov test
(3D-KS), which requires no binning, we can make a statistical
comparison which interrogates the maximum information content of the
sample. However, as shown in figure \ref{fig:nh_det_vs_nh_inp}, our
absorption estimation method only weakly constrains $N_H$ for sources
with very small absorbing columns.  Therefore, in order to reduce the
effect on the 3D-KS test of the large scatter at low absorption
levels, all sources with absorption below a threshold of $N_H =
10^{21}$~cm$^{-2}$, are taken to lie at this threshold.

The conversion from the 3D-KS statistic, to a probability, is rather
dependent on the size of the sample and the correlations within it
\citep{fasano87}.  This is important for the test we wish to apply
here because of the strong correlation between $z$ and $L_{2-10}$ in
the sample.  Therefore, we have adapted the method described in
\citet{dwelly05} in which the conversion from the 3D-KS statistic to a
probability is calculated numerically for the actual correlations, and
real number of sources in the tested data set.  The results of the
3D-KS tests are shown in table \ref{tab:stat_test}. We see that only
the \gillia\ $N_H$ model is able to reproduce the distribution of the
\xcdfs\ sample in $N_H,z,L_{2-10}$ space with better than 1\%
probability.

\begin{table*}
\caption{Statistical comparison of the \xcdfs\ sample with the
predictions of the AGN population models. Columns 2 and 3 show the
results of two $\chi^2$ tests of the ability of the simulated $N_H$
models to reproduce the redshift dependence, and the luminosity
dependence of the absorbed fraction measured in the \xcdfs\ sample.
The redshift test is computed using the 0$<z<$1, 1$<z<$2, 2$<z<$3, and
3$<z<$4 bins, and the luminosity test is calculated using the $10^{40}
< L_{2-10} < 10^{42}$, $10^{42} < L_{2-10} < 10^{43}$, $10^{43} <
L_{2-10} < 10^{44}$, and $10^{44} < L_{2-10} < 10^{45}$~erg~s$^{-1}$
bins.  Column 4 shows the Kolmogorov-Smirnov (KS) probabilities that
the $N_H$ distribution of the \xcdfs\ sample and the $N_H$
distribution predicted by each model population follow the same
underlying distribution.  Column 5 shows the three dimensional KS test
probability that the distributions of the sample and model sources in
$N_H$, $z$ and $L_{2-10}$ space follow the same underlying
distribution.  }
\label{tab:stat_test}
\begin{tabular}{@{}l@{\hspace{20mm}}cc@{\hspace{20mm}}cc}
\hline
            &   \multicolumn{2}{l}{Distribution of the absorbed fraction}  & \multicolumn{2}{c}{Overall source distribution} \\
$N_H$ model &        in redshift  & in $L_{2-10}$           &   in $N_H$      & in $N_H,z,L_{2-10}$  \\
            &   $\chi^2/dof$ (prob.) &  $\chi^2/dof$ (prob.) & KS prob.     & 3D-KS prob.  \\
\hline
\betatwo   &       19 / 4 (0.0004) &   18 / 4  (0.001) &   0.002              &  0.0009     \\
\betafive  &      9.6 / 4 (0.05)   &  8.5 / 4  (0.07)  &   0.06               &  0.001      \\
\betaeight &      6.0 / 4 (0.20)   &  5.5 / 4  (0.24)  &   0.001              &  0.005      \\
\treister  &      16  / 4 (0.003)  &  14  / 4  (0.007) &   $3\times 10^{-10}$ &  $<$0.00002 \\
\gillia    &      6.4 / 4 (0.17)   &  4.8 / 4  (0.31)  &   0.35               &  0.03      \\
\gillib    &       15 / 4 (0.005)  &   12 / 4  (0.02)  &   0.001              &  0.003      \\
\ueda      &      18  / 4 (0.001)  &   17 / 4  (0.002) &   0.01               &  0.00002  \\
\hline
\end{tabular}
\end{table*}

\section{Discussion}
\label{sec:discussion}

We have carried out statistical comparisons of the absorption
distribution found in the \xcdfs\ sample to the absorption
distributions predicted by a number of $N_H$ models.  A simple measure
of the relative importance of absorbed and unabsorbed AGN over a
range of redshifts and luminosities is found by measuring the fraction
of AGN which are significantly absorbed.  This tells us about the
relative numbers of absorbed and unabsorbed AGN, and the
redshift/luminosity dependence of this distribution.  In addition, the
high fidelity of our $N_H$ estimation process means that an analysis
of the shape of the recovered absorption distribution in the \xcdfs\
sample is informative. This is important for XRB synthesis models
such as that of \citet{treister05} in which the shape of the
$N_H$ distribution is closely related to the geometry of some ``typical''
absorbing torus.

\subsection{Ability of AGN population models to reproduce the absorption distribution in the \xcdfs\ sample}

The total $N_H$ distribution of the \xcdfs\ sample (see
fig. \ref{fig:nh_histo}) reveals that there is a wide range of
absorbing columns present in the AGN population.  The $N_H$ models we
have tested reproduce the observed distribution with varying degrees
of success (see table \ref{tab:stat_test}).  In section
\ref{sec:nh_dist_kstest}, the \gillia\ $N_H$ model was seen to
provide the best match to the shape of the observed $N_H$
histogram.  It is remarkable that the distribution of absorption in
local Seyfert-2 galaxies, (on which the \gillia\ model is based),
provides a good match to a sample which reaches to QSO
luminosities and to $z = 3.7$. However, this test
only compares the {\em total} $N_H$ distribution.  So if the
underlying $N_H$ distribution of AGN is actually dependent on redshift and/or
luminosity, then the total observed $N_H$ distribution will depend
on where in redshift/luminosity space the sample lies.

Therefore, in section \ref{sec:absorbed_fraction} we divided the
\xcdfs\ sample into several bins, firstly in luminosity, and then in
redshift, and tested how well the $N_H$ models matched the observed
absorbed fraction in each bin. Because it examines the {\em fraction}
of absorbed sources, this comparison should not depend strongly on
differences between the {\em total} redshift/luminosity distributions
seen in the \xcdfs\ sample and predicted by the population models.  We
found that there was a marked contrast in the ability of the different
$N_H$ models to reproduce the redshift and luminosity dependence of
the absorbed fraction measured in the \xcdfs\ sample.  The \betatwo,
\ueda, \gillib, and \treister\ $N_H$ models are all unable to
reproduce the pattern seen in the \xcdfs\ sample.  However, the
\betaeight\ and \gillia\ $N_H$ models provide statistically adequate
(probabilities of order 0.20), fits to both the redshift and
luminosity dependence of the absorbed fraction in the \xcdfs.  The
\betafive\ model also provides a statistically adequate match, but
less well than the latter two models.

The \betatwo\ and \treister\ $N_H$ models are both poor descriptions
of the AGN in the \xcdfs\ sample, and are rejected with high
confidence in all of the statistical tests. The former significantly
under-predicts the number of absorbed AGN, and the latter significantly
over-predicts the number.

The strong downturn in the absorbed fraction at high luminosities
predicted by the \ueda\ $N_H$ model is not seen in the \xcdfs\
sample. In addition, the \ueda\ $N_H$ model predicts a total absorbed
fraction of 30.1\% compared to $\ge 38\%$ seen in the \xcdfs\
sample.  This implies that if indeed there are relatively fewer
absorbed AGN at high luminosities, then the downturn can only be
important at higher luminosities ($> 10^{45}$~erg~s$^{-1}$) than 
are covered by this sample.

We do not see evidence for the increase in the absorbed fraction from
$z=0$ to $z=1.3$ predicted by the \gillib\ model. There is a
suggestion that the absorbed fraction in the \xcdfs\ sample does
increase at much higher redshifts ($z>3$).  However, as there are only
seven \xcdfs\ sources in this redshift range, no definitive conclusion
can be drawn from this dataset alone.  At such high redshifts it is
difficult to measure even large absorbing columns ($N_H \le
10^{23}$~cm$^{-2}$) because most of the effects of the absorption are
shifted out of the \xmm-EPIC bandpass.  Five of the \xcdfs\ objects at
$z>3$ have been spectroscopically identified by \citet{szokoly04}; two
broad-line AGN (BLAGN), and three ``high excitation line'' galaxies
(HEX). The optical classifications of the high redshift objects tally
with the X-ray determinations of their properties; the BLAGN have
X-ray colours consistent with little or no absorption, but the three
HEX objects are heavily absorbed.  The two other \xcdfs\ objects at
$z>3$ have only photometrically determined redshifts, their X-ray
colours indicate they both have significant absorption.

Of the seven $N_H$ models examined in this study, the two which incorporate
some redshift or luminosity dependence are both strongly rejected.
The luminosity dependent \ueda\ $N_H$ model predicts that there are
few ``type-2'' QSOs, whereas the redshift dependent \gillib\
$N_H$ model suggests that nearly all of the accretion power at $z > 1.3$
is obscured.  Neither of these scenarios fits the pattern
seen in the \xcdfs\ sources, which is much better described by models
in which a similar distribution of absorption is found in the AGN
population at all redshifts and luminosities (namely the \betaeight\ 
and \gillia\ $N_H$ models).

\subsection{The \xcdfs\ sources without redshift determinations}

\begin{figure}
\begin{center}
\includegraphics[angle=270,width=80mm]{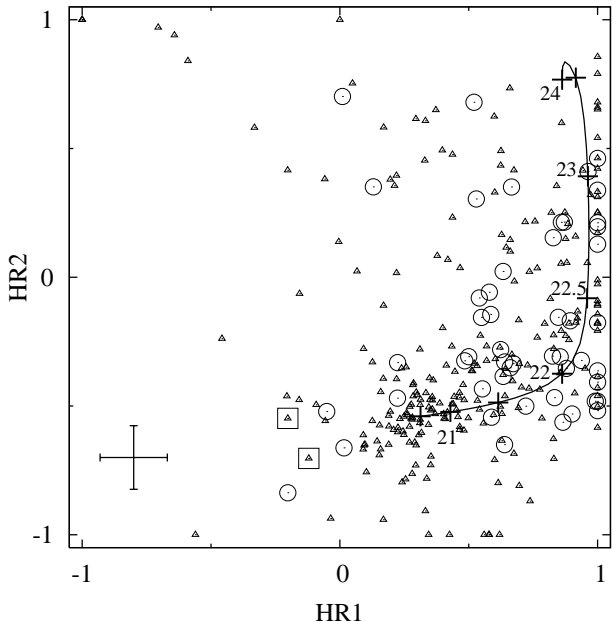}
\end{center}
\caption{The X-ray hardness ratio distribution of the \xcdfs\
sample. The optically identified sources are shown with triangles. 
Sources without optical identifications are marked with
circles. The two sources which do not match any objects in the
simulated library are highlighted with boxes. For clarity, we show the
median hardness ratio errors for the whole sample rather than the
errors for each source.  The solid line is the path in hardness ratio
space for a model AGN lying at $z=1$, with $\Gamma=1.9$ and with
absorbing columns ranging from zero to $10^{24}$~cm$^{-2}$
(graduations are marked, and labelled where space permits, 
for absorbing columns of log$N_H$ = 24, 23.5,
23, 22.5, 22, 21.5, 21 and 19). The hardness ratios, $HR1$ and $HR2$
are defined in the text. }
\label{fig:hardness_ratios}
\end{figure}

There are 50 \xcdfs\ sources for which we do not have a spectroscopic
or photometric redshift.  Here we discuss the nature of these X-ray
detections.  For ``type-1'' quasars at high redshift, optical
identification is made relatively easy by prominent, broad emission
lines in the rest frame UV.  However, for absorbed AGN, in which the
optical spectrum is primarily that of the host galaxy, determination
of redshifts is much more difficult.  In particular, the so called
``photo-z desert'' ( $1.5 < z < 2.5$) occurs where typical galaxy
spectra have no easily identifiable spectroscopic features in the
observed optical band.  However, with the addition of NIR data, this
problem can be attenuated.  For example, by utilising deep NIR
photometry, \citet{mainieri05b} were able to photometrically identify
faint ($R > 25$) counterparts to 1Ms \chandra\ sources in the \cdfs\
sample.  Indeed, these authors showed that these faint objects lie on
average at higher redshift than the optically brighter counterparts.
The unidentified objects in our sample are nearly all optically faint;
43/50 of the unidentified sources have $R > 24.5$ (see fig.
\ref{fig:flux_vs_R}).  The three optically brightest unidentified
sources do not have \combo\ redshift estimates because they lie close
to bright stars.  It is reasonable to assume that most of the
unidentified sources have no redshift estimates because they lie at $z
>1.5$, that is, they are beyond the upper redshift limit for galaxies
in the \combo\ survey (where the $4000$\AA\ break has left the reddest
band).

It is difficult to allow for such a bias against high redshift objects
being identified in our \xcdfs\ sample, because the relationship
between X-ray and optical properties of absorbed and unabsorbed
sources is far from clear cut.  The effect is mitigated by the high
completeness ($84\%$) of optical identification in our sample.
However, figure \ref{fig:hardness_ratios} shows that on average, the
unidentified sources have harder X-ray spectra than the identified
sources. For example, the median $HR1$ value is $0.44$ for the
optically identified extragalactic sources, whereas it is $0.70$ for
the unidentified sources.  We have tested the significance of this
difference by making a two-dimensional Kolmogorov--Smirnov comparison
of the distributions of the identified and unidentified sources in
($HR1,HR2$), space.  The probability that the unidentified and
identified extragalactic sources follow the same underlying
distribution in ($HR1,HR2$) is $6\times 10^{-5}$.

As an experiment, we assign the unidentified \xcdfs\ sources a nominal
redshift of $z=2$ (approximately the middle of the ``photo-z
desert''), and then use the $N_H/L_X$ estimation technique in the same
way as for the identified sources.  We find that 38/50 of the objects
have $N_H>10^{22}$~cm$^{-2}$, and their median intrinsic luminosity is
$10^{44.0}$erg~s$^{-1}$.  Figs. \ref{fig:z_lum_histos} and
\ref{fig:nh_histo} show the resulting luminosity and $N_H$
distributions in the \xcdfs\ sample if we make this assumption.  The
number of \xcdfs\ sources in the $N_H \ge 10^{22}$~cm$^{-2}$,
$L_{2-10} \ge 10^{44}$erg~s$^{-1}$ regime (a common definition for a
``type-2'' QSO) is doubled from 23 to 46 if the unidentified objects
are placed at $z=2$.  If in fact the unidentified sources lie at an
average redshift greater than 2, then obviously the numbers of
absorbed sources and their median luminosity will be higher. With the
addition of these luminous, high redshift AGN, the observed
redshift/luminosity distribution is closer to that predicted by the
XLF of \citet{miyaji00}. However, the unidentified sources cannot
fully produce the numbers of high luminosity, high redshift AGN
predicted by the XLF model of \citet{miyaji00}.

\subsection{The luminous absorbed AGN population}
Rather few luminous absorbed (type-2) QSOs have been found in X-ray
surveys to date. There are certainly fewer than predicted by XRB
synthesis models in which a large fraction of the XRB is made up of
luminous but highly absorbed AGN at redshifts 1.5--2.5
\citep{gilli01,comastri95}.  In response to the lack of type-2 QSOs,
models have been devised in which the fraction of AGN with significant
absorption declines with increasing luminosity
\citep{barger05,ueda03}. The physical interpretation is that highly
luminous QSOs have the power to remove a significant fraction of the
surrounding material, effectively increasing the opening angle of any
circumnuclear structure, a so called ``receding torus'' model
\citep[e.g.][]{lawrence91}.  The X-ray samples detected in recent
\chandra\ pencil-beam studies {\em do} contain large numbers of
absorbed AGN, but they lie at lower redshifts, and have lower
luminosities than the peak in type-1 QSO activity ($L_X \sim
10^{44}$~erg~s$^{-1}$, $z\sim1.5-2$).  It is this low redshift, low
luminosity absorbed population which is postulated to take the place
of the type-2 QSOs in making up the hard spectrum of the XRB. These
findings, if they are taken at face value, raise important questions
about the cosmic history of the AGN population.  The implication is
that, with soft X-ray and optical surveys, we have already detected
the majority of intrinsically luminous accretion powered objects in
the form of type-1 QSOs.  In this scheme, because the absorbed
population lies at low redshifts and luminosities, the total energy
output that is powered by accretion, integrated over cosmic
timescales, is significantly reduced from the predictions of the
simplest ``unified'' schemes.

A few rare examples of type-2 QSOs, selected in X-ray and/or optical
surveys have been reported in the last few years
\citep[e.g.][]{norman02,dellaceca03,gandhi04,mainieri05a,ptak06}.  In
fact, two of these, the \citet{norman02} and \citet{mainieri05a}
objects, appear in our \xcdfs\ sample. The type-2 QSOs appear to be
far less numerous than the large population predicted by say the
\citet{gilli01} XRB synthesis model.  However, recent mid-infrared
surveys with \spitzer\ have started to reveal the existence of
significant numbers of type-2 QSOs.  Surveys in the mid-infrared are
sensitive to emission originating in dusty material that is heated by
a compact heat source, i.e. a powerful AGN.
\citet{martinezsansigre05} have demonstrated that type-2 QSOs at $z>2$
can be selected efficiently by choosing objects bright at 24\micron,
but faint at near-infrared and radio wavelengths.  Using these
criteria, they identified a population of luminous obscured QSOs
between 1 and 3 times as numerous as the population of luminous type-1
QSOs found by optical surveys \citep[e.g.][]{croom04}.  The absorbed
fraction of 50--75\% found by \citet{martinezsansigre05} is comparable
to the absorbed fraction at high luminosities of $\sim 75\%$ predicted
by the two $N_H$ models (namely the \betaeight\ and \gillia\ $N_H$
models) which best match the pattern of absorption in the \xcdfs\
sample.

\subsection{AGN with complex X-ray spectra}

For the purposes of this study, we have considered only a rather
simple AGN spectral model; an absorbed or unabsorbed power law (having
a range of intrinsic spectral slopes) with a component of reflected
radiation.  However, high signal-to-noise X-ray spectra of brighter
samples \citep[e.g.][]{piconcelli03,mateos05a,mateos05b,page06}, have
revealed that a small fraction of AGN have ionised rather than neutral
absorbers, and that many AGN have additional spectral components. In
particular, in the sample of \citet{piconcelli03}, the spectral fits
to $\sim 35\%$ of the absorbed AGN were improved by the addition of an
extra soft component.  The most common hypotheses for the origin of
this soft component are that it is either due to strong star formation
activity in the host galaxy, or that it is reprocessed emission from
the AGN itself.

It is possible that soft X-ray emission powered by intense star
formation in the host galaxy could contribute to the spectra of some
AGN, causing us to underestimate their absorbing columns.  The single
{\em most} X-ray luminous starburst galaxy in the local Universe,
NGC3256, has a 0.5--10~keV luminosity of $\sim 5 \times
10^{41}$~erg~s$^{-1}$ in our assumed cosmology \citep{lira02}.  If
similarly powerful starbursts are common in AGN host galaxies, then we
will expect to underestimate $N_H$ for some AGN. However, in the
\xcdfs\ sample, nearly all (85\%) of the optically identified sources
have observed 0.5--10~keV fluxes $\ge 10$ times the flux expected from
NGC3256 if it were placed at the same redshift (K corrected assuming
$\Gamma = 2.5$).  Any contribution to the X-ray colours from star
formation is expected to be dwarfed by the primary AGN component in
these luminous objects.

X-ray emission from AGN can reach the observer indirectly via
reprocessing in a body of photo-ionised plasma which extends well
beyond the obscuring ``torus'' \citep[e.g.][]{turner97}.  For AGN
where much of the direct X-ray continuum is absorbed, the reprocessed
emission may constitute a large fraction of the observed soft X-ray
flux.  For example, in the archetypal Seyfert-2 galaxy, NGC1068,
virtually all of the soft X-ray flux can be attributed to
reprocessed emission from the AGN \citep{kink02}. However, the
intensity of the scattered component is generally only a small
fraction of the primary power law component \citep[][]{turner97,page06}.
So unless the AGN is very heavily absorbed (like NGC1068), the
direct power law component will still dominate the observed spectrum.

%A very small number of the AGN in the \xmm\ samples of
%\citet{piconcelli03} and \citet{page06} have X-ray spectra which are
%best fitted by a spectral model in which the absorbing material is
%ionised.  The net effect is an X-ray spectrum which has a deficit of
%flux at intermediate energies compared to a simple power law.  Even
%with high signal to noise X-ray spectra, it is often hard to
%distinguish between an ionised absorber, and an AGN with a neutrally
%absorbed power-law component plus an additional soft component.  We do
%not expect to be able to differentiate between these cases in this
%work because we have examined only broad band X-ray colours.

%As discussed in appendix \ref{app:unusual}, there are two bright
%unabsorbed AGN in the \xcdfs\ sample with X-ray colours indicative of
%a very steep spectrum at low photon energies, but a normal slope at
%harder energies. The redshifts, count rates and hardness ratios of
%these two objects were not matched by any of the simulated library
%sources.  There may be other examples of these soft, unabsorbed AGN in
%the \xcdfs\ sample, which have less pronounced spectral shapes and/or
%are measured with lower signal to noise.  However, the soft spectra of
%such objects will not prevent our $N_H/L_X$ estimation method from
%correctly determining these AGN to be unabsorbed.

We do include a reflection component in our spectral model. This has
the effect of hardening the spectrum at high photon energies
($E>5$~keV).  Therefore, the addition of a reflection component to the
spectral model reduces the absorbing column that is required to
reproduce the X-ray spectrum of an absorbed AGN. For example, consider
an example AGN at high redshift (where the additional reflection
component becomes more important), and let us assume that for this
AGN, we measure a hardness ratio between the 2--5 and 5--10~keV bands
of $HR3 = -0.5$.  For a simple power-law model with slope $\Gamma =
1.9$, $HR3 = -0.5$ corresponds to an absorbing column of
$10^{23.39}$~cm$^{-2}$ at $z=2$, in comparison, a slightly smaller
column of $10^{23.17}$~cm$^{-2}$ is required to get the same $HR3$
value with the power-law plus reflection spectral model. Our
$N_H/L_X$ estimation technique uses information from all three $HR$
measures, including the $HR1$ and $HR3$ bands where the addition of a
reflection component in the model spectrum typically makes only a
small difference.  Thus we do not expect the results of this study to
be strongly affected by our particular choice of spectral model.

In summary, we expect these extra spectral components to have only a
small influence on our analysis of the $N_H$ function in the \xcdfs\
sample. Although additional soft components may be a common feature
in absorbed AGN, their amplitudes are small in comparison to the
primary power law component. The net effect will be that we
underestimate the absorbing columns of some AGN.  
%In some sources,
%an ionised absorber could cause our $N_H$ estimation scheme to
%miscalculate the absorption, but the expected numbers of such objects
%are small.

\subsection{Why have other X-ray surveys arrived at different conclusions?}
Many authors have reported a lack of X-ray selected absorbed AGN at
high redshifts and with high luminosities, and have developed
luminosity dependent absorption schemes to explain this phenomenon
\citep[e.g.][]{franceschini02,steffen03,ueda03,barger05,lafranca05,lamastra06}.
In the following subsections we discuss possible reasons why these
studies have arrived at different conclusions to our own.  Perhaps
because of its excellent point source sensitivity, the majority of
recent deep X-ray survey studies have relied on data from
\chandra. The AGN population studies which have used \xmm\
data have typically been limited to relatively bright fluxes
\citep[e.g.][]{piconcelli03,caccianiga04,dellaceca04,mateos05a}.
Therefore here we pay particular attention to the differences
between the capabilities of \chandra\ observations with respect to
the \xmm\ data used in this work.

\subsubsection{Spectroscopic incompleteness}
The 2--8~keV band luminosity function derived by \citet{barger05}
relies at its faint end on an X-ray sample which is only 50-60\%
spectroscopically identified.  Photometric redshifts raise their
identified fraction, but will systematically miss objects in the
1.5--2.5 redshift interval.  We have shown that the 50 sources without
redshifts in our sample have harder than average colours, and
therefore could be intrinsically luminous but heavily absorbed QSOs.
It is possible therefore that the \citet{barger05} sample is
underestimating the size of the population of such objects. This
effect could explain the lack of objects in their sample lying at
$z>1$, and having {\em observed} 2--8~keV luminosities below
$10^{44}$~erg~s$^{-1}$.  

%We have also illustrated the importance of
%high angular resolution in correctly identifing optical counterparts to 
%X-ray sources.

%Manual examination of the optical and \chandra\ imaging in the \cdfs\
%(in particular, the high spatial resolution GEMS z-band images),
%reveals that of the 167 1Ms \chandra\ sources which are matched to
%\xcdfs\ sources, 10\% of the optical counterparts chosen by
%\citet{zheng04} are probably incorrect (see section
%\ref{sec:optical_counterparts}).  We have checked the optical
%identifications only for the $\sim$half of the full 1Ms \chandra\
%catalogue sources which have \xmm\ counterparts, so there may be
%additional cases of misidentification in the \citet{zheng04} sample.
%If the true redshifts of the misidentified sources are similarly
%distributed to the rest of the \citet{zheng04} sample, then the 16
%misidentifications will not be important.  However,
%nearly all of the preferred optical counterparts are fainter than the
%counterparts chosen by \citet{zheng04}, and so may well lie at higher
%redshifts. If this is the case, then the 16 misidentified sources
%could significantly affect the size of the high-redshift population.
%For example, of the 167 1Ms \chandra\ sources manually examined here,
%only 7 lie at very high redshifts ($z>3$); only a handful of the 
%misidentified sources would be required to increase significantly 
%the size of this population.

\subsubsection{X-ray selection function}
The faintest sources in the sample of \citet{ueda03} are taken from
the first 1Ms observations of the \chandra\ Deep Field North.
A 2--8~keV flux limit of $3.0 \times 10^{-15}$~\cgs\ was
applied to define the \citet{ueda03} sample (much shallower than the
limit of the \chandra\ data).  Many of the high redshift, absorbed
sources in our sample would not have been selected with this
criterion. Of the \xcdfs\ sources with $z>1$ and $N_H >
10^{22}$~cm$^{-2}$, most (29/52) have 2--8~keV fluxes below this limit
(calculated from their 2--5~keV fluxes, assuming a power law slope of
$\Gamma = 1.4$).  Therefore it is not surprising that by extrapolating
the XLF and $N_H$ model of \citet{ueda03} to fainter flux limits, we
are unable to reproduce fully the \xcdfs\ sample.

\subsubsection{The broad energy range of \xmm\ EPIC}
The combined EPIC detectors have an effective area (mirror
plus detector quantum efficiency) of more than 1000~cm$^2$ at 7~keV
compared to only $\sim$100~cm$^2$ for ACIS-I.  The
additional sensitivity of EPIC compared to ACIS-I at hard energies has
two important effects for surveys of heavily absorbed AGN.  Firstly,
absorbed AGN are detectable with EPIC because of its sensitivity to
the high-energy unabsorbed part of their spectrum. Secondly, the wide
spectral range of EPIC provides better constraints on the spectral
shape of sources, and hence allows a better measurement of absorbing
column and intrinsic luminosity.

\xmm\ EPIC is also usefully sensitive over a much broader energy range
(0.2--10~keV) than \chandra\ ACIS-I (0.7--7~keV).  
Section \ref{sec:nh_est_method} describes how we took advantage of the soft
X-ray sensitivity of EPIC by including data from the 0.2--0.5~keV band
in our $N_H$ estimation technique.  At low redshifts, we are sensitive
to columns of $N_H \ge 10^{21.1}$~cm$^{-2}$.  More importantly, for
redshifts up to $\sim 3$, we can reliably detect columns of $N_H \ge
10^{22}$~cm$^{-2}$, which is the traditional dividing line between
absorbed and unabsorbed AGN (see section \ref{sec:individual}).
However, X-ray classification schemes that are based on \chandra\
hardness ratios may miss considerable absorption in many high-$z$ AGN,
where even substantial absorbing columns can be shifted out of the
ACIS-I sensitivity range.  For example, the X-ray classification
scheme of \citet{zheng04} uses a fixed \chandra\ hardness ratio as a
dividing line between ``type-1'' and ``type-2'' AGN/QSOs.  We find
that there are nine high redshift ($z > 2$) absorbed ($N_H \ge
10^{22}$~cm$^{-2}$) sources in the \xcdfs\ which are classified as
``type-1'' AGN/QSOs by \citet{zheng04}. The \chandra\ XID numbers
\citep[see][]{giacconi02} of these sources are 6, 62, 85, 122, 159,
179, 225, 506 and 517.

% What is more, the steep
%decline of the ACIS-I response at high energies means that even for
%very hard spectrum sources, most of the counts in the 2--7~keV band
%will be caused by photons with energies $<$5~keV. For example, for a
%$\Gamma = 1.0$ power law spectrum, 81\% of counts are due to photons
%in the 2--5~keV band (calculated using the online
%PIMMS\footnote{http://asc.harvard.edu/toolkit/pimms.jsp} tool). 

\subsubsection{The effects of large scale clustering}
\label{sec:clusters}
The redshift distribution of X-ray sources in the 1Ms \chandra\
catalogue in the \cdfs\ is dominated by narrow over-densities at
$z=0.67,0.73$ containing $\ge 38$ sources \citep{gilli03}.  These
over-densities are also seen in our sample with at least 26 \xcdfs\
sources lying in these redshift spikes.  We note that large redshift
spikes at $z \le 1$ also appear in the 2Ms \chandra\ Deep Field North
catalogue \citep{barger03,gilli05}. It is not yet clear whether such
clustering of AGN at $z \le 1$ is a ubiquitous feature of the
Universe, and therefore common over the whole sky. Despite the
enhancements at low redshifts, the \cdfs\ field has a total sky
density of X-ray sources somewhat lower than other deep X-ray fields
\citep{rosati02,manners03,nandra04,loaring05}. This suggests that the
\cdfs\ is under dense at higher redshifts.  As described in section
\ref{sec:model_populations}, for each of the simulated model AGN
populations the normalisation of the XLF model was adjusted so that
the simulated source counts matched the observed integral
extragalactic source counts above a 0.5--2~keV flux of $2 \times
10^{-15}$~\cgs.  We found that an XLF normalisation 0.7 times that
given in \citet{ueda03} is required in order for the simulated source
counts predicted by their full population model (XLF and \ueda\ $N_H$
model) to match the \xcdfs\ extragalactic source counts.

\begin{figure}
\begin{center}
\includegraphics[angle=0,width=80mm]{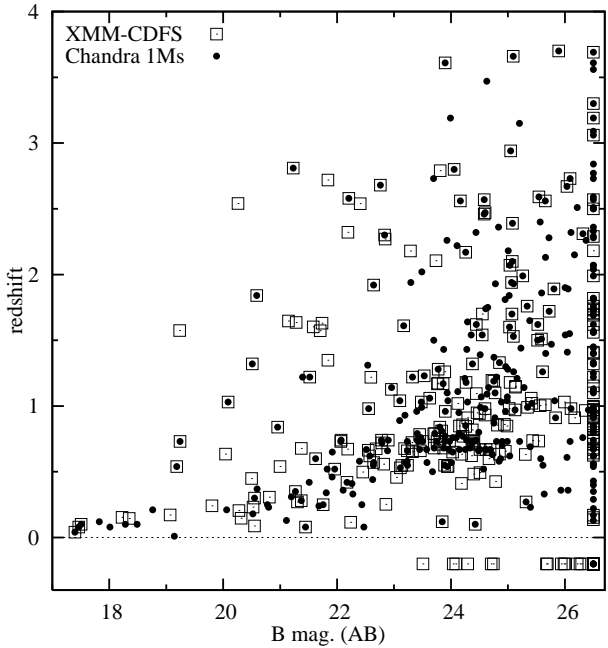}
\end{center}
\caption{The redshifts and B band magnitudes of the extragalactic
objects in the \xcdfs\ sample (squares), and the 1Ms \chandra\ \cdfs\
catalogue (filled circles). The optical magnitudes are
calculated by matching the X-ray sources to objects 
in the EIS catalogue \citep{arnouts01}.}
\label{fig:opt_vs_z}
\end{figure}

In figure \ref{fig:opt_vs_z} we compare the redshift versus
B~magnitude distribution of our sample with the distribution of the
sources in the 1Ms \cdfs\ catalogue.  We see that the \cdfs\ has
rather few of the optically bright, high redshift objects typically
detected in optically selected quasar surveys \citep[e.g. 2QZ/SDSS
][]{croom04,abaz03}.  For example, there is only one optically
bright ($B_{AB} < 21$) AGN at $z > 2$ in the \xcdfs\ sample, and none
in the region covered by the \chandra\ 1Ms observations.  The
predicted numbers of such objects from the optical QSO luminosity
function of \citet{croom04} are 3.1 in the $\sim 0.18$~deg$^2$ covered
by the \xmm\ observations, and 1.9 in the $\sim 0.11$~deg$^2$ of the
1Ms \chandra\ coverage.  Given this mean sky density, there is an 82\%
probability that we would observe more than one $B_{AB} < 21$ quasar
at $z > 2$ in the \xcdfs. This is again consistent with the \xcdfs\
field being an under-dense region of the sky at high redshifts.

%\subsubsection{Contribution of the \xcdfs\ sample to the XRB}
Another way to investigate whether the \cdfs\ is an underdense region
of the sky is to compare the integrated emission from resolved X-ray
sources to that found in other deep fields, and to the mean intensity
of the extragalactic X-ray background.  We calculate the contribution
to the XRB for just the sources detected in the inner 10\arcmin\ of
the \xcdfs\ field, where the \xmm\ observations are most
sensitive. Note that we have excluded from our sample X-ray sources
associated with Galactic stars and groups/clusters of galaxies (see
section \ref{sec:optical_counterparts}).

Following \citet{worsley04}, we use
the \citet{deluca04} 2--10~keV XRB model to compute the expected
extragalactic XRB intensity in the 0.5--2, 2--5, and 5--10~keV bands.
\citet{deluca04} found that in the 2--10~keV band, the extragalactic
XRB is well described by a power law with slope $1.41 \pm 0.06$, and
total intensity of $2.24 \pm 0.16 \times 10^{-11}$~\cgsd, in close
agreement with the result of \citet{lumb02}. We do not
consider the 0.2--0.5~keV band here because the extragalactic XRB
intensity at these energies is poorly constrained.

The integrated flux of the \xcdfs\ sources and the mean extragalactic
XRB intensity are shown in table \ref{tab:contribution_to_xrb}.  For
comparison, \citet{worsley04,worsley05} have made similar calculations
for the \xmm\ observations in the Lockman Hole, and the \chandra\ deep
fields North and South.  For the Lockman Hole \citet{worsley04} find
that the integrated fluxes from their sources are equivalent to $90\pm
6\%$, $73 \pm 7\%$, $53 \pm 7\%$ and $42 \pm 7\%$ of the extragalactic
XRB intensity in the 0.5--2, 2--4.5, 4.5--7.5 and 7.5--12~keV bands
respectively.  Although the \xmm\ observations in the Lockman Hole are
somewhat deeper than in the \cdfs\ (680~ks of good pn time compared to
340~ks), the flux limits of the \citet{worsley04} sample are similar
to those of the \xcdfs\ sample.  More importantly, \citet{worsley04}
calculate their countrate-to-flux conversion factor assuming a power
law spectral model with slope $\Gamma = 1.4$, rather harder than the
$\Gamma = 1.7$ used here. Converting from countrate to flux using
$\Gamma = 1.4$ would change our calculated fluxes by factors of 1.092,
0.978 and 0.962 in the 0.5--2, 2--5 and 5--10~keV bands
respectively. Even using $\Gamma = 1.4$, the integrated 0.5--2~keV
flux of sources in the \xcdfs\ amounts to only $\sim 75\%$ of that
reported for the Lockman Hole by \citet{worsley04}. This result is
consistent with the \citet{worsley05} study which showed that the
integrated 0.5--2~keV emission from sources in the 1Ms \chandra\
\cdfs\ was significantly lower than that found in the \cdfn\ and
Lockman Hole.

We also note that
the 28 \xcdfs\ objects in the redshift spikes at $z=0.67,0.73$ account
for a significant part of the summed flux from the \xcdfs\ sample:
around 20\% of the summed 0.5--10~keV flux of sources within 10\arcmin
of the aim point. These findings are consistent with the hypothesis
that the \cdfs\ is a relatively under-dense region of the sky.

\begin{table}
\caption{The contribution of the \xcdfs\ sources to the intensity of
the extragalactic XRB. $I_{XRB}$ is the
extragalactic XRB intensity in units of $10^{-12}$~\cgsd\ (see text)
$\Sigma S_i/A$ is the summed flux from the
individual \xcdfs\ sources, divided by the geometric area, in units of
$10^{-12}$~\cgsd, and ``frac'' is the fraction of $I_{XRB}$ that this
constitutes.}
\label{tab:contribution_to_xrb}
\begin{tabular}{@{}cccc}
\hline
Energy band  & $I_{XRB}$ & $\Sigma S_i/A$  & frac \\
\hline
0.5--2~keV   & 8.1       &  $  5.15 \pm 0.02 $ & 0.64 \\
2--5~keV     & 10.3      &  $  5.11 \pm 0.04 $ & 0.50 \\
5--10~keV    & 12.6      &  $  5.78 \pm  0.1 $ & 0.46 \\
\hline
\end{tabular}
\end{table}

It is important to point out that in our statistical comparison of the
models and the \xcdfs\ sample in section \ref{sec:absorbed_fraction},
we first grouped the AGN into a number of bins in redshift/luminosity,
and then calculated the fraction (rather than the absolute number) of
AGN with significant absorption in each bin.  Therefore, even if the
AGN in the \cdfs\ field {\em do} have a distribution in
redshift/luminosity space which is unrepresentative of the Universe at
large, we still expect our findings about the AGN absorption
distribution to be valid.

\subsection{Comparison to the \chandra\ spectral analysis of \citet{tozzi06}}
\begin{figure*}
\begin{center}
\includegraphics[angle=0,width=72mm]{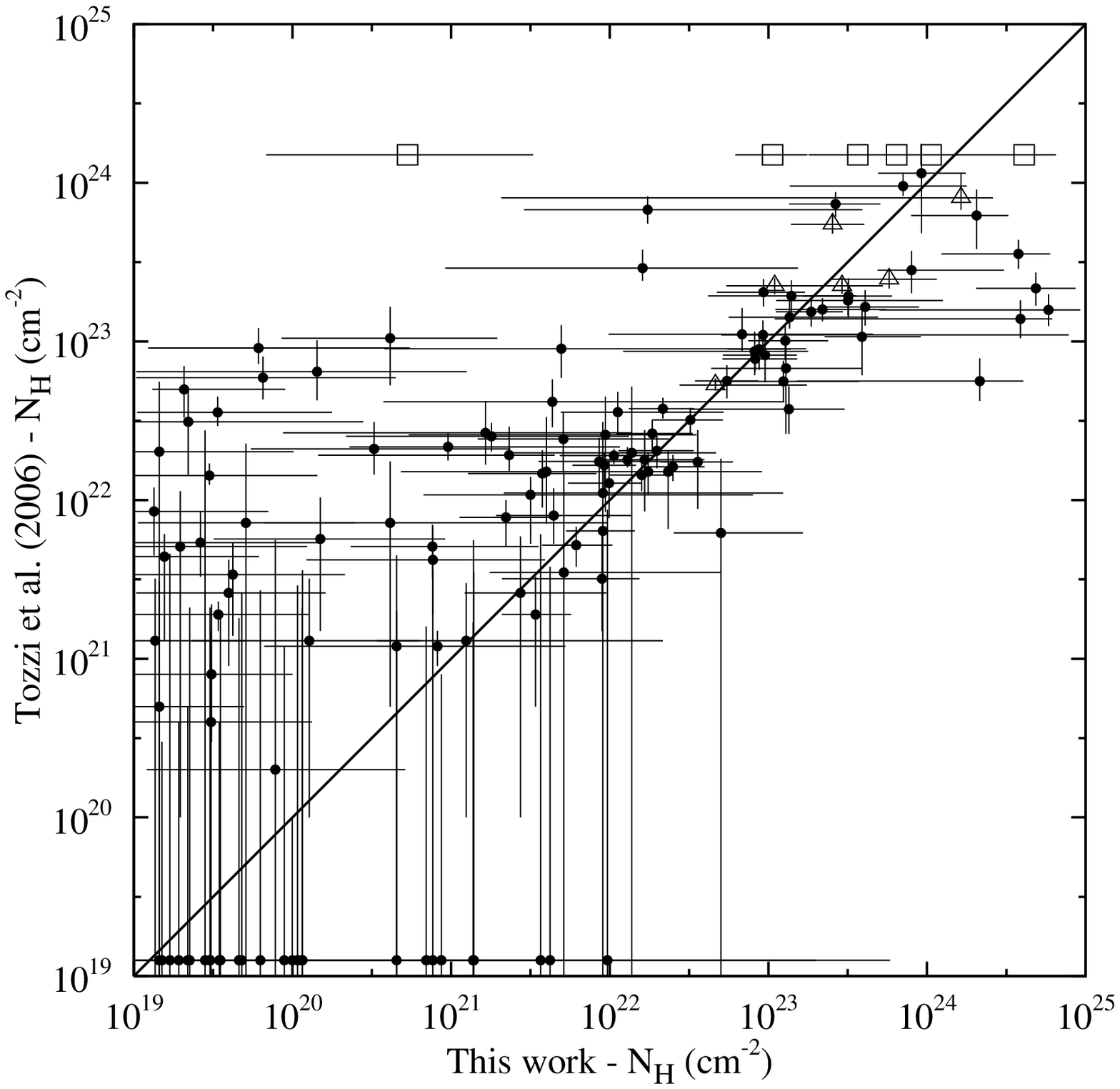}
\includegraphics[angle=0,width=72mm]{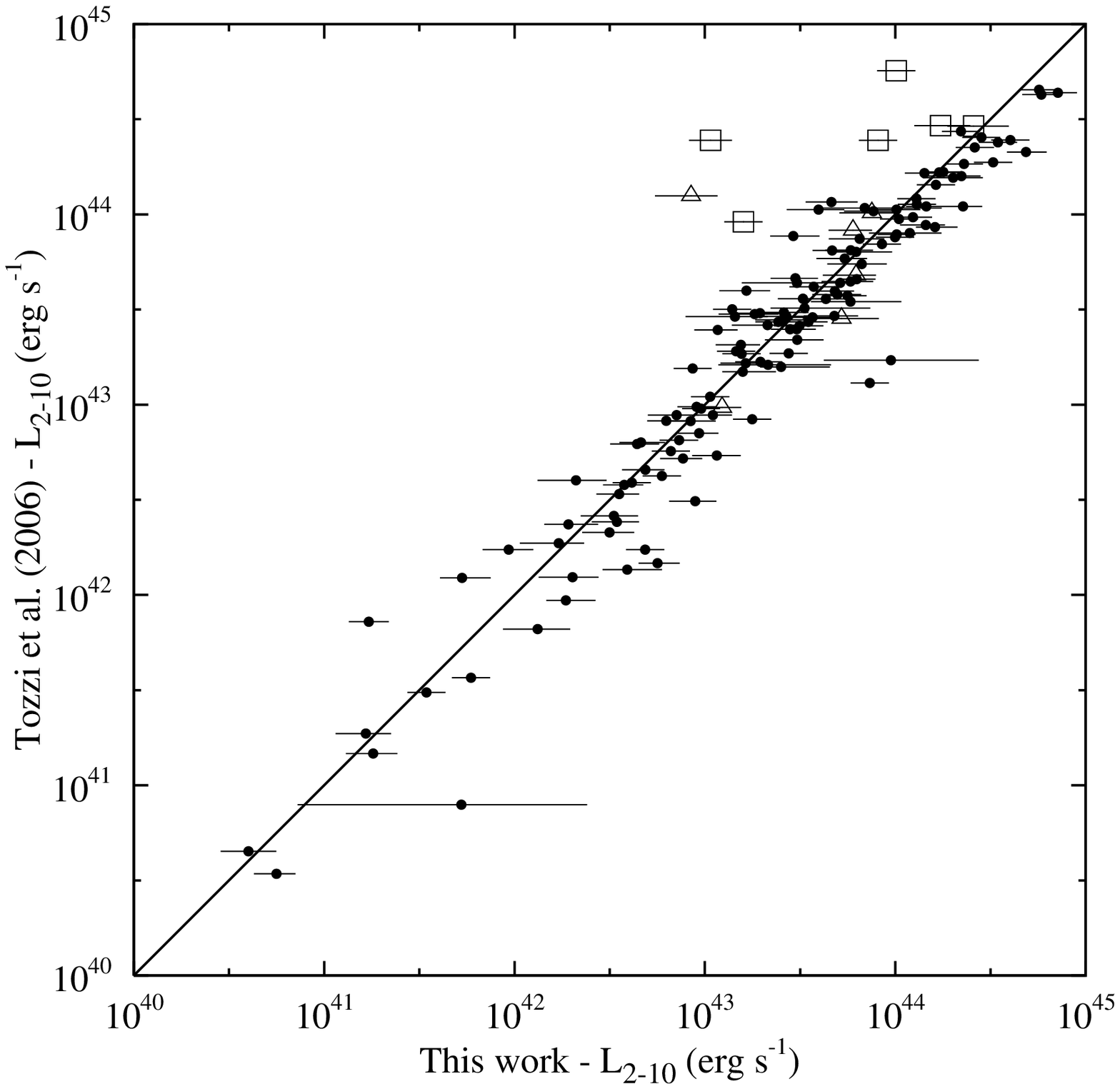}
\includegraphics[angle=0,width=72mm]{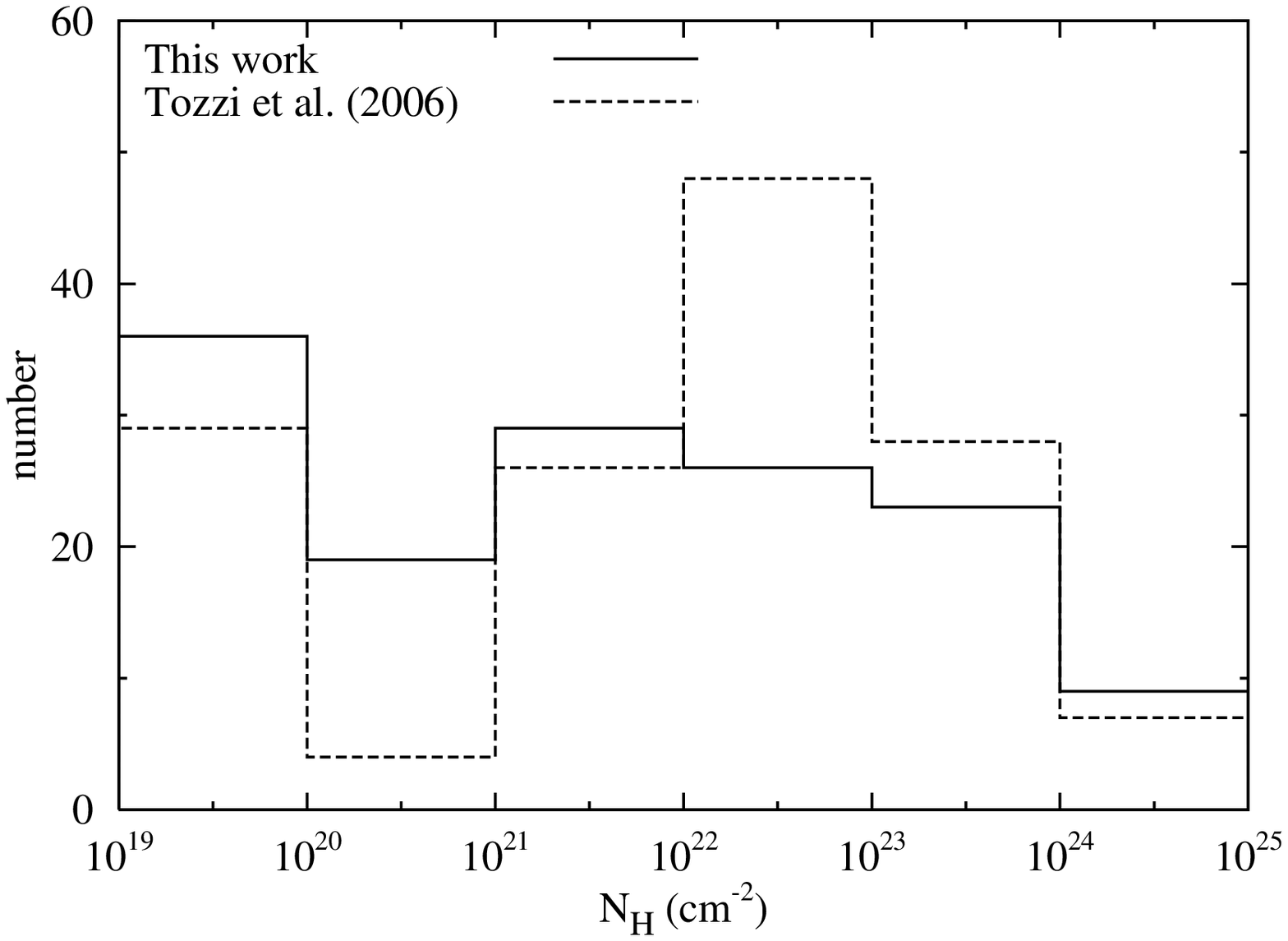}
\includegraphics[angle=0,width=72mm]{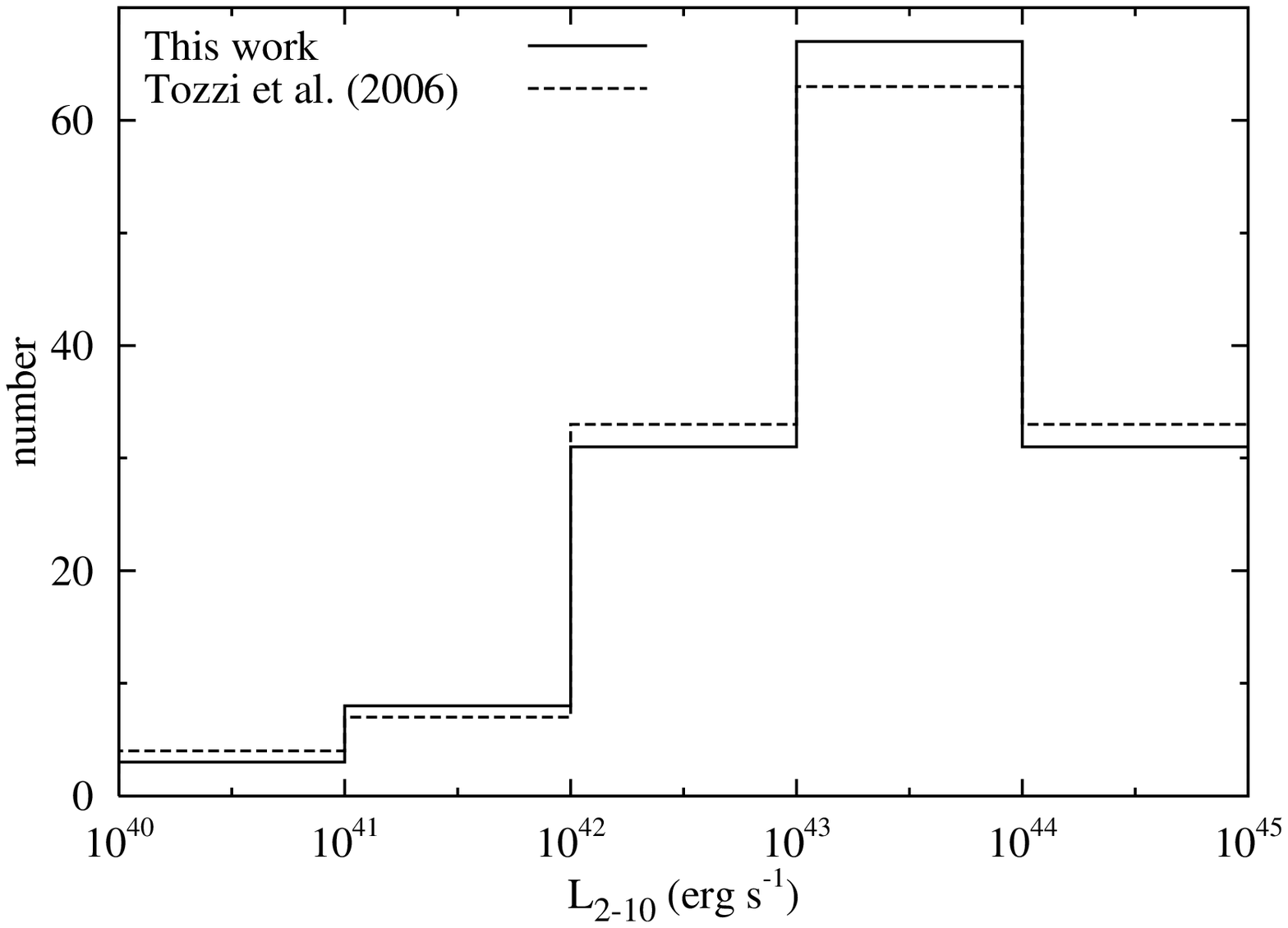}
\end{center}
\caption[Comparison with \citet{tozzi06}]{\small A comparison of the
$N_H$ (top left) and $L_{2-10}$ (top right) measurements for the 142
AGN which appear in both the \xcdfs\ sample, and the \citet{tozzi06}
sample, and for which we agree a redshift.  The sources determined to
have zero $N_H$ by \citet{tozzi06} are plotted at $N_H =
10^{19.1}$~cm$^{-2}$.  Sources treated by \citet{tozzi06} as ``Compton
Thick'', and so fitted with a pure reflection spectrum, are marked
with open boxes.  Sources determined to have an additional soft
component by \citet{tozzi06} are marked with triangles.  The lower two
panels show the equivalent histograms of absorption (left) and
luminosity (right) for the sources common to both samples. The sources
determined to have zero $N_H$ by \citet{tozzi06} are placed in the
leftmost $N_H$ bin.}
\label{fig:compare_to_tozzi}
\end{figure*}

Recently, \citet{tozzi06} have published the results of a \chandra\
X-ray spectral analysis of sources detected in the 1Ms \chandra\
imaging of the \cdfs. Here we briefly compare the results of the
latter study with our own. After careful correction for selection
effects \citet{tozzi06} find no evidence for a correlation between
absorption and intrinsic luminosity, in agreement with our findings.
The intrinsic (that is before selection effects) $N_H$ distribution
derived by \citet{tozzi06} is a log-normal distribution broadly
similar (at least below $10^{24}$~cm$^{-2}$) to the \treister\
model. However, the \treister\ model is strongly
rejected as a good description of the \xcdfs\ population (see table
\ref{tab:stat_test}), primarily because it
under-predicts the number of unabsorbed sources in the \xcdfs\ and
predicts too many sources with $N_H \sim 10^{22}$~cm$^{-2}$ (see
fig. \ref{fig:nh_histo}).

We now compare the absorption and luminosity measurements for the
sources common to the \xcdfs\ sample and the \citet{tozzi06} study.
Redshifts for some of the \citet{tozzi06} sources are taken from
\citet{zheng04}, several of which we have determined to have incorrect
optical counterparts. Therefore, for the purposes of this comparison,
we consider only 142 sources out of the 158 sources which appear in
both the \xcdfs\ and the \citet{tozzi06} samples.  In the left hand
panels of figure \ref{fig:compare_to_tozzi} we compare the $N_H$
measurements of these 142 AGN. It can be seen that there are some
differences in the absorbing columns determined by the two
studies. The most marked difference is for \xcdfs\ sources which we
determine to have effectively zero absorption ($N_H <
10^{21}$~cm$^{-2}$), but which are found to have a high ($N_H >
10^{22}$~cm$^{-2}$) absorbing column by \citet{tozzi06}. This effect
is apparent in both the $N_H$ scatter plot and the $N_H$ histogram,
with the \citet{tozzi06} study finding $\sim 2$ times as many sources
in the $10^{22} < N_H < 10^{23}$~cm$^{-2}$ bin.  The \citet{tozzi06}
spectral fitting has been carried out over the 0.7--7~keV
ACIS-I sensitivity range. As discussed by
\citet{tozzi06}, the limited soft energy range of the \chandra\ data
sometimes means that unabsorbed sources are spuriously scattered into
the high absorption ($N_H > 10^{22}$~cm$^{-2}$) regime.  This effect
could explain the differences between the $N_H$ distributions measured
for objects in common to \xcdfs\ and the \citet{tozzi06} study.

The luminosity measurements agree very well as demonstrated by the
right hand panels of figure \ref{fig:compare_to_tozzi}.  The
exceptions are the few sources for which \citet{tozzi06} have chosen
to fit with a pure reflection model. For such objects, the
luminosities we determine are typically lower than
determined by \citet{tozzi06}.

\subsection{Obscured black hole growth over cosmic timescales}
The absorption distribution of AGN in the \xcdfs\ sample is best
matched by population models in which obscured AGN are $\sim3$ times
as populous as unobscured AGN at all redshifts and luminosities, implying 
that most ($\sim 75\%$) supermassive black hole
growth was obscured. Various studies have attempted to reconcile the
observed accretion powered luminosity density at high redshifts, with
the locally observed relic black hole mass function
(e.g. \citealt{fabian99,yu02,elvis02,marconi04}).  \citet{marconi04}
found that the AGN population model of \citet[][XLF and $N_H$
function]{ueda03} was consistent with the local black hole mass
function if the mean accretion efficiency is $\sim 0.08$.  However, as
we have shown in this study, the observed numbers of luminous,
absorbed AGN are substantially higher than the predictions of the
\ueda\ $N_H$ model.  Coupled with the \citet{martinezsansigre05}
findings, the implication is that the mean accretion efficiency is
higher than the \citet{marconi04} result, or alternatively that the
local black hole mass density has been underestimated.

%\citet{yu02} found
%that even with a high accretion efficiency ($\sim 0.1$), the
%%luminosity density of quasars seen in optical surveys was sufficient
%to have produced the locally measured black hole mass density.  This
%result left little room for any additional obscured accretion (and
%hence black hole growth) as required by XRB synthesis models
%\citep[e.g.][]{gilli01}. However, such studies are dependent on two
%crucial assumptions; the correction factor used to convert optical
%luminosity to bolometric luminosity and the local black hole mass
%density. More recent studies have revisited this question with a
%revised set of assumptions.  For example, the study of
%\citet{marconi04} uses a more detailed bolometric correction method,
%and a higher estimate of the local black hole mass density than that
%calculated by \citet{yu02}. 

\section{Conclusions and summary}
We have analysed the distribution of absorption in a sample of AGN
detected in the very deep \xmm\ observations of the \cdfs.
Importantly, spectroscopic or photometric redshift determinations are
available for 84\% of the X-ray sample. We determined the absorption
and intrinsic luminosity of each AGN using a novel method which takes
advantage of the high photon throughput and broad bandpass of \xmm\
EPIC.  The AGN in the \xcdfs\ sample were compared to the predictions
of a number of model AGN populations, using Monte Carlo simulations
which allow for the selection function and particulars of the \xmm\
observations.  We find no evidence for a decline in the fraction of
AGN with significant absorption at high luminosities that has been
reported by many authors.  The $N_H$ distribution models which most
closely match the pattern of absorption in the \xcdfs\ sample are
independent of redshift and luminosity.  Our sample contains at least
23 heavily absorbed AGN with QSO-like luminosities.  We postulate that
nearly half of the objects without redshift determinations are also
absorbed QSOs; the reason they are without \combo\ redshift estimates
is because they lie in the ``photo-z desert'' ($1.5 < z < 2.5$).  In
order to confirm this hypothesis, the redshifts of these optically
faint objects must be determined; photo-$z$s incorporating near- and
mid-infrared photometry should make this possible.

\section*{Acknowledgements}
Based on observations obtained with XMM-Newton, an ESA science mission
with instruments and contributions directly funded by ESA Member
States and NASA. TD acknowledges the support of a PPARC Studentship.
We thank the \combo, \cdfs, \cdfn, \ecdfs, VVDS, GEMS and GOODS teams 
for making X-ray and optical images, sourcelists and spectra
publicly available.

\appendix

\section[]{Positional accuracy of the \xmm\ detections}
\label{app:pos_diffs}
We have tested the efficacy of the X-ray position matching method used
to pair \xmm\ detections with sources in the \chandra\ catalogues, and
to optical counterparts (see section \ref{sec:other_data}). The \xmm\
imaging has been tied to the 1Ms \chandra\ sourcelist, which was in
turn tied to the optical frame, and so systematic offsets have already
been removed.  We have examined the differences between input and
output position of sources in the simulated library discussed in
section \ref{sec:library}.  We find that the fraction of output
detections having an input source within our variable matching ellipse
is 98.4\%. However, the input sources are simulated to a much fainter
limit than that reached by the faintest output detections.  Therefore,
we add the requirement that for an input source to be considered
``valid'', it must have an input 0.2--10~keV countrate of at least
half the output 0.2--10~keV countrate.  The fraction of output
detections having a ``valid'' input source within the variable
matching ellipse is 97.5\%.  Hence, for the 309 point-like detections
in the \xcdfs\ sample, we expect the true astrophysical position to be
within the matching ellipses for all but $\sim 8$ detections.

\section[]{Regions of extended emission in the \xmm\ observations of the \cdfs}
\label{app:clusters}
There are four prominent regions of diffuse soft X-ray emission in the
\xmm\ observations of the \cdfs, these are highlighted in figure
\ref{fig:soft_image}.  It is beyond the scope of this paper to
investigate these sources in depth, as this diffuse emission is most
likely due to clusters of galaxies.  We have compared the redshift
distributions of \combo\ galaxies in each of these four ellipsoidal
regions to the redshift distribution over the whole field.  Region \#2
(located at 03~32~45.8 -27~40~57.3) has a very clear peak in its
redshift distribution at $z \sim 0.75$.  Region \#4 (located at
03~31~49.7 -27~49~21.7) has several weak peaks, but with the most
prominent lying at $z \sim 0.67$. This extended source was also found
in the 1Ms \chandra\ imaging \citep[source \#645 of][]{giacconi02}.
For regions \#3 and \#1 (located at 03~32~25.8 -27~58~55.6 and
03~33~21.2 -27~48~53.4 respectively), the strongest peaks lie at $z
\sim 0.14$. In fact this redshift bin is enhanced over the entire
\combo\ field, suggesting that these diffuse objects are embedded in a
``sheet'' structure of angular extent equal to or greater than the
$0.5^{\circ} \times 0.5^{\circ}$ \combo\ field. Extended source \#1
matches CXOECDFS~J033320.3-274836 of \citet{lehmer05}.

\section[]{\xcdfs\ sources with unusual spectra}
\label{app:unusual}

We find that two of the \xcdfs\ sources are not matched to any objects
in the simulated source library, and so for these objects we are
unable to apply our $N_H/L_X$ estimation technique.  These two objects
have X-ray colours which are not well matched to the ``power law plus
reflection'' model spectrum, possibly because they have more complex
spectra.  We can estimate the number of such non-detections which are
a result of the finite length of our simulated source library
(i.e. not all of $z$,$L_X$,$N_H$,$\Gamma$ space is populated with
library sources).  In the simulated test populations there are on
average 0.8 sources per field which remained unmatched, consistent with
our finding two unmatched objects in the \xcdfs\ sample. 

The two unmatched sources are among the X-ray brightest in our sample,
they both have \chandra\ matches and bright ($R<20$) optical
counterparts. Their spectra are rather soft compared to the other
sources in the \xcdfs\ sample (see fig. \ref{fig:hardness_ratios}).
The first source is identified with a broad line AGN at z=1.031
(\cdfs-044, \citealt{giacconi02,szokoly04}).  The second source is
identified with an optically bright galaxy with photometric redshift
$0.539 \pm 0.03$ (\ecdfs-381, \citealt{lehmer05,wolf04}).  These two
unmatched sources both have $HR1 \le -0.1$, softer than the bulk of
the unabsorbed objects in the sample.  However, the two unmatched
sources have relatively normal HR2 and HR3 values, indicating that
their spectral slopes flatten toward higher energies.  For the
purposes of this study, we assume that these two objects have zero
absorption and calculate their rest frame 2--10~keV luminosities from
their observed 2--5~keV flux assuming a photon index of 1.9.

\bsp

\label{lastpage}

\end{document}